\documentclass[aps,prb,reprint,superscriptaddress]{revtex4-1}

\usepackage{subfigure}
\usepackage{amsfonts, amssymb, amsmath}
\usepackage{bm}
\usepackage{graphics}
\usepackage{graphicx}
\usepackage{wasysym}

\usepackage{color}
\definecolor{darkblue}{rgb}{0,0,0.6}
\definecolor{darkred}{rgb}{0.6,0,0}

\usepackage[colorlinks=true,urlcolor=darkblue,citecolor=darkblue,linkcolor=darkred,hyperfootnotes=false]{hyperref}

\newcommand{\sref}[1]{Sec.~\ref{#1}}
\newcommand{\fref}[1]{Fig.~\ref{#1}}
\newcommand{\aref}[1]{Appendix~\ref{#1}}
\newcommand{\figwidth}{0.9\columnwidth}

\newcommand{\argc}[1]{\left[#1\right]}
\newcommand{\arga}[1]{\left\lbrace #1\right\rbrace }
\newcommand{\argp}[1]{\left(#1\right)}
\newcommand{\valabs}[1]{\vert #1\vert}
\newcommand{\moy}[1]{\left\langle  #1 \right\rangle }
\newcommand{\moydes}[1]{\overline{#1}}

\newcommand{\gdD}{\mathcal{D}}
\newcommand{\gdO}{\mathcal{O}}
\newcommand{\gdH}{\mathcal{H}}
\newcommand{\gdR}{\mathbb{R}}

\newcommand{\gdF}{\mathcal{F}}

\newcommand{\dbar}{d\mkern-6mu\mathchar'26 \!}
\newcommand{\deltabar}{\delta \mkern-8mu\mathchar'26}

\newcommand{\dlamb}{\dbar \lambda}
\newcommand{\dqpi}{\dbar q}
\newcommand{\dqpit}{\dbar \tilde{q}}

\newcommand{\gdHel}{\gdH_{\text{el}}}
\newcommand{\gdHdis}{\gdH_{\text{dis}}}
\newcommand{\gdHtil}{\widetilde{\mathcal{H}}}
\newcommand{\gdHelt}{\widetilde{\gdH}_{\text{el}}}
\newcommand{\gdHdist}{\widetilde{\gdH}_{\text{dis}}}

\newcommand{\sigc}[1]{\argc{\sigma}(#1)}

\begin{document}

\title{Temperature-induced crossovers in the static roughness of a one-dimensional interface}

\author{Elisabeth Agoritsas}
\email[]{Elisabeth.Agoritsas@unige.ch}
\affiliation{DPMC-MaNEP, University of Geneva, 24 Quai Ernest-Ansermet, 1211 Geneva 4, Switzerland}
\author{Vivien Lecomte}
\affiliation{DPMC-MaNEP, University of Geneva, 24 Quai Ernest-Ansermet, 1211 Geneva 4, Switzerland}
\affiliation{Laboratoire de Probabilit\'es et Mod\`eles Al\'eatoires (CNRS UMR 7599), Universit\'e Paris Diderot, 2 Place Jussieu, 75251 Paris cedex 05, France}
\author{Thierry Giamarchi}
\affiliation{DPMC-MaNEP, University of Geneva, 24 Quai Ernest-Ansermet, 1211 Geneva 4, Switzerland}

\date{\today}


\begin{abstract}

At finite temperature and in presence of disorder, a one-dimensional elastic interface displays different scaling regimes at small and large lengthscales. 
Using a replica approach and a Gaussian Variational Method (GVM), we explore the consequences of a finite interface width $\xi$ on the small-lengthscale fluctuations. We compute analytically the static roughness $B(r)$ of the interface as a function of the distance $r$ between two points on the interface. We focus on the case of short-range elasticity and random-bond disorder.
We show that for a finite width $\xi$ two temperature regimes exist. At low temperature, the expected thermal and random-manifold regimes, respectively for small and large scales, connect via an intermediate `modified' Larkin regime, that we determine.  This regime ends at a temperature-independent characteristic `Larkin' length.
Above a certain `critical' temperature that we identify, this intermediate regime disappears. The thermal and random-manifold regimes connect at a single crossover lengthscale, that we compute. This is also the expected behavior for zero width.
Using a directed polymer description, we also study via a second GVM procedure and generic scaling arguments, a modified toy model that provides further insights on this crossover.  We discuss the relevance of the two GVM procedures for the roughness at large lengthscale in those regimes. In particular we analyze the scaling of the temperature-dependent prefactor in the roughness $B(r)\sim T^{2 \text{\thorn}} r^{2 \zeta}$ and its corresponding exponent $\text{\thorn}$.
We briefly discuss the consequences of those results for the quasistatic creep law of a driven interface, in connection with previous experimental and numerical studies.

\end{abstract}


\maketitle


\section{Introduction} \label{intro}

Almost everyone has already unwittingly spilt coffee on her/his work
table and observed the inexorable progression of the liquid
into her/his favorite article.
However inconvenient may be the consequences of such a simple home
experiment, it shares in fact a lot of common physical features with a
variety of systems and phenomena ranging from
ferromagnetic\cite{lemerle_1998_PhysRevLett80_849,metaxas_2007_PhysRevLett99_217208} and
ferroelectric\cite{tybell_2002_PhysRevLett89_097601,paruch_2005_PhysRevLett94_197601} domain
walls, to growth surfaces\cite{krim_1995_IntJModPhysB9_599}, contact line in
wetting experiments\cite{moulinet_2002_EurPhysJE8_437,ledoussal_2009_EPL87_56001}, or crack propagation
in paper\cite{alava_2006_RepProgPhys69_669}.
All those systems display different coexisting phases separated by an
interface, whose shape and dynamics are determined from two competing tendencies:
the elastic cost of the interface which tends to flatten it while
disorder in the environment induces deformations adapting the
interface to the local energetic valley and hills.

To describe these phenomena, a successful theoretical approach is that
of disordered elastic systems\cite{kardar_1998_PhysRep301_85,giamarchi_1998_Young321}
(DES), in which the bulk details of the interface are summarized in
its mere position, seen as a fluctuating manifold whose energy is the
sum of elastic and disorder contributions.
Such a description also encompasses periodic systems such as charge
density waves\cite{gruner_1988_RevModPhys60_1129,brazovskii_2004_AdvPhys53_177} or vortex
lattices\cite{blatter_1994_RevModPhys66_1125} occurring in type II superconductors.
This approach accounts for a complex free-energy landscape which
exhibits metastability and explains the dynamical glassy properties --~such as
hysteresis, creep or ageing~-- observed in experimental realizations.
Such systems display similar features
independently of the scale at which they are observed: in other words,
they present ranges of scale on which they are statistically
scale-invariant\cite{barabasi_book}.
A simple way to characterize this property is to study the
\emph{roughness} of the interface (denoted $B(r)$), defined as the
variance of the relative displacements of the manifold at a given
lengthscale $r$.
Scale invariance translates into having the roughness behaving as a
power law $B(r)\sim r^{2\zeta}$ characterized by a roughness exponent
$\zeta$.

In addition to its direct experimental relevance e.g. in ferromagnetic
domain walls, the 1D interface shares many universal features with
other distinct physical problems, such as the directed polymer (DP) in
random media (for reviews,
see Refs.~\onlinecite{halpin_zhang_1995_PhysRep254,book_polymers_Giacomin}),
the noisy Burgers' equation in
hydrodynamics\cite{forster_nelson_stephen_1977_PhysRevA16_732,book-burgers}, or bosons with
attractive hard-core interaction in one
dimension\cite{kardar_1987_NuclPhysB290_582}.  These systems are all falling in
the so-called Kardar-Parisi-Zhang (KPZ) class \cite{kardar_1986_originalKPZ_PhysRevLett56_889,kardar_1987_PhysRevLett58_2087}.
This problem has now been studied for many decades, the attention being
focused on the large-scale --~or so-called `random-manifold' (RM)~--
properties of the roughness: many approaches have been used to
establish the non-trivial exponent $\zeta_{\text{RM}}=2/3$ for
$B(r)\sim r^{2\zeta}$ in the large $r$ limit,
ranging from dynamical renormalization group\cite{forster_nelson_stephen_1977_PhysRevA16_732,ioffe_vinokur_1987_JPhysC20_6149}, to
hidden symmetries\cite{huse_henley_fisher_1985_PhysRevLett55_2924} and Bethe Ansatz computations\cite{kardar_1987_NuclPhysB290_582}.

In spite of the versatility of the previous studies, there has been a
recent uprise of interest in different directions: from a mathematical
point of view it is only very recently\cite{balazs_arXiv:0909.4816} that the
$2/3$ RM 
exponent has been proven, the method itself making the link with
another class of systems --~transport models known as asymmetric
exclusion processes.  Besides, in a recent series of works in the
mathematics\cite{sasamoto_2010_NuclPhysB_834_523,amir_arXiv:1003.0443} and
physics\cite{dotsenko_2010_EPL90_20003,calabrese_2010_EPL90_20002} communities, it has been
shown that the free-energy distribution for the polymer endpoint is
obtained from the Tracy-Widom distribution, when the random potential
is delta-correlated. Similarly, another interesting and related
question which seems to have been neglected up to recently is the role
of temperature in the large scale behaviour of the roughness: although
physically one could \textit{a priori} expect the roughness to be
temperature-independent at large lengthscales, since it is dominated by the
disorder, it happens that scaling arguments show this shall not be the case
at small enough disorder correlation length $\xi$, and that there is a subtle
interplay between low temperature and small $\xi$
limits\cite{bustingorry_arXiv:1006.0603,calabrese_2010_EPL90_20002}.

Despite these recent progress several questions remain. In particular, in these works, 
the width of the interface or the disorder correlation
length $\xi$ is assumed to be zero. It is thus important to ascertain how keeping $\xi$ finite
--~as always the case in physical realizations~-- influences the
physics at stake.
The role of $\xi$ in the large scale limit is exemplified specially in
the context of functional renormalization
group (FRG)\cite{fisher_1986_PhysRevLett56_1964,chauve_2000_ThesePC_PhysRevB62_6241}, where at large
scale the roughness arises from the zero-temperature fixed point of
the renormalization flow, hence displaying no temperature dependence~-- in contrast to
pure scaling predictions (in other words, having $\xi$ finite
seems to influence temperature scaling exponents at large
distance\cite{bustingorry_arXiv:1006.0603}).

In this paper, we address the issue of the role of $\xi$ concerning the roughness scaling
and the temperature regimes. We focus on the roughness of a one-dimensional static
interface subjected to `random-bond' disorder, at finite
temperature. It is known that at short lengthscale disorder plays no
role and the interface is in a thermal regime
($\zeta_{\text{th}}=\frac 12$), while at large lengthscale the
interface is in the RM regime, characterized
by a roughness exponent $\zeta_{\text{RM}}=\frac 23$
(see~Refs.~\onlinecite{forster_nelson_stephen_1977_PhysRevA16_732,huse_henley_fisher_1985_PhysRevLett55_2924,ioffe_vinokur_1987_JPhysC20_6149}). We examine in
detail the way the roughness crosses over from thermal to RM
behaviour, and in particular whether there
is one or several crossover lengthscales, possibly allowing for a
non trivial `intermediate regime'.
We tackle these problems by using a Gaussian variational method
(GVM)\cite{giamarchi_ledoussal_1995_PhysRevB52_1242,mezard_parisi_1991_replica_JournPhysI1_809,ledoussal_2008_PhysRevB77_064203,book_beyond-MezardParisi}.
Although this method is only approximate, it allows rather complete calculations 
for the roughness $B(r)$. The alternative methods are not so well suited
to analyse this issue for a one dimensional interface. Even though numerical methods 
such as Monte-Carlo\cite{yoshino_1996_JPhysA29_1421}, Langevin\cite{bustingorry_2008_epl-81-2-26005} or transfer
matrix\cite{medina_1992_PhysRevB46_9984} methods are efficient to deal with the interface, one 
would need to implement them for very large system sizes, in order to tackle the issue of crossover lengthscales. As for the functional renormalization
group (FRG) approach\cite{fisher_1986_PhysRevLett56_1964,chauve_2000_ThesePC_PhysRevB62_6241}, it is based on an expansion 
in $\epsilon=4-d$ for a manifold
of internal dimension $d$ and may not be suited for our case of interest ($d=1$).

The plan of the paper is as follows. In \sref{model} we
introduce the DES description of a 1D interface. We show how
the replica trick allows to make the average over disorder. In
\sref{fullGVM} we detail the variational procedure, originally
introduced by M{\'e}zard and Parisi\cite{mezard_parisi_1991_replica_JournPhysI1_809}, that we use in the paper to
obtain an approximation of the various physical properties of
the interface, and in particular the correlation function
$B(r)$. The corresponding roughness is computed in
\sref{full_roughness}, along with the
crossover lengthscales separating its different power-law
regimes, and their temperature dependence. In \sref{toymodel}
we present a second GVM procedure, this time in direct-space
representation, on a `toy model' which has been
argued to be an effective model for the study of a directed polymer's
end-point fluctuations, and on which the model of a 1D
interface can carefully be mapped. 
In \sref{scaling_analysis} we examine how generic scaling arguments
shed light on the interface and the toy-model results.
The two GVM results are
compared and discussed in \sref{discussion},
and we finally conclude in \sref{conclusion}.
We present for reference standard Flory arguments in \aref{A-Flory},
and we recall in \aref{A-RSB_inversion} some useful properties of
hierarchical matrices.


\section{Model of a 1D interface} \label{model}

\subsection{Model} \label{model_model}

\begin{figure}
 \subfigure[\label{fig:description_interface}]{\includegraphics[width=0.6\columnwidth]{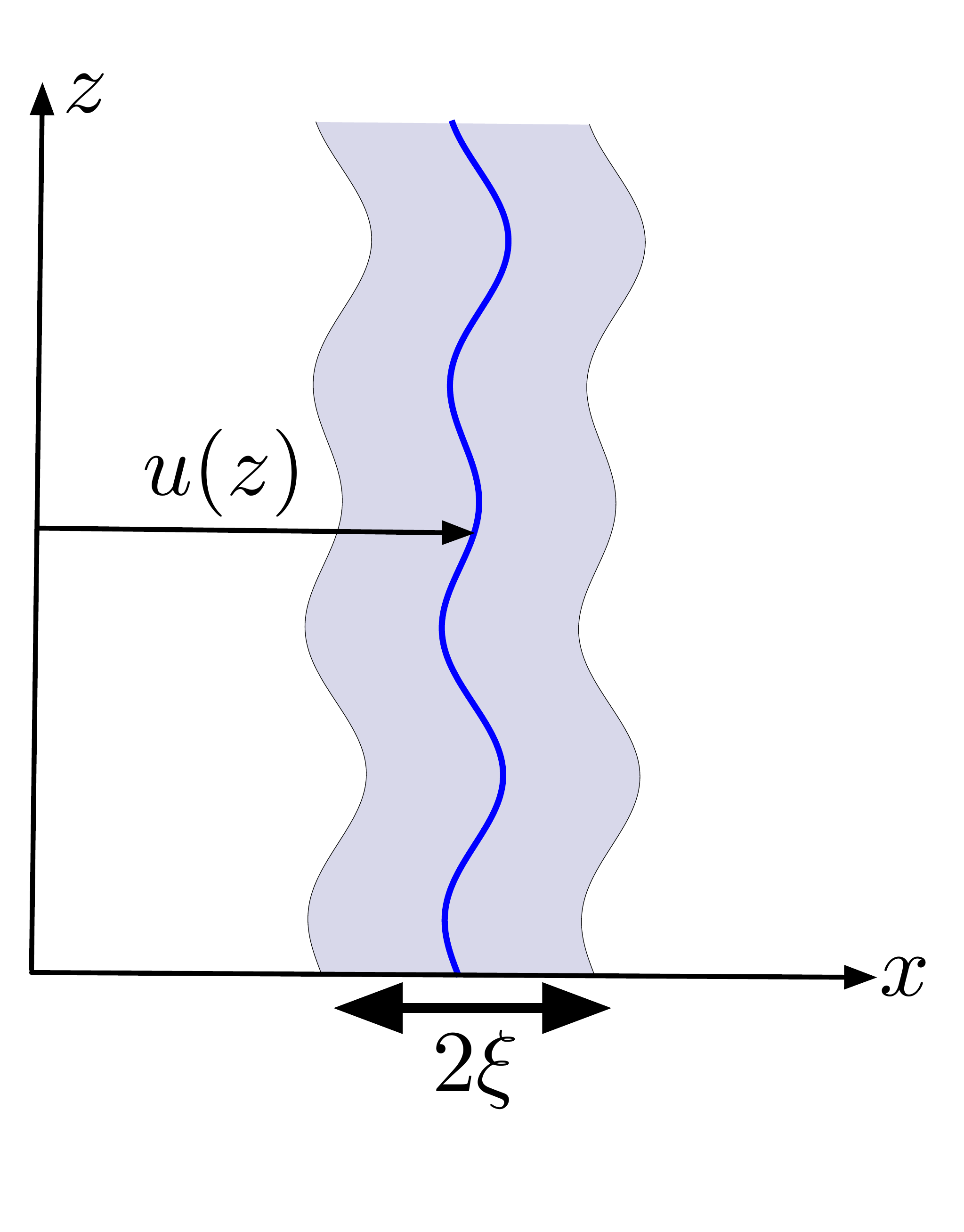}}
 \subfigure[\label{fig:density3D}]{\includegraphics[width=\columnwidth]{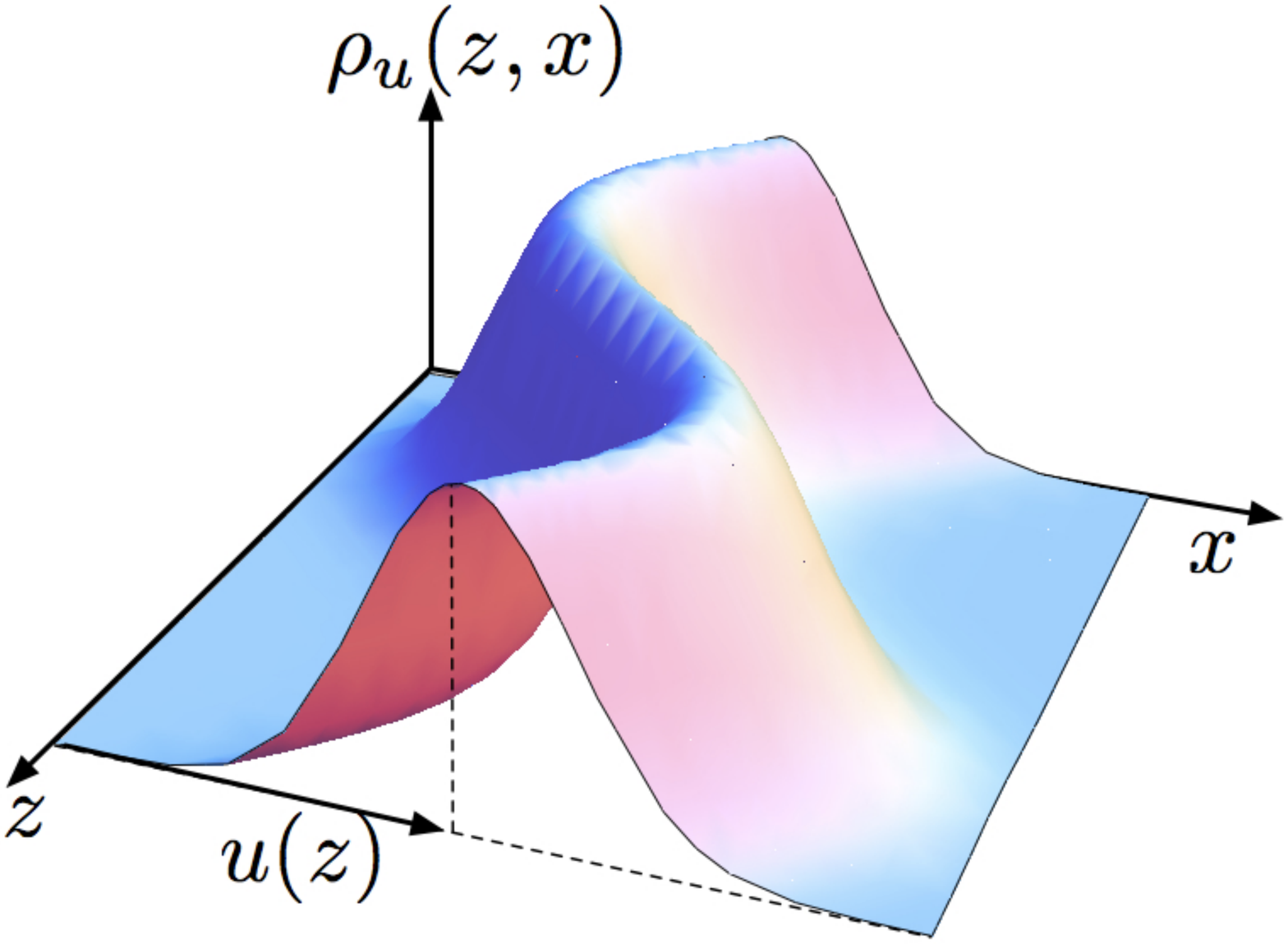}}
 \caption{
	\subref{fig:description_interface} Displacement from a given flat configuration described by the $z$ axis.
	\subref{fig:density3D} Gaussian density $\rho_u (z,x)$ centered on the position $(z,u(z))$ of the interface.}
\end{figure}

In the DES framework, an interface can be described as an
elastic manifold of dimension $d$ with $m$ transverse
components, submitted to the random potential of a physical
space of dimension $D=d+m$. The coordinates in this space can
thus be split between the \emph{internal} and \emph{transverse}
coordinates of the interface, respectively $\argp{z,x} \in
\gdR^d \times \gdR^m = \gdR^D$. In the case $d=m=1$ (simply
denoted `1+1' for DPs), a one-dimensional interface is thus
described as an elastic line living in a bidimensional plane,
with a disordered energy landscape that will be defined below.
Restricting ourselves to the case without bubbles nor
overhangs, each configuration of the interface can be indexed
by a univalued displacement field $u(z)$ which parametrizes the
position $\argp{z,u(z)}$ of the interface along its internal
direction $z$. This is schematically indicated in
\fref{fig:description_interface}.

In disordered systems there is no strict spatial translational
invariance, but it can be recovered statistically once the
disorder is averaged out. Thereafter, we will thus mostly work
in Fourier-transform representation and denote the Fourier
modes along $z$ and $x$ respectively $q$ and $\lambda$. However
there are in fact both infra-red and ultra-violet cutoffs in
those Fourier modes, since a physical realization of an
interface lives in a finite bidimensional plane of typical size
$L$ and is supported by a sublattice (e.g. a crystal in a
solid) whose spacing $1/\Lambda$ defines the smallest physical
lengthscale in the system. The cutoffs $1/L< q,\lambda
<\Lambda$ will be conveniently reintroduced whenever needed to
cure non-physical divergences in Fourier-transform integrals.
Thereafter we use the notation $\dbar^d q \equiv \frac{d^d q}{(
2 \pi)^d}$, as well as $\deltabar^{(d)} (q) \equiv (2\pi)^d
\delta^{(d)} (q)$, and similarly for other Fourier modes.

More importantly, in order to implement a finite interface width $\xi$ into this model, we introduce the density of the interface $\rho_{u} \argp{z,x}$ which is positive and
normalized at fixed $z$ by $\int_{\gdR} dx \cdot \rho_{u} \argp{z,x}=1$,
decreases significantly for $\valabs{x-u(z)}>\xi$
and tends to a Dirac $\delta$-function when $\xi \to 0$. As schematized in \fref{fig:density3D} we
choose the simplest smooth density which is a Gaussian centered
on $(z,u(z))$, of uniform standard deviation $\xi$, whose
Fourier representation is thus:
\begin{equation}
 \label{density1D}
	\rho_{u} (z,x)
	= \int_{\gdR} \dlamb \cdot e^{i \lambda \argp{x-u(z)}} e^{-\lambda^2 \xi^2 /2}
\end{equation}

The interface is subject to a random potential $V(z,x)$ whose
distribution is Gaussian and uncorrelated in space. This
corresponds to the limit of the collective pinning by many weak
impurities. Denoting the statistical average over disorder by
an overline, since it is Gaussian it is fully defined by:
\begin{equation}
 \label{distrV}
	\moydes{V(z,x)}=0,
	\, \moydes{V(z,x) \, V(z',x')} = D \cdot \delta (z-z') \, \delta (x-x')
\end{equation}
where $D$ is the strength of disorder.

We can now construct the DES Hamiltonian of the interface,
which is the sum of the energetic cost of its distortions
$\gdHel \argc{u}$ and the contribution of the disordered energy
landscape $\gdHdis \argc{u,V}$. We assume that the elastic
limit $\valabs{\nabla_z u(z)} \ll 1$ is realized and that the elasticity
is short range, so that the energy per Fourier mode
$u_q \equiv \int_{\gdR} dz \cdot u(z) \, e^{i q z}$ is $c q^2$,
where $c$ is the elastic constant. The mode $u_{q=0}$
corresponds to the mean position of the interface, and
introduces an additive constant in $\gdHel$ which will
disappear in the Boltzmann weight and moreover does not
contribute to the roughness. It can thus be put equal to zero
directly in the elastic Hamiltonian by the \emph{ad hoc}
redefinition of $u$. We assume a random-bond disorder, i.e.
that the interface couples locally to the random potential.
Thus the full Hamiltonian $\gdH$ is given by:
\begin{align}
 \label{Ham_DES}
	\gdH \argc{u,V}
	&= \gdHel \argc{u} + \gdHdis \argc{u,V} \\
 \label{Ham_el}
	\gdHel \argc{u}
	&= \frac{c}{2} \int_{\gdR} dz \cdot \argp{\nabla_z u(z)}^2
	= \frac{1}{2} \int_{\gdR} \dqpi \cdot c q^2 \, u_{-q} u_q \\
 \label{Ham_dis}
	\gdHdis \argc{u,V}
	&= \int_{\gdR^2} dz \, dx \cdot \rho_{u} (z,x) \, V(z,x)
\end{align}

Note the alternative formulation of the $\gdHdis$ with an
effective random potential $\widetilde{V}$ coupled to a
zero-width interface:
\begin{equation}
 \label{Ham_dis_bis}
	\gdHdis \argc{u, \widetilde{V}} = \int_{\gdR} dz \cdot \widetilde{V} \argp{z,u(z)}
\end{equation}
and its associated disorder distribution
\begin{equation}
 \label{distrV_bis}
	\moydes{\widetilde{V} (z,x)}=0, \,
	\moydes{\widetilde{V} (z,x) \, \widetilde{V}(z',x')} = \delta (z-z') \, R_{\xi}(x-x')
\end{equation}
where $R_{\xi} (u)$ is up to an additive constant
the usual correlator of the disorder which is the key function
renormalized in FRG procedures on DES \cite{chauve_2000_ThesePC_PhysRevB62_6241,ledoussal_2008_PhysRevB77_064203}. It corresponds to the
overlap between two densities $\rho_u$ in our formulation and
encodes the statistical translational invariance of the random
potential. Note that $R_{\xi} (u_z,u'_z) \equiv D \int_{\gdR}
dx \cdot \rho_{u} (z,x) \rho_{u'} (z,x)$ inherits both the
translational invariance and the symmetry of the density.
The Gaussian density \eqref{density1D} translates into the following Gaussian
correlator:
\begin{equation}
 \label{disorder_correlator}
	R_{\xi} \argp{u-u'}
	= D \cdot \int_{\gdR} \dbar \lambda \cdot e^{i \lambda \argp{u-u'}} e^{-\lambda^2 \xi^2}
\end{equation}

So the parameter $\xi$ can be seen either as the width of the
interface or as the disorder correlation length (or a convolution of both).
In previous
GVM computations on DES
\cite{mezard_parisi_1991_replica_JournPhysI1_809,ledoussal_2008_PhysRevB77_064203} the
correlator $R(u)$ was assumed to exhibit an asymptotic power-law behaviour
whose exponent depended on the university class of the disorder.
Here we focus on the role of a finite $\xi$ rather than on this
asymptotic behaviour, following closely a similar approach of
periodic DES \cite{giamarchi_ledoussal_1995_PhysRevB52_1242}.

\subsection{Statistical averages} \label{model_statistical_average}

The static properties of an interface are accessible by
averaging over its thermal fluctuations and the stochastic
variable $V$ which is associated to each configuration of
quenched disorder, so two statistical averages have to be
successively performed for a given observable $\gdO$. The first
one is the thermal average at fixed disorder $\moy{\gdO}_V$
using the Boltzmann weight $e^{- \beta \gdH \argc{u,V}}/Z_V$,
where $\beta=1/T$ is the inverse of the temperature (taking the
Boltzmann constant $k_B=1$) and $Z_V$ the canonical partition
function:
\begin{align}
 \label{partition_functionV}
	Z_V
	&= \int \gdD u \cdot e^{-\beta \gdH \argc{u,V}} \\
 \label{def_moyth}
	\moy{\gdO}_V
	&= \frac{1}{Z_V} \int \gdD u \cdot \gdO \argc{u} \cdot e^{-\beta \gdH \argc{u,V}}
\end{align}
where the functional integral $\int \gdD u$ sums over all
possible configurations $u$ of the interface.

The second one is the average over disorder $\moydes{\gdO}$
which has already been introduced in \eqref{distrV}, assuming
that the system is large enough to be disorder self-averaging
(the equivalent of ergodicity for thermal averages)\cite{book_beyond-MezardParisi}. To recover
a translational invariance and be able to work in Fourier
space, one would like to average technically first over
disorder. This can in fact be done using the well known replica
trick \cite{book_beyond-MezardParisi}. Indeed, introducing $n$
replicas of the partition function at fixed disorder ${Z_V}$,
$n$ being an arbitrary integer, we can replace $1/Z_V$ in the
thermal average \eqref{def_moyth} by $\lim_{n \to 0}
{Z_V}^{n-1}$ under the (strong) assumption that at the end of
our computations the analytical continuation $n \to 0$ is
well-defined and physically meaningful. This gives:
\begin{equation}
	\moy{\gdO}_V
	= \lim_{n \to 0} \int \gdD u_1 (\cdots) \gdD u_n \cdot \gdO \argc{u_1} \cdot e^{- \beta \sum_{a=1}^{n} \gdH \argc{u_a,V} }
\end{equation}
Using the linearity in $V$ of $\gdHdis$ \eqref{Ham_dis} and the
Gaussian distribution of disorder \eqref{distrV}, we can then
explicitly average over disorder and define an effective
replicated Hamiltonian $\gdHtil$ which couples all the replicas
($\vec{u} \equiv \argp{u_1, \dots, u_n}$):
\begin{equation}
 \label{moyth_rep}
 \begin{split}
	\moydes{\moy{\gdO}}
	&= \lim_{n \to 0} \int \gdD u_1 (\cdots) \gdD u_n \cdot \gdO \argc{u_1} \cdot \moydes{e^{- \beta \sum_{a=1}^{n} \gdH \argc{u_a,V}}} \\
	&\equiv \lim_{n \to 0} \int \gdD u_1 (\cdots) \gdD u_n \cdot \gdO \argc{u_1} \cdot e^{-\beta \gdHtil \argc{\vec{u}}}
\end{split}
\end{equation}
where
\begin{equation}
 \label{Ham_DES_effectif}
	\gdHtil \argc{\vec{u}}
	= \sum_{a=1}^{n} \gdHel \argc{u_a} - \frac{\beta}{2} \int_{\gdR} dz \cdot \sum_{a,b=1}^{n} R_{\xi} \argp{u_a (z)-u_b(z)}
\end{equation}
We have thus reformulated the problem of one interface in a
random potential $V$ into a system of $n$ coupled interfaces without
disorder, in the limit $n \to 0$.

Finally, using \eqref{Ham_el} and \eqref{disorder_correlator},
we explicit the effective replicated Hamiltonian which is exact
up to this point, diagonal in Fourier space due to
translational invariance and depends only on the parameters $\arga{\xi,c,D,T}$:
\begin{equation}
 \label{Hrepl_exact}
 \begin{split}
	\gdHtil \argc{\vec{u}}
	&= \gdHelt \argc{\vec{u}} + \gdHdist \argc{\vec{u}} \\
	\gdHelt \argc{\vec{u}}
	&= \frac{1}{2} \int_{\gdR} \dqpi \cdot c q^2 \sum_{a=1}^n u_a (-q) u_a^{} (q) \\
	\gdHdist \argc{\vec{u}}
	&= - \frac{\beta D}{2} \int_{\gdR} \dlamb \cdot e^{-\lambda^2 \xi^2} \int_{\gdR} dz \cdot \sum_{a,b=1}^n e^{i \lambda \argp{u_a (z) - u_b (z)}}
 \end{split}
\end{equation}

\subsection{Roughness and displacement correlation function} \label{model_roughness}

In order to study the static fluctuations of the position of
the interface, we compute the variance of the relative
displacements of two points of the interface: $B (z_1,z_2)
\equiv \moydes{\moy{\argp{u(z_1)-u(z_2)}^2}}$. The disorder
average leads back to translational invariance
$B(z_1,z_2)=B(z_1-z_2,0)$.  One can thus define the roughness as a
function of the lengthscale $r$, formally the two-point
correlation function of our system:
\begin{equation}
 \label{def_roughness_Br}
	B(r) \equiv \moydes{\moy{\argp{u(r)-u(0)}^2}}
\end{equation}
which is the Fourier transform of the structure factor $S(q)$:
\begin{equation}
 \label{def_structure_factor}
 \begin{split}
	S(q) &\equiv \int \dqpit \cdot \moydes{\moy{u_{-\tilde{q}} u_q^{}}} \\
	B(r) &= \int \dqpi \cdot 2 \argp{1-\cos \argp{q r}} S(q)
 \end{split}
\end{equation}

If the DES displays a scale invariance in its fluctuations, the roughness is expected to follow a power-law behaviour $B(r) \sim r^{2 \zeta}$, with a corresponding \emph{roughness exponent} $\zeta$.
To describe the possible interplay between different terms in $B(r)$ we generalize the definition of $\zeta$:
\begin{equation}
 \label{def_roughness_exponent_loglog}
	\zeta(r) \equiv \frac{1}{2} \frac{\partial \log B(r)}{\partial \log r}
\end{equation}
whose different values characterize the different regimes of fluctuations depending on the lengthscale considered. Along with the $r$-independent prefactor of $B(r)$ it probes the physics at different lengthscales and a roughness regime is actually defined by a constant value of $\zeta$.

If the exact replicated Hamiltonian could be put into a
quadratic replicated form, diagonal in Fourier space, such as:
\begin{equation}
 \label{def_quadr_repli_Ham}
	\gdH_0 \argc{\vec{u}} = \frac{1}{2} \int \dqpi \sum_{a,b=1}^{n} u_a (-q) G^{-1}_{ab} (q) u_b (q)
\end{equation}
then the corresponding structure factor would be:
\begin{equation}
 \label{quadr_structure_factor}
	S_0 (q)
	=\beta^{-1} \cdot \lim_{n \to 0} G_{aa} (q)
\end{equation}
%
In the absence of disorder ($\gdH = \gdHel$) the replicas are
uncoupled and $G^{-1}_{ab} (q)= c q^2 \cdot \delta_{ab}$, so
the structure factor is proportional to the `thermal
propagator' $\frac{1}{c q^2}$ which leads to the purely thermal
roughness of a 1D interface $B_{\text{th}} (r)= \frac{T
r}{c}$. The exact Hamiltonian \eqref{Hrepl_exact} is diagonal
in Fourier space, but it cannot be put into a quadratic form
because of the coupling of all replicas $\sum_{a,b=1}^n e^{i
\lambda \argp{u_a (z) - u_b (z)}}$. This forbids the direct use
of \eqref{quadr_structure_factor} for the computation of
$S(q)$. To be able to do so, we thus approximate $\gdHtil$ in
the Boltzmann weight \eqref{moyth_rep} by a quadratic
replicated Hamiltonian $\gdH_0$ optimized by GVM, as explained
in the next section.


\section{GVM and full-RSB Ansatz}\label{fullGVM}

The GVM has already been applied to DES, periodic systems
\cite{giamarchi_ledoussal_1995_PhysRevB52_1242} as well as manifolds
\cite{mezard_parisi_1991_replica_JournPhysI1_809,ledoussal_2008_PhysRevB77_064203}, to study the
temperature dependence of observables at thermodynamic
equilibrium. Here we extend these computations specifically to
the case of a one-dimensional interface of finite width $\xi$,
in order to explore its small-lengthscales behaviour.

We follow in this derivation the main steps outlined in
Ref.~\onlinecite{giamarchi_ledoussal_1995_PhysRevB52_1242} for periodic systems. One
important difference comes from the fact that in our case the
variable $\lambda$ in \eqref{Hrepl_exact} is continuous, while
it takes discrete values for periodic systems. As we will see
this has drastic consequences for the physical properties of
the system as well as for the calculation itself.

\subsection{GVM in Fourier representation} \label{fullGVM_introGVM_Fourier}

The variational method consists in replacing, in the Boltzmann
weight of statistical averages, the exact Hamiltonian $\gdHtil$
or more generally the exact action of a system by a trial
Hamiltonian $\gdH_0$  with variational parameters. The
criterion chosen to optimize this approximation is given by the
Gibbs-Bogoliubov inequality, which states that the free energy
$\mathcal{F}$ of a system is minimum at equilibrium, i.e. when
the probability measure of the system is precisely described by
its exact Boltzmann weight:
\begin{equation}
 \label{Gibbs-Bogoliubov_inequality}
	\gdF \leq \gdF_{\text{var}} \equiv \gdF_0 + \moy{\gdHtil - \gdH_0}_0
\end{equation}
where $\gdF_{\text{var}}$ is the variational free energy
associated to the trial Hamiltonian $\gdH_0$ and which has to be
minimized, $\gdF$ and $\gdF_0$ the free energies corresponding
respectively to $\gdHtil$ and $\gdH_0$ (they are defined with
respect to their corresponding partition function $Z \equiv
e^{-\beta \gdF}$), and $\moy{\gdO}_0$ the statistical average
defined over $\gdH_0$.

For the replicated Hamiltonian \eqref{Hrepl_exact}, the trial
Hamiltonian $\gdH_0$ is chosen quadratic of the generic form
\eqref{def_quadr_repli_Ham}, parametrized\cite{mezard_parisi_1991_replica_JournPhysI1_809}
by a $n \times n$ matrix $G^{-1}_{ab} (q)$.
Its inverse matrix gives thus directly access to the correlation functions:
\begin{equation}
 \label{def_Greenfunction}
	\moy{u_a \argp{-\tilde{q}} u_b (q)}_0
	= \beta^{-1} \cdot G_{ab} (q) \cdot \deltabar \argp{\tilde{q} - q}
\end{equation}
and in particular to the structure factor $S(q)$ and the roughness $B(r)$ itself. The minimization of $\gdF_{\text{var}}$ with respect to the variational parameters $G_{ab} (q)$ (the Green function of $\gdH_0$) gives a saddle point equation for the optimal matrix $G^{-1}_{ab} (q)$. Besides, the replica trick constrains the structure of $G^{-1}_{ab}$, which must be a \emph{hierarchical} matrix. In \aref{A-RSB_inversion} we recall some useful properties of such matrices, including their inversion formulas in the limit $n \to 0$ which will be used extensively thereafter.

The extremalization condition $\partial \gdF_{\text{var}} / \partial G_{ab} (q) = 0$ can be reformulated as:
\begin{equation}
 \label{def_sigma_ab}
 \begin{split}
	G^{-1}_{ab} (q) &= c q^2 \cdot \delta_{ab} - \sigma_{ab} \\
	\sigma_{ab} & \equiv - \beta \frac{\partial}{\partial G_{ab} (q)} \moy{\gdHdist}_0
 \end{split}
\end{equation}
Note that $\sigma_{ab}$ is independent of the Fourier mode $q$ since $\gdHdist$ is itself purely local in $z$.

We point out that although the GVM approach is only approximate in our context, it becomes exact\cite{ledoussal_2008_PhysRevB77_064203} in the limit of an infinite number of transverse components $m\to\infty$. The extremalization equations then appear as genuine saddle-point equations with $1/m$ playing the role of a small parameter. By extension we extensively use thereafter this improper denomination to refer to \eqref{def_sigma_ab}.
Furthermore, one could for completeness check the stability of the GVM solution by considering its associated Hessian matrix \cite{mezard_parisi_1991_replica_JournPhysI1_809}. This problem can be quite complicated, so we simply check the physical consistency of our results in what follows.

\subsection{Saddle point equation in the full-RSB formulation} \label{fullGVM_saddle_pt_equation}

To compute explicitly $\sigma_{ab}$ \eqref{def_sigma_ab} we start from \eqref{Hrepl_exact} and performing the Gaussian statistical average
\begin{equation}
	\moy{e^{i \lambda (u_a (z)- u_b (z)}}_0 = e^{-\frac{\lambda^2}{2} \moy{(u_a (z) - u_b (z))^2}_0}
\end{equation}
we have:
\begin{equation}
 \label{Gauss_moy_Hdis}
	\moy{\gdHdist}_0 = -\frac{\beta D}{2} \int_{\gdR} dz \int_{\gdR} \dlamb \cdot  \sum_{a,b=1}^{n} e^{- \lambda^2 (\xi^2 + \frac{1}{2}\moy{\argp{u_a(z) - u_b(z)}^2}_0)}
\end{equation}
where the variance of the relative displacement between two replicas at fixed $z$ is obtained by applying \eqref{def_Greenfunction} on its Fourier-transform representation. This gives the translational-invariant quantity:
\begin{equation}
 \label{Gauss_moy_2replica}
	\moy{\argp{u_a (z) - u_b (z)}^2}_0
	= T \int_{\gdR} \dqpi \argp{G_{aa} (q) + G_{bb} (q) - 2 G_{ab} (q)}
\end{equation}
Performing $\partial / \partial G_{ab} (q)$ and denoting $G_{aa} \equiv \widetilde{G}$ we eventually obtain $\sigma_{ab}$, separating off-diagonal $a \neq b$ terms
\begin{equation}
 \label{saddle-pt-equation_ab}
	\sigma_{a \neq b}
	= \frac{D}{T} \int_{\gdR} \dlamb \cdot \lambda^2 \cdot e^{- \lambda^2 \argp{ \xi^2 +
	T \int \dqpit \argp{\widetilde{G} (q) - G_{a \neq b} (q)}}}
\end{equation}
from the diagonal $a=b$:
\begin{equation}
 \label{sigmatilde_ab}
	\sigma_{aa} = - \sum_{a'} \sigma_{a' \neq a} \equiv \tilde{\sigma}
\end{equation}
Note that $\moy{\gdHdist}_0$ is extensive in the interface size (see~\eqref{Gauss_moy_Hdis})
whereas $\sigma_{a b}$ is intensive as expected since $\int dz$ has disappeared.

We can compute the connected part of the hierarchical matrix $\widehat{G}^{-1} (q)$:
\begin{equation}
 \label{GVM-Gc-1}
	G^{-1}_c (q) \equiv \sum_{a'} G^{-1}_{aa'} (q)
	= c q^2 - \tilde{\sigma} - \sum_{a'} \sigma_{a' \neq a}
	= c q^2
\end{equation}
Using \eqref{A_inversion_connected_part} we know that:
\begin{equation}
 \label{GVM-Gc-2}
	G_c (q) = \frac{1}{c q^2}
\end{equation}
In this GVM framework the connected part of the hierarchical matrices $\widehat{G}^{-1} (q)$ and $\widehat{G} (q)$ have thus a straightforward physical meaning: in the Hamiltonian it represents the elastic energy per Fourier mode $G^{-1}_c (q)= c q^2$ and in the Green function it gives back the thermal propagator $G_c (q) = \frac{1}{c q^2}$. On one hand the irruption of disorder populates by construction the off-diagonal elements of $\widehat{G}^{-1} (q)$ with the $q$-independent coupling terms $-\sigma_{a \neq b}$ under the constraint \eqref{GVM-Gc-1}. On the other hand the invariance of $G_c(q)$ is a consequence of the statistical tilt symmetry of the DES description \cite{giamarchi_ledoussal_1995_PhysRevB52_1242,schulz_1988_JStatPhys51_1, hwa_1994_PhysRevB49_3136,fisher_huse_1991_PhysRevB43_10728,chauve_2000_ThesePC_PhysRevB62_6241}.

To determine the actual propagators we consider the generic case of a full replica-symmetry breaking (RSB) Ansatz, the off-diagonal term being parametrized by $u \in \argc{0,1}$:
\begin{equation}
 \label{hierarchical_param_fullRSB}
	\widehat{G}^{-1} (q) \equiv \argp{G_c^{-1} (q)-\tilde{\sigma}, -\sigma (u)}	
	\Leftrightarrow
	\widehat{G} (q) \equiv \argp{\widetilde{G} (q), G(q,u)}
\end{equation}
where the definition of the connected part $G^{-1}_c$ is consistent with the continuous version \eqref{A_sigma_tilde} for the definition of $\tilde{\sigma}$ \eqref{sigmatilde_ab} .
The saddle point equation \eqref{saddle-pt-equation_ab} becomes:
\begin{align}
 \label{saddle-pt-equation_u_lambda}
	\sigma (u)
	&= \frac{D}{T} \int_{\gdR} \dlamb \cdot \lambda^2 \cdot e^{-\lambda^2 \argp{\xi^2 + T \int \dqpi \argp{\tilde{G} (q)-G(q,u)}} } \\
 	\label{saddle-pt-equation_u_powerlaw}
		&= \frac{D}{4 \pi T} \argc{\xi^2 + T \int_{\gdR} \dqpi \argp{\tilde{G} (q)-G(q,u)}}^{-3/2}
\end{align}
and the relation between $\argp{\widetilde{G} (q) - G(q,u)}$ and $\sigma (u)$ can be made explicit using the inversion formulas \eqref{A_replic_Gt-Gu1} and \eqref{A_replic_Gt-Gu2} and the definition \eqref{A_replic_sigc} of the self-energy $\sigc{u}$ which appears in them. $\sigma (u)$ must thus satisfy this self-consistent saddle-point equation, but we will work alternatively with $\sigma (0)$ and $\sigc{u}$.

\subsection{Determination of $\sigma (u)$ and $\sigc{u}$}\label{fullGVM_determination_ansatz}

We assume that $\sigma (u)$ is continuous by sector, and aim to point out its power-law behaviour using the following implication of the definition \eqref{A_replic_sigc}:
\begin{equation}
 \label{derivee_sigc}
	\argc{\sigma}' (u) = u \cdot \sigma ' (u)
\end{equation}
We first apply $\partial_u$ on \eqref{saddle-pt-equation_u_powerlaw}, then identify a power of $\sigma (u)$ and use \eqref{A_replic_Gt-Gu2} and \eqref{A_propagator2} to make explicit the following term:
\begin{equation}
 \label{partial_u_Gt-G}
 \begin{split}
	& \partial_u \argc{T \int_{\gdR} \dbar q \, \argc{\widetilde{G} (q) - G(q,u)}} \\
	&= T \int_{\gdR} \dbar q \cdot \partial_u \argc{\int_u^{1} \frac{\sigma '(v)}{\argp{cq^2 + \sigc{v}}^2}} \\
	&= T \int_{\gdR} \dbar q \, \frac{\sigma ' (u)}{\argp{cq^2 + \sigc{u}}^2}
	= \frac{T}{4 \sqrt{c}} \cdot \sigma ' (u) \cdot \sigc{u}^{-3/2}
 \end{split}
\end{equation}
We thus obtain the new equation:
\begin{equation}
 \label{derivee_sigma_powerlaw}
	\sigma '(u)
	= \sigma ' (u) \cdot \argp{T \, \sigma (u)}^{5/3} \, \sigc{u}^{-3/2} \, \frac{3 \pi^{1/3}}{2^{5/3} D^{2/3} c^{1/2}}
\end{equation}
We have either $\sigma ' (u)=0$ which corresponds to plateaux in $\sigma (u)$ and possibly to a full replica-symmetric (RS) solution, or if $\sigma '(u) \neq 0$ we have a strictly monotonous segment that should satisfy the new equation:
\begin{equation}
 \label{relation_sigma_sigc}
	T \, \sigma (u)
	= \sigc{u}^{9/10} \cdot 2 \cdot \argp{\frac{D^{4} c^{3}}{3^{6} \pi^{2}}}^{1/10}
\end{equation}
Note that by definition of the self-energy \eqref{A_replic_sigc} we have $\sigc{0}=0$, so the relation \eqref{relation_sigma_sigc} implies in particular that $\sigma (0)=0$, property that can be checked a posteriori by plugging our full-RSB solution directly into the initial saddle point equation \eqref{saddle-pt-equation_u_powerlaw}.

Differentiating again \eqref{relation_sigma_sigc} and using \eqref{derivee_sigc} to reintroduce a $u$ dependence,
we finally obtain for the monotonous segments of the self-energy $\sigc{u}$:
\begin{equation}
 \label{fullRSB_sigc_solution}
	\sigc{u} = \frac{3^{14}}{5^{10} \pi^2} \cdot c^3 D^4 \cdot \argp{u/T}^{10}
	\equiv A_{\argp{c,D}} \cdot \argp{u/T}^{10}
\end{equation}
It is important to note that the power-law form $\argp{u/T}^{\nu}$, which will actually condition the asymptotic temperature dependence of the roughness, is totally constrained by the GVM procedure. Indeed, we can track down the temperature factors in the previous procedure:
the extremalization condition of $\gdF_{\text{var}}$ introduces a first $\beta$ in \eqref{def_sigma_ab} via $\partial \gdF_0 / \partial G_{ab} (q) =  \beta^{-1}  G_{ab}^{-1} (q)$. The derivative involving $\moy{\gdHdist}_0$ introduces no new $T$-dependence because its overall $\beta$ factor is canceled through the derivative $\partial / \partial G_{ab} (q)$ of its exponential argument.
The only $T$ factor which remains in the relation \eqref{relation_sigma_sigc} between $\sigma (u)$ and $\sigc{u}$ eventually leads to the $\argp{u/T}^{\nu}$ dependence once the $u$ dependence is made explicit.
We will go back to that property in \sref{GVM_rescaling}, and compare its implications for the asymptotic $T$-dependence of the roughness for the 1D interface versus the prediction of another model in \sref{discussion}.

We have obtained so far that for the values of $u$ for which $\sigma (u)$ is not constant, it must obey
\begin{equation}
 \label{fullRSB_sigma_solution}
	\sigma (u) = \frac{10}{9} \frac{A_{\argp{c,D}}}{T} \cdot \argp{u/T}^9 + \text{cte}
\end{equation}
where this last constant has to be determined in relation to the possible plateaux and cutoffs imposed on the Ansatz $\sigma (u)$, using the saddle point equation \eqref{saddle-pt-equation_u_powerlaw} to fix them.

In addition to the above power-law behaviour we must consider the possibility of a plateau in $\sigma(u)$. One such trivial example
would be a totally replica-symmetric (RS) solution for which $\sigma(u)$ is independent of $u$.
In that case, in the saddle-point equation \eqref{saddle-pt-equation_u_lambda} the term $\widetilde{G} (q) - G(q,u) = G_c (q) = \frac{1}{c q^2}$ as explained in \eqref{A_RS-inversion-formulas} and thus:
\begin{equation}
 \label{Ansatz_RS_sigma0_0}
	\sigma_{\text{RS}} (u) = \sigma_{\text{RS}} (0)
	= \frac{D}{T} \int_{\gdR} \dbar \lambda \cdot \lambda^2 \cdot e^{-\lambda^2 \xi^2} e^{-\lambda^2 \int_{\gdR} \dbar q \cdot G_c(q)} =0
\end{equation}
The RS solution completely eliminates all effects of disorder, it is thus clearly unphysical, as was the case in the higher dimensional cases \cite{mezard_parisi_1991_replica_JournPhysI1_809} and the periodic ones \cite{giamarchi_ledoussal_1995_PhysRevB52_1242}.

\begin{figure}[htbp]
 \subfigure[\label{fig:interface1D_sigmacrochet_sigma0}]{\includegraphics[width=\figwidth]{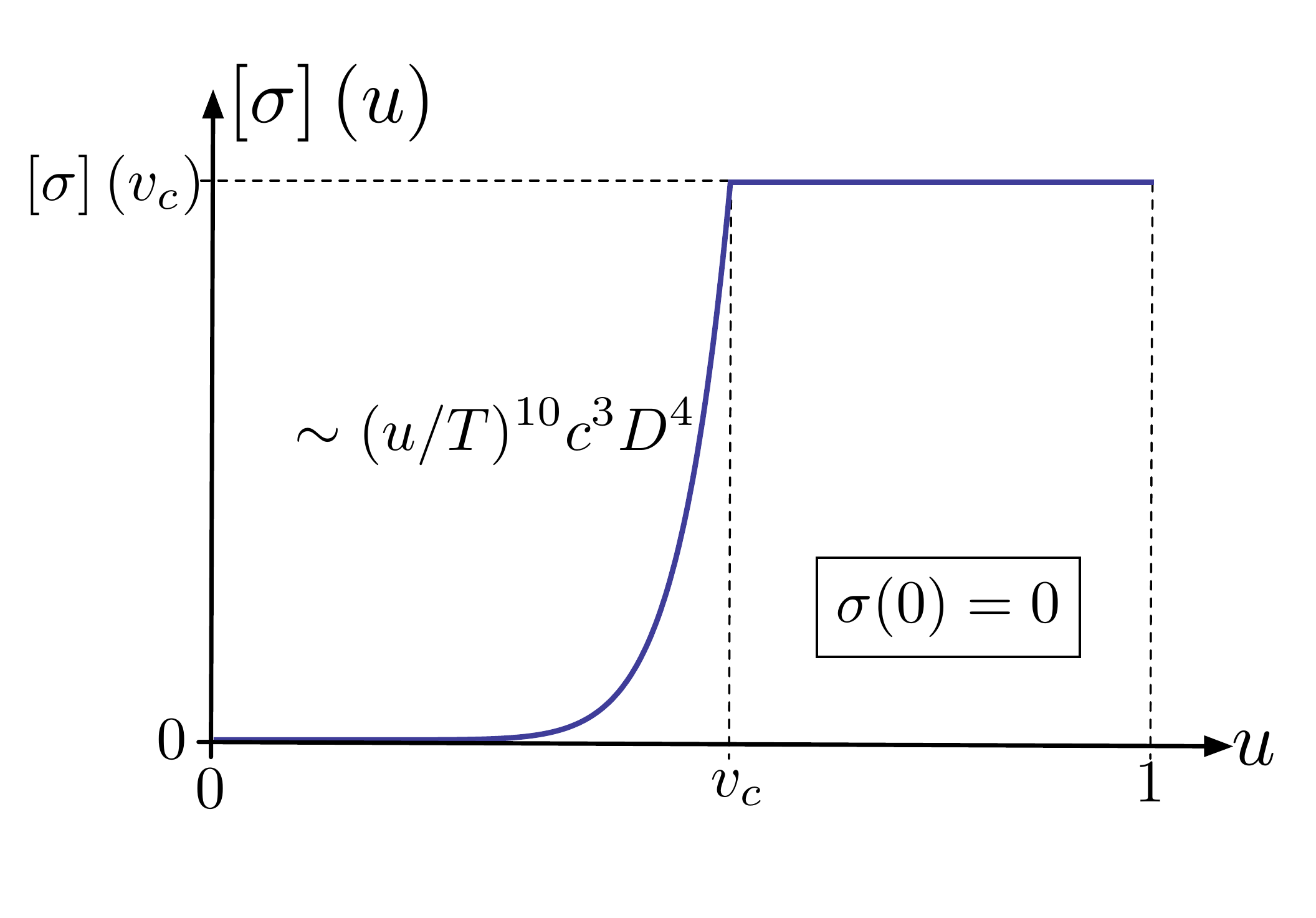}}
 \subfigure[\label{fig:interface1D_vcAtilde}]{\includegraphics[width=\figwidth]{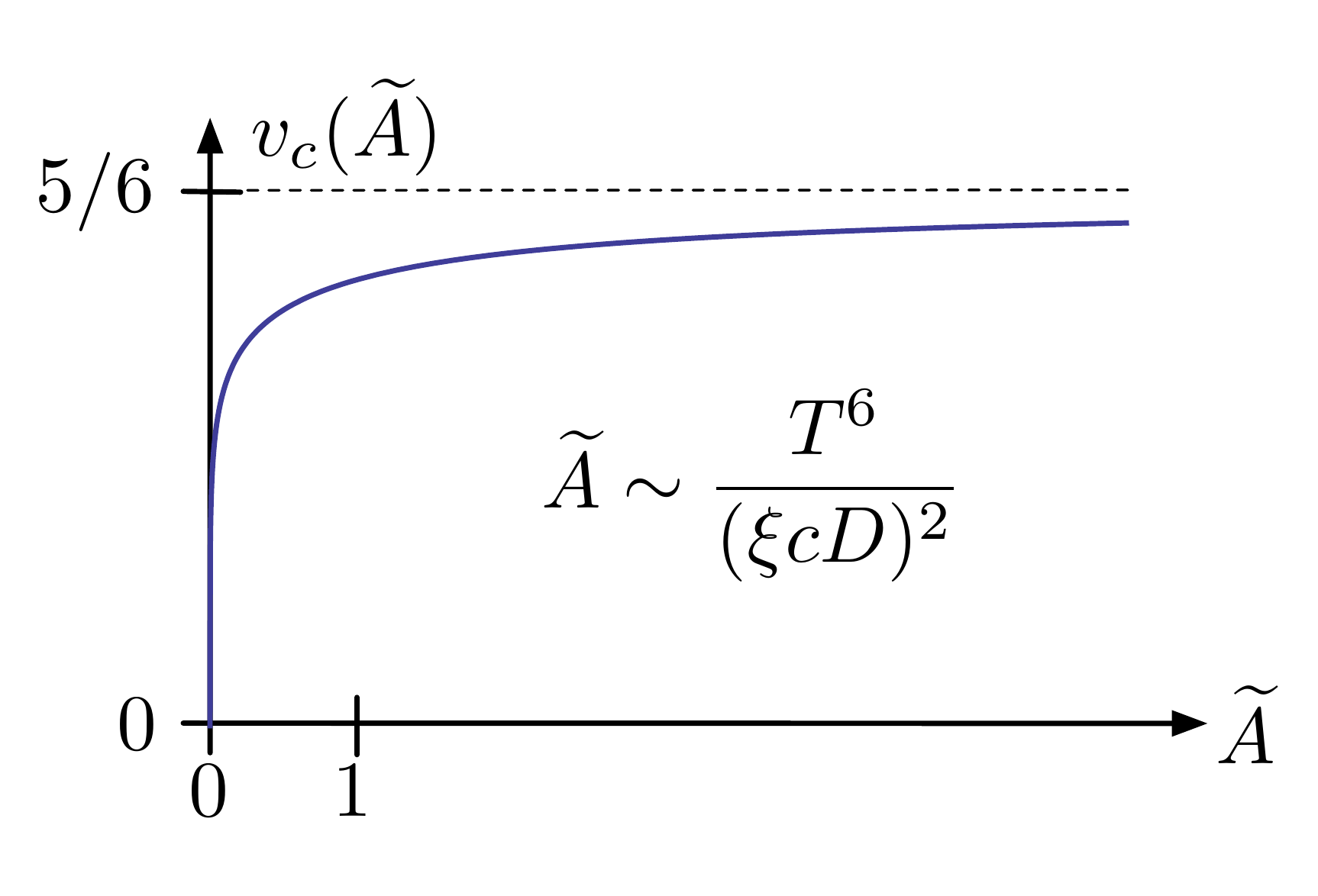}}
 \caption{\label{fig:interface1D_GVMsolution}
 	GVM solution for the 1D interface.
	\subref{fig:interface1D_sigmacrochet_sigma0} Self-energy $\sigc{u}$ given by \eqref{Ansatz_fullRSB_sigmacrochet}; using $\left[ \sigma \right] ' (u) = u \sigma ' (u)$ and $\sigma (0)=0$ the expression \eqref{Ansatz_fullRSB_sigma} for $\sigma (u)$ can be recovered.
	\subref{fig:interface1D_vcAtilde} Full-RSB cutoff $v_c$ as a function of $\widetilde{A} \sim \frac{T^6}{(\xi c D)^2}$, obtained by solving the polynomial equation \eqref{equa_vc_xi-2}. It starts linearly at $\widetilde{A} \to 0$ and saturates to $5/6$ at $\widetilde{A} \to \infty$, yielding in particular the low-temperature dependence given by \eqref{vc_small_T}.}
\end{figure}

To take into account the possible presence of plateaux we thus look for a full-RSB solution $\sigma (u)$ with a single cutoff $v_c \in \argp{0,1}$ of the form:
\begin{equation}
 \label{Ansatz_fullRSB_sigma}
 \begin{split}
	& \sigma \argp{u \leq v_c} = \frac{10}{9} \frac{A_{\argp{c,D}}}{T} \cdot \argp{u/T}^9 + \sigma (0) \\
	& \sigma \argp{u \geq v_c} = \sigma \argp{v_c}
 \end{split}
\end{equation}
and for the self-energy:
\begin{equation}
 \label{Ansatz_fullRSB_sigmacrochet}
 \begin{split}
	& \sigc{u \leq v_c} = A_{\argp{c,D}} \cdot (u/T)^{10} \\
	& \sigc{u \geq v_c} = \sigc{v_c} \\
	& A_{\argp{c,D}} \equiv \frac{3^{14}}{5^{10} \pi^2} \cdot c^3 D^4
 \end{split}
\end{equation}
Up to this point the width $\xi$ does not appear in these expressions. It is in fact encoded into the single full-RSB cutoff $v_c \argp{\xi}$. The equation for this cutoff is obtained by checking the consistency of the solution \eqref{Ansatz_fullRSB_sigmacrochet} with the initial saddle point equation \eqref{saddle-pt-equation_u_powerlaw}. Using the inversion formula \eqref{A_replic_Gt-Gu2} adapted to the presence of a cutoff $\sigma ' \argp{u \geq v_c}=0$, then integrating over the $q$ modes with \eqref{A_propagator1} and \eqref{A_propagator2} and inserting our full-RSB solution \eqref{Ansatz_fullRSB_sigmacrochet} we have:
\begin{equation}
 \label{int_q_Gt_G}
 \begin{split}
	& T \int_{\gdR} \dbar q \cdot \argp{\widetilde{G} (q) - G(q,u)} \\
	&= T \frac{\sigc{v_c}^{-1/2}}{2 \sqrt{c}} + T \int_u^{v_c} dv \cdot \sigma ' (v) \cdot \frac{\sigc{v}^{-3/2}}{4 \sqrt{c}} \\
	&= \frac{5^5}{2 \cdot 3^7} \pi \frac{1}{\argp{c D}^2}\cdot \argc{\argp{\frac{v_c}{T}}^{-6} \argp{v_c - 5/6} + \frac{5}{6} \argp{\frac{u}{T}}^{-6}}
 \end{split}
\end{equation}
Substituting this expression into \eqref{saddle-pt-equation_u_powerlaw}, the full-RSB Ansatz $\sigma \argp{u \leq v_c}$ \eqref{Ansatz_fullRSB_sigma} is self-consistent if the $\xi$-dependence cancels the $v_c$ terms:
\begin{equation}
 \label{equa_vc_xi-1}
	\xi^2 + \frac{5^5}{2 \cdot 3^7} \pi \frac{1}{\argp{c D}^2}\cdot \argp{\frac{v_c}{T}}^{-6} \argp{v_c - 5/6} =0
\end{equation}
and also if $\sigma (0) =0$. This last condition is enforced by \eqref{relation_sigma_sigc} and can be checked a posteriori by combining \eqref{saddle-pt-equation_u_powerlaw}, \eqref{int_q_Gt_G} and \eqref{equa_vc_xi-1}:
\begin{equation}
 \label{check_sigma0_0}
    \sigma (0) = \frac{D}{T} \int_{\gdR} \dbar \lambda \cdot \lambda^2 \cdot \lim_{u \to 0} e^{-\lambda^2 \frac{5^5}{2 \cdot 3^7} \pi \frac{T^6}{\argp{c D}^2}\cdot u^{-6}} = 0
\end{equation}
The single full-RSB cutoff is thus given by the polynomial equation:
\begin{equation}
 \label{equa_vc_xi-2}
	v_c^6 = \widetilde{A} \argp{5/6 - v_c} \, , \quad
	\widetilde{A} \equiv \frac{5^5}{2 \cdot 3^7} \pi \frac{T^6}{\argp{\xi c D}^2}
\end{equation}
whose solution is plotted in \fref{fig:interface1D_vcAtilde}.
The factor $5/6$ can be shown to be closely related to the Flory exponent $\zeta_F$ in $d=1$ (cf. \aref{A-Flory}); the presence of a cutoff $v_c < 5/6 = \argp{2 \zeta_F \vert_{d=m=1}}^{-1}$ in the full-RSB Ansatz $\sigc{u}$ is in fact necessary for $\sigc{u}$ to be a solution of the GVM saddle point equation.

All the above results are summarized in \fref{fig:interface1D_GVMsolution}.


\subsection{Low versus high temperature limits of $\sigc{u}$} \label{fullGVM_limits}

Before computing the roughness $B(r)$, which we will do in the next section, we have to explicit the low and the high temperature limits of this solution, since we also aim to probe the $T$-dependence of the roughness.

An explicit analytical expression for $v_c \in \argp{0,5/6}$ can be obtained for the two opposite limits of $\widetilde{A}$, which at fixed $\xi, D >0$ correspond respectively to $T \to 0$ and $T \to \infty$ and yield the following asymptotic behaviour:
\begin{align}
 \label{vc_small_T}
	v_c & \stackrel{\widetilde{A} \to 0}{\approx} \frac{5}{6} \argp{\frac{4}{3} \sqrt{\pi}}^{1/3} \cdot \frac{T}{ (\xi c D)^{1/3}} \to 0 \\
 \label{vc_large_T}	
	v_c & \stackrel{\widetilde{A} \to \infty}{\approx} 5/6 + 0^-
\end{align}

	The crossover between the two regimes of high and low temperature is in fact conditioned by the value of the dimensionless parameter $\widetilde{A}$ in the equation for $v_c (\xi)$ \eqref{equa_vc_xi-2}; an arbitrary definition of a `critical' temperature is naturally given by the crossover between the two opposite limits $\widetilde{A} \to \infty$ and $\widetilde{A} \to 0$, which happens at:
	\begin{equation}
	\label{Atilde_1_def_Tc}
		\widetilde{A} =1
		\Longleftrightarrow
		\frac{T_c}{(\xi c D)^{1/3} } = \argp{\frac{2 \cdot 3^7}{5^5 \pi}}^{1/6} \approx 0.87
	\end{equation}
and this last constant is of order 1.
Note that this critical temperature depends explicitly on the width $\xi$, in such a way that the limits $T \to 0$ and $\xi \to 0 $ cannot be exchanged with impunity: imposing $\xi = 0$ from the beginning is equivalent to considering exclusively the `high' temperature regime, and the somehow non-physical regime at simultaneously zero-temperature and zero-width has to be carefully handled.

The first limit \eqref{vc_small_T} implies that when the thermal fluctuations are suppressed, $\sigc{u}$ tends to a non-zero RS solution, since $v_c \to 0$ and
\begin{equation}
 \label{sigc_vc_lowT}
	\sigc{v_c}
	\stackrel{\widetilde{A} \to 0}{\approx}
	\frac{1}{4} \argp{\frac{3}{4 \sqrt{\pi}}}^{2/3} \cdot \xi^{-10/3} c^{-1/3} D^{2/3}
\end{equation}
Increasing the strength of disorder $D$ is compatible with this limit, and indeed the self-energy is intuitively expected to increase with $D$.

On the contrary the decrease of $D$ or $\xi$ leads to the second limit \eqref{vc_large_T}, in which the $\xi$-dependence has been washed out from the GVM solution by the relatively large thermal fluctuations:
\begin{equation}
 \label{sigc_vc_highT}
	\sigc{v_c}
	\stackrel{\widetilde{A} \to \infty}{\approx}
	\frac{3^4}{2^{10} \pi} \frac{c^3 D^4}{T^{10}}
\end{equation}


\section{GVM roughness and crossover lengthscales of the 1D interface} \label{full_roughness}

Using the definition of the structure factor \eqref{def_structure_factor} in relation with the Green function of a quadratic Hamiltonian \eqref{quadr_structure_factor}, we can now compute the corresponding roughness as a function of the lengthscale $r$ along the internal coordinate $z$ of the interface:
\begin{equation}
	B(r)
	= T \int_{\gdR} \dbar q \cdot 2 (1-\cos (qr)) \cdot \lim_{n \to 0} \widetilde{G} (q)
\label{Br_widetildeG}
\end{equation}
The inversion formula for $\lim_{n \to 0} \widetilde{G}(q)$ \eqref{A_replic_Gt} contains two contributions (since $\sigma (0) =0$) which yield respectively a purely thermal and a disorder-induced roughness:
\begin{align}
 \label{Br_Bth_Bdis}
	B(r)
	& = B_{\text{th}} (r) + B_{\text{dis}} (r) \\
 \label{Br_Bth}
	B_{\text{th}} (r)
	& = T \int_{\gdR} \dbar q \frac{2(1-\cos (qr))}{c q^2}
	= \frac{T r}{c}  \\
 \label{Br_Bdis}
	B_{\text{dis}} (r)
	& = T \int_{\gdR} \dbar q \frac{2(1-\cos (qr))}{c q^2} \int_0^1 \frac{dv}{v^2} \frac{\sigc{v}}{c q^2 + \sigc{v}}
\end{align}
The structure factor in $B_{\text{dis}}(r)$ is a combination of propagators $\frac{(cq^2/\sigc{v} + 1)^{-1}}{c q^2}$
organized by the RSB parameter $v$, with an increasing self-energy $\sigc{v}$ bounded by its value at the full-RSB cutoff $\sigc{v_c}$. The corresponding roughness is eventually computed by integrating explicitly over the Fourier modes $q$.

\subsection{GVM roughness of a 1D interface} \label{full_roughness_GVM_1Dinterface}

Using the identity \eqref{A_propagator_cos} we obtain an analytical expression for a generic $\sigc{v}$:
\begin{equation}
 \begin{split}
	B_{\text{dis}} (r)
	& = T \int_0^1 \frac{dv}{v^2} \cdot \sigc{v} \cdot \int_{\gdR} \dbar q \frac{2(1-\cos (qr))}{cq^2 (cq^2+ \sigc{v})} \\
	& = \frac{T}{\sqrt{c}} \sum_{k=2}^{\infty} \frac{(-r/\sqrt{c})^k}{k!} \int_0^1 \frac{dv}{v^2} \cdot \sigc{v}^{\frac{k-1}{2}}
 \end{split}
\end{equation}
which simplifies for our full-RSB solution \eqref{Ansatz_fullRSB_sigmacrochet} into:
\begin{equation}
 \label{roughness_dis_full}
	B_{\text{dis}} (r)
	= \frac{T r_0}{c} \cdot v_c^{-1} \sum_{k=2}^{\infty} \frac{(r/r_0)^k}{k!} \argp{\frac{1}{5 k -6} + (1-v_c)}
\end{equation}
where $r_0$ is a characteristic lengthscale which appears naturally in the formalism in order to obtain dimensionless quantities. It is defined by:
\begin{equation}
 \label{def_Larkin_r0}
	\sigc{v_c} \equiv c q_0^2
	\, \stackrel{q_0 \equiv 1/r_0}{\Longleftrightarrow} \,
	r_0 = \sqrt{c/\sigc{v_c}}
\end{equation}
and can be made explicit using \eqref{Ansatz_fullRSB_sigmacrochet}:
\begin{equation}
 \label{Larkin_r0_T_vc}
	r_0 = \frac{5^5 \pi}{3^7} \frac{1}{c D^2} \argp{\frac{T}{v_c}}^5
\end{equation}
$B_{\text{dis}}(r)$ is thus composed on one hand of the prefactor $\frac{T r_0}{c}$ which fixes its dimensions and is actually the thermal roughness at the scale $r_0$, and on the other hand of a dimensionless series in $(r/r_0)$ including the parameter $v_c$ \eqref{equa_vc_xi-2}.

The whole displacement correlation function $B(r)$ is plotted in \fref{fig_graph_roughness_compil-1} where we distinguish the low versus high temperature cases. In \fref{fig_graph_roughness_compil-2} we show the evolution of the roughness at increasing $T$ and fixed $D$ versus the other way around.
A summary of the different roughness regimes along with their corresponding exponent $\zeta$ and their crossover lengthscales is given by \fref{fig:1Dinterface_roughness_summary}.

\subsubsection{At small lengthscales: thermal regime} \label{full_roughness_regime_thermal}

At small lengthscales the linear term of $B_{\text{th}} (r)$ dominates the whole roughness, and the 1D interface fluctuates as expected as if there were no disorder:
\begin{equation}
 \label{roughness_thermal}
	B(r)
	\stackrel{r \to 0}{\approx}
	B_{\text{th}} (r)
	= \frac{T r}{c}
	\sim r^{2 \zeta_{\text{th}}}
\end{equation}
This defines the thermal regime of the roughness, of corresponding thermal exponent $\zeta_{\text{th}}=1/2$.

\subsubsection{At large lengthscales: RM regime} \label{full_roughness_regime_RM}

To make explicit the asymptotic behaviour at large lengthscales, we can reformulate the truncated alternated series \eqref{roughness_dis_full} using the following definition of the Euler $\Gamma$-function and its generalization:
\begin{equation}
 \begin{split}
	\Gamma (a)
	&= \int_0^{\infty} dt \cdot t^{a-1} \cdot e^{-t} \\
	\Gamma (a,z)
	&= \int_z^{\infty} dt \cdot t^{a-1} \cdot e^{-t}
 \end{split}
\end{equation}
which yield:
\begin{equation}
 \begin{split}
	& B_{\text{dis}}(r)
	= \frac{T r_0}{c} v_c^{-1} \left[ (1-v_c) \argp{e^{-r/r_0} - r/r_0 -1} - r/r_0 \right. \\
	& + \left. \frac{1}{6} + (r/r_0)^{6/5} \argp{-\frac{1}{6} \Gamma \argp{-\frac{1}{5}}-\frac{1}{5} \Gamma \argp{-\frac{6}{5},r/r_0} } \right]
 \end{split}
\end{equation}
Since $\Gamma (-6/5,a)$ tends exponentially fast to $0$ for increasing $a$, at large lengthscales the total roughness is dominated by the following power-law behaviour:
\begin{equation}
 \label{roughness_asympt_zetaF}
	B(r)
	\stackrel{r \to \infty}{\approx}
	B_{\text{asympt}}
	\equiv
	\frac{T r_0}{c} v_c^{-1} \cdot \argp{\frac{r}{r_0}}^{6/5}
	\sim r^{2 \zeta_F}
\end{equation}
with an overall numerical factor $\argp{-\frac{1}{6} \Gamma \argp{-\frac{1}{5}}} \approx 0.970191$, and the Flory roughness exponent $\zeta_F = \frac{4-d}{4+m}\vert_{d=m=1}= 3/5$ (see \aref{A-Flory}). This asymptotic exponent which is not the exact one $\zeta_{RM} = 2/3$, is a consequence of the variational approximation. It leads in particular to an asymptotic \emph{temperature independence} when $r_0$ is made explicit using \eqref{Larkin_r0_T_vc}:
\begin{equation}
 \label{roughness_asympt_thorn0}
	B_{\text{asympt}}
	= \frac{3}{5} \argp{\frac{9}{\pi}}^{1/5} \cdot c^{-4/5} D^{2/5} \cdot r^{6/5}
\end{equation}
We will examine this point in more details in \sref{GVM_rescaling}.

\subsection{Larkin length, RSB and effective width} \label{full_roughness_Larkin_xieff}

The RS section $u \geq v_c$ corresponds to small lengthscales (large $q$) in which the interface fluctuates purely thermally, whereas the full-RSB section $u \leq v_c$ corresponds to large lengthscales (small $q$) and encodes the disorder-induced metastability experienced by the interface in the RM regime. So the cutoff $v_c$, and consequently $r_0$ must correspond to the definition of a characteristic crossover lengthscale between two roughness regimes, namely the Larkin length \cite{Larkin_model_1970-SovPhysJETP31_784}, which marks the beginning of the RM regime.

Using the low- and high-temperature limit of $\sigc{v_c}$, respectively \eqref{sigc_vc_lowT} and \eqref{sigc_vc_highT}, analytical expressions of the characteristic lengthscale $r_0$ are accessible at finite $c$ and $\xi$:
\begin{align}
 \label{larkin_r0_low_T}	
	r_0
	& \stackrel{T \to 0}{\approx} 2 \argp{\frac{4 \sqrt{\pi}}{3}}^{1/3} \cdot \xi^{5/3} c^{2/3} D^{-1/3} \\
 \label{larkin_r0_high_T}
	r_0
	& \stackrel{T \to \infty}{\approx} \frac{2^5 \pi}{3^2} \frac{T^5}{c D^2}
\end{align}
At low temperatures it is temperature-independent and depends explicitly on the width $\xi$, whereas at high temperature it is the opposite, and the microscopic parameter $\xi$ is completely washed out.
This leads in particular to a radical simplification of the prefactor $\frac{T r_0}{c} v_c^{-1}$ in $B_{\text{dis}} (r)$ at low temperature:
\begin{equation}
 \label{roughness_dis_lowT}
	B_{\text{dis}}(r)
	\stackrel{T \to 0}{\approx}
	\frac{12}{5} \xi^2 \, \sum_{k=2}^{\infty} \frac{(-r/r_0)^k}{k!} \argp{\frac{1}{5 k -6} +1}
\end{equation}
which gives at $r=r_0$:
\begin{equation}
 \label{roughness_r0_low_T}
	B_{\text{dis}} (r_0)
	\stackrel{T \to 0}{\approx}
	\frac{6}{5} \xi^2
\end{equation}
It follows from this last expression that $r_0$ satisfies the usual definition of the Larkin length $L_c$, i.e. the lengthscale at which the mean displacement of the interface is of the order of its width :
\begin{equation}
	u (L_c) \sim \sqrt{B(L_c)} \sim \xi
\end{equation}
At higher temperatures, if we want to keep $r_0$ as the Larkin length, i.e. associate the beginning of the RM regime and the breaking of the replica symmetry by imposing $r_0 \equiv L_c$, we have then to define an effective width $\xi_{\text{eff}}$ with respect to the roughness at $r=r_0$:
\begin{align}
 \label{roughness_r0_high_T}
	& B_{\text{dis}} (r_0)
	\stackrel{T \to \infty}{\approx}
	\frac{T r_0}{c} \cdot \frac{1}{5}
	\equiv \xi_{\text{eff}}^2 \\
 \label{def_width_effective}
	& \xi_{\text{eff}}
	= \frac{3}{2} \frac{T^3}{c D}
\end{align}
so at high temperature the interface `thickens' due to the increasing interplay of thermal fluctuations with disorder.

\subsection{Crossover at low temperature: modified Larkin regime} \label{full_roughness_cross_lowT}

The end of the thermal regime can be defined as the lengthscale at which the quadratic term $\sim r^2$ in the sum of power-law corrections of $B_{\text{dis}} (r)$ becomes of the same order as the linear term of the thermal roughness:
\begin{align}
 \label{def_r1_vc}	
	& r_1 = r_0 \frac{2 v_c}{5/4 - v_c}
\end{align}
Note that $r_1$ does never diverge since $v_c < 5/6 <5/4$.
At low temperature it can be made explicit using \eqref{larkin_r0_low_T}:
\begin{equation}
 \label{fin_therm_r1_low_T}	
	r_1
	\stackrel{T \to 0}{\approx} \frac{16}{9} (6 \pi)^{1/3} \cdot T \cdot \xi^{4/3} c^{1/3} D^{-2/3}
\end{equation}
This increases linearly with the temperature and at some point will even merge with $r_0$, defining an upper critical temperature with the help of \eqref{equa_vc_xi-2}:
\begin{equation}
 \label{1Dinterface_collapse_r1r0}
	r_1 = r_0
	\Longleftrightarrow
	v_c = \frac{5}{12}
	\Longleftrightarrow
	\frac{T_c}{(\xi c D)^{1/3}} = \argp{\frac{3^2}{2^9 \pi}}^{1/6} \approx 0.42
\end{equation}
Up to a factor 1/2 this criterion is equivalent to \eqref{Atilde_1_def_Tc}.

\begin{figure}
 \subfigure[\label{fig:graph_roughness_lowT} Low-temperature regime]{\includegraphics[width=\figwidth]{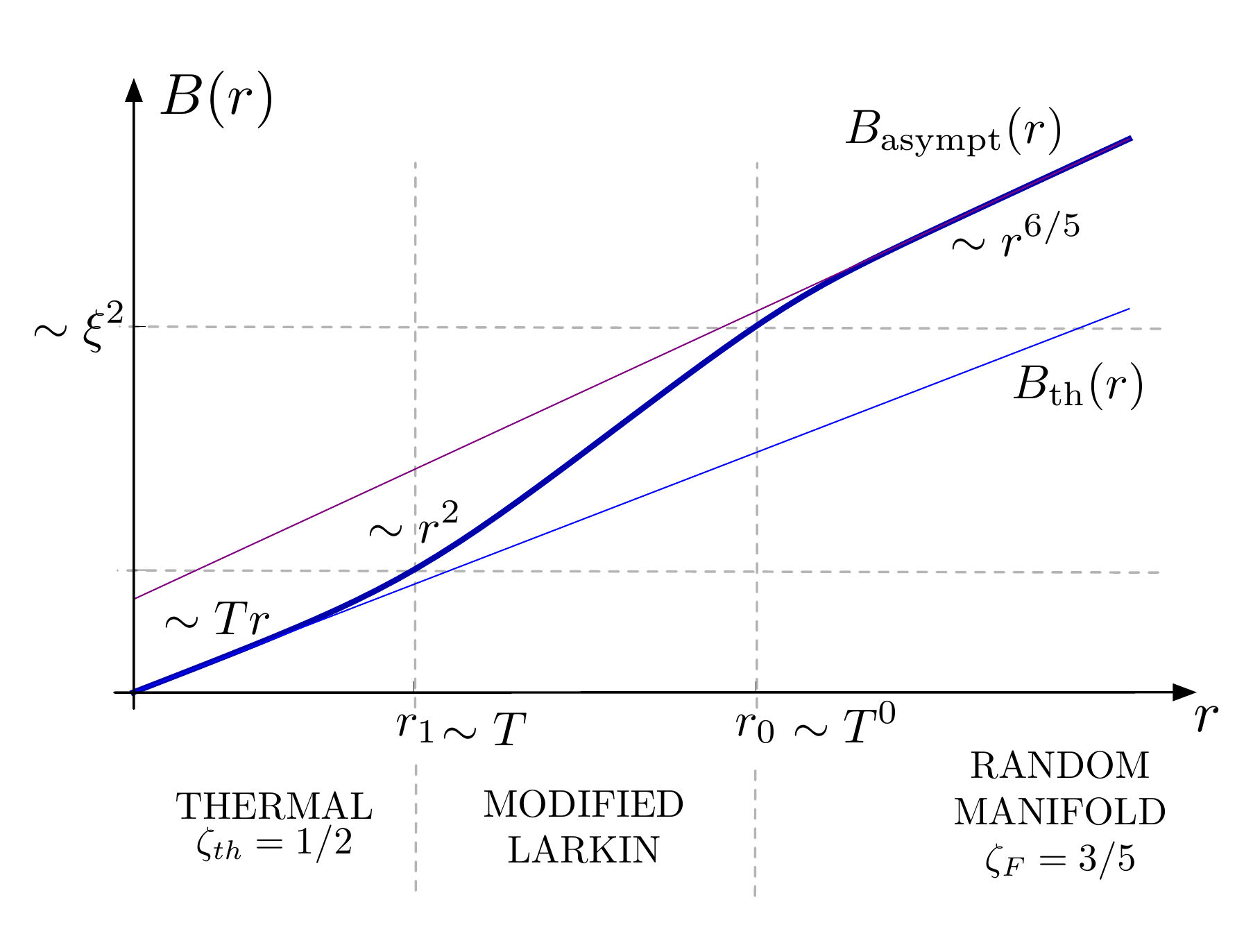}}
 \subfigure[\label{fig:graph_roughness_highT} High-temperature regime]{\includegraphics[width=\figwidth]{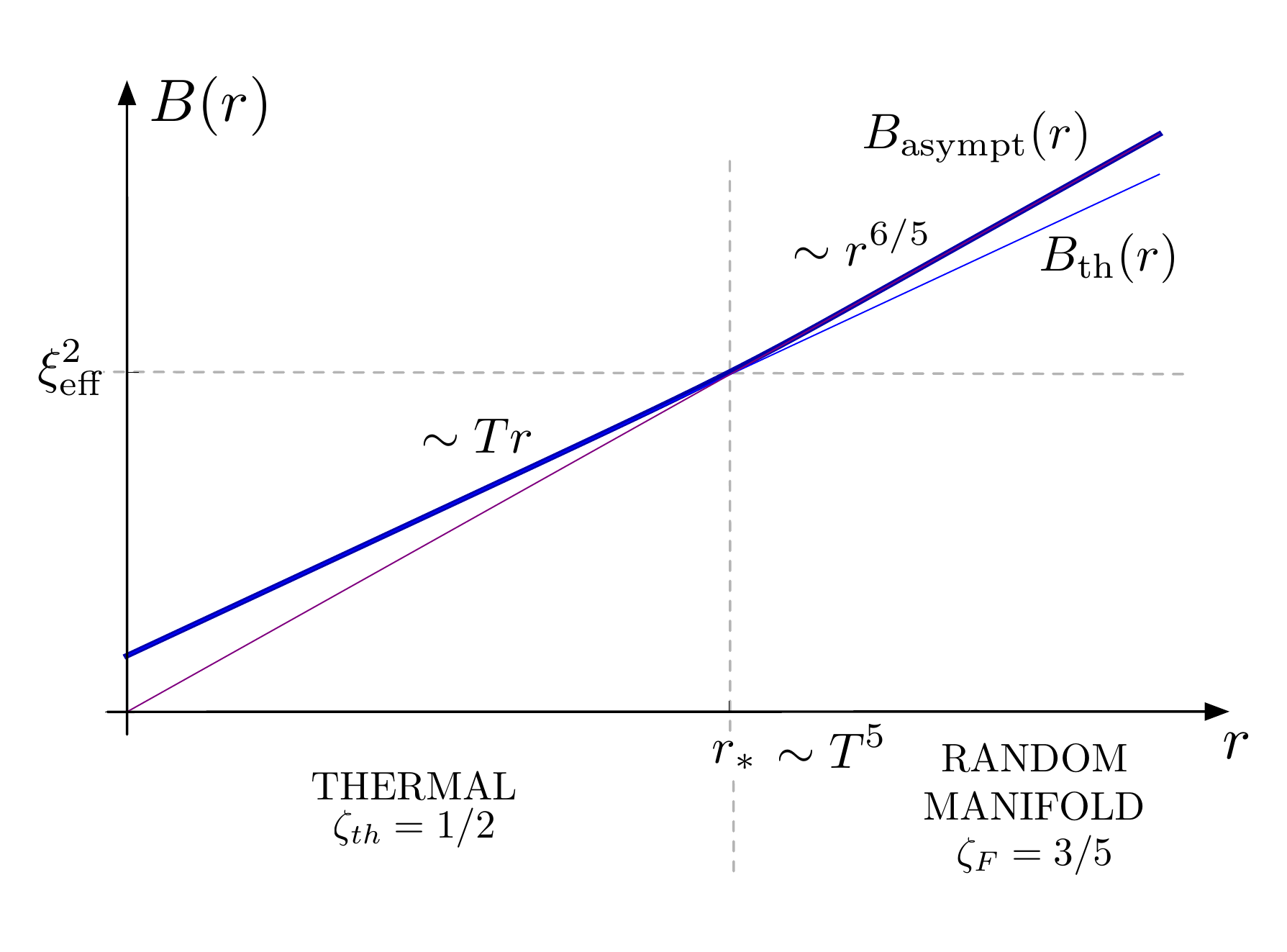}}
 \caption{ \label{fig_graph_roughness_compil-1} GVM prediction for the
   1D-interface static roughness $B(r)$, in $\log-\log$
   representations; the slope of the curves corresponds to $2
   \zeta(r)$ as defined by \eqref{def_roughness_exponent_loglog}
   ($\xi=c=D=1$). \subref{fig:graph_roughness_lowT_toymodel} At low
   temperature ($T=10^{-3}$) an intermediate regime appears between
   the small and large lengthscales regimes, while
   \subref{fig:graph_roughness_highT_toymodel} at high temperature
   ($T=10$) no intermediate regime occurs. The scalings of the different
   quantities, including crossover lengthscales, are recalled directly on the figures.}
\end{figure}

\begin{figure}
 \subfigure[\label{fig:graph_roughness_fixedT} At fixed temperature]{\includegraphics[width=\figwidth]{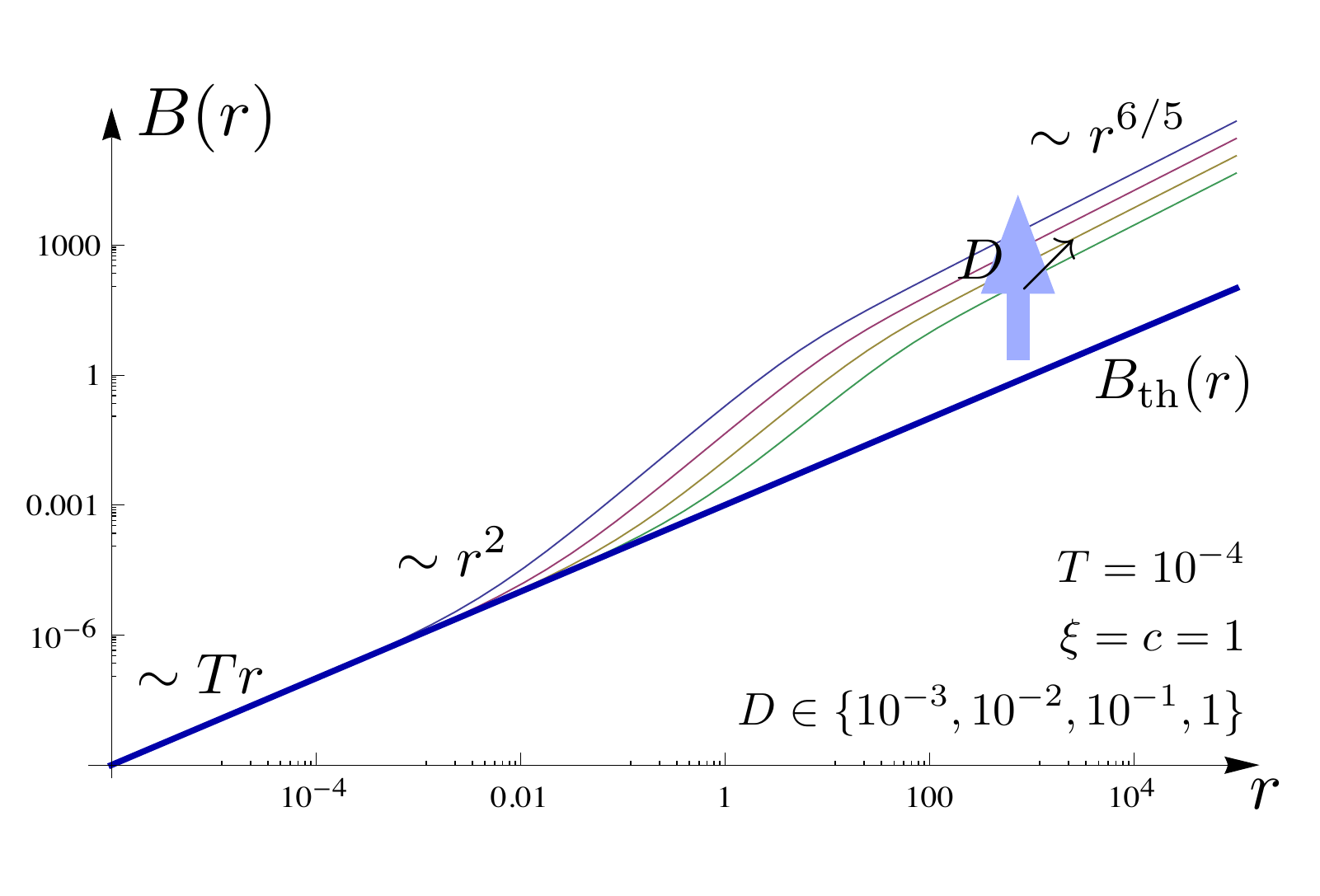}}
 \subfigure[\label{fig:graph_roughness_fixedD} At fixed disorder]{\includegraphics[width=\figwidth]{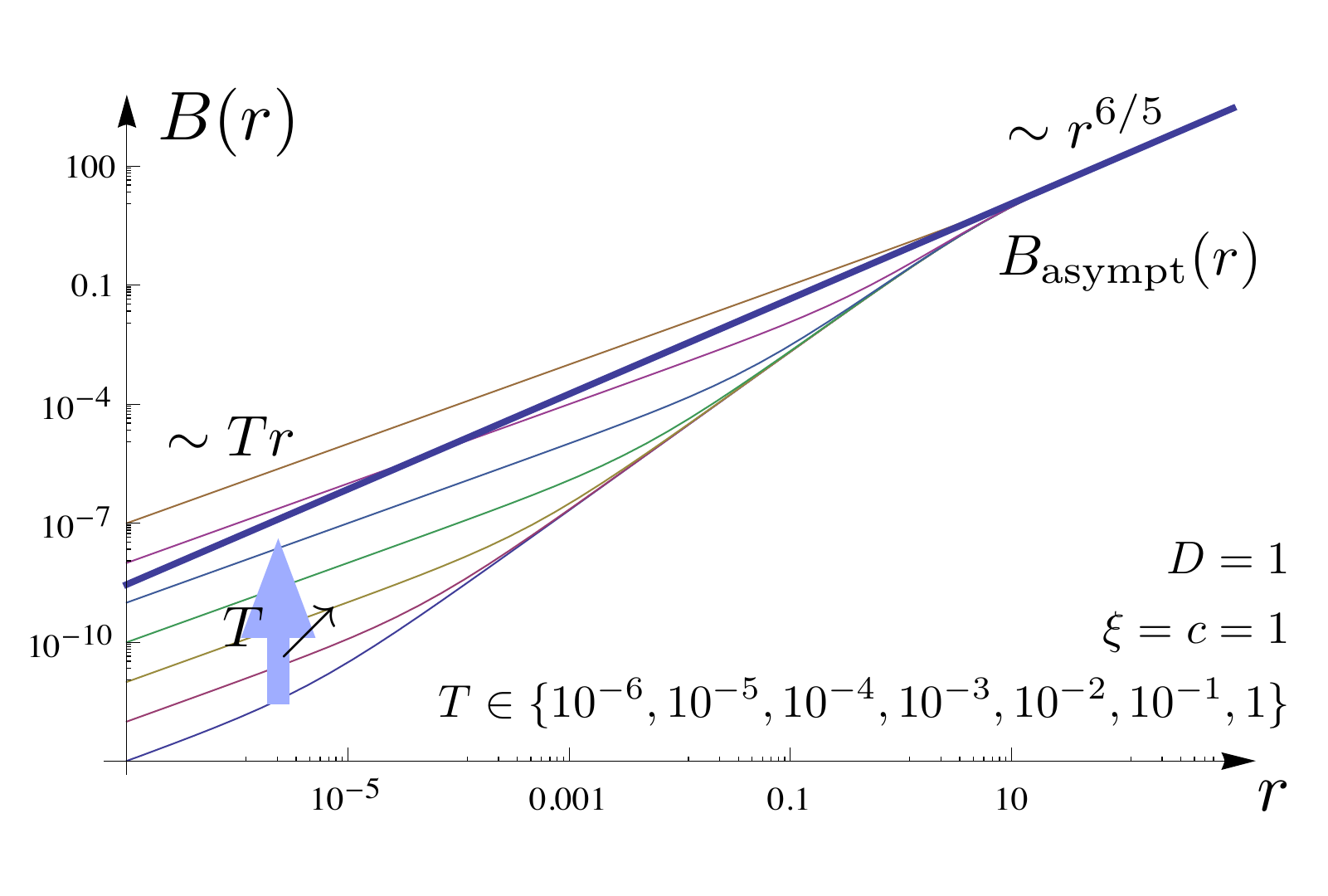}}
 \caption{\label{fig_graph_roughness_compil-2}
 	Evolution of the 1D-interface static roughness at finite width $\xi$.
 	\subref{fig:graph_roughness_fixedT}
		At fixed $T$ (low-temperature regime) all the roughness curves start from the same $B_{\text{th}}(r)=\frac{Tr}{c}$ at small lengthscales, and behave asymptotically in $r^{6/5}$ with a $T$-independent prefactor increasing with disorder.
	\subref{fig:graph_roughness_fixedD}
		At fixed $D$ all the roughness curves collapse on the same $B_{\text{asympt}} (r) \sim T^0 r^{6/5}$ at large lengthscales (see \eqref{roughness_asympt_zetaF}), and start from a microscopic thermal roughness which is boosted by increasing thermal fluctuations.
 }
\end{figure}

For intermediate lengthscales between $r_1$ and $r_0$ the roughness function shows a smooth connection between the thermal regime at $r<r_1$ and the RM regime at $r>r_0$, as can be seen in the low temperature case \fref{fig:graph_roughness_lowT}.
At fixed strength of disorder $D$ and increasing $T$ the crossovers $r_1$ and $r_0$ are getting closer and squeeze this intermediate regime, as can be seen in \fref{fig:graph_roughness_fixedD}.
In other dimensionalities this intermediate regime would be described by the Larkin model \cite{Larkin_model_1970-SovPhysJETP31_784}, which predicts a power-law behaviour of exponent $\zeta_L = \frac{4-d}{2}$.
For the 1D interface ($d=m=1$) the perturbative expansion of the Larkin model is not valid, and this regime is described by a sum of power-law corrections in $r$, starting from the quadratic term $\sim r^2$ in $B_{\text{dis}} (r)$ \eqref{roughness_dis_full}. By extension we thus call it the \emph{modified Larkin regime}.

In the zero-temperature limit the linear roughness $B_{\text{th}} (r) = \frac{Tr}{c}$ is suppressed and the small lengthscales are dominated by the quadratic term $\sim r^2$, which corresponds to an exponent $\zeta =1$ which is again not the Larkin exponent $\zeta_L \vert_{d=1} =3/2$.

\subsection{Crossover at high temperature} \label{full_roughness_cross_highT}	
		
At higher temperatures the modified Larkin regime has disappeared, and the intersection of the thermal and the asymptotic roughness, respectively \eqref{roughness_thermal} and \eqref{roughness_asympt_zetaF}, defines another characteristic crossover length $r_*$:
\begin{equation}
 \label{def_rstar}
	r_* = r_0 \cdot v_c^5 \frac{1}{-\frac{1}{6} \Gamma (-1/5)}
\end{equation}	
and since the value of the cutoff $v_c$ saturates at $5/6$ at high $T$, up to a factor $r_*/r_0 \approx 0.414225$ the crossover $r_*$ is equivalent to the Larkin length $r_0$. Using \eqref{larkin_r0_high_T} we eventually obtain:
\begin{equation}
 \label{expr_rstar_1Dinterface}
    r_*
    \stackrel{T \to \infty}{\approx} \frac{2}{5} \cdot \frac{T^5}{c D^2}
\end{equation}
This result can be predicted by scaling arguments assuming that there is no intermediate regime between the thermal and the asymptotic Flory-like regimes. This hypothesis is true only in the high-temperature regime, e.g. imposing $\xi=0$ from the beginning (see \sref{scaling_analysis}). This last approximation actually wrongs the low-temperature physics below the Larkin length since it misses the intermediate modified Larkin regime, whereas the GVM clearly points out its existence.

\subsection{Alternative formulation of the GVM roughness} \label{GVM_rescaling}

We finally give an alternative formulation of the GVM roughness \eqref{roughness_dis_full}, starting from \eqref{Br_widetildeG} and \eqref{A_replic_Gt}, and rescaling in the propagators the energy per Fourier mode $G_c^{-1} (q) = c q^2$ with respect to the mass term at the RSB cutoff $\sigc{v_c}$.
The \textit{ad hoc} changes of variable make explicit the `crossover function' from the thermal to the RM roughness regimes and provide a straightforward way of predicting the temperature dependence of the asymptotic GVM roughness.

The following argument being quite generic, we start with an internal dimension $d$ and a generalized isotropic elasticity indexed by the exponent $\mu>0$ ($\mu=2$ for the usual short-range elasticity \eqref{Ham_el}), which translates into:
\begin{equation}
	G_c^{-1} (q) = c \valabs{q}^{\mu}
	\Rightarrow
	\gdHel \argc{u} = \frac{1}{2} \int \dbar^d q \cdot G_c^{-1}(q) \cdot u_{-\vec{q}} u_{\vec{q}}
\end{equation}
As seen in \sref{fullGVM_determination_ansatz}, the GVM procedure constrains the form of a full-RSB mass term $\sigc{u}$ into the following power law ($\nu>0$):
\begin{equation}
 \label{Br_sigc_generic}
	\sigc{u \leq v_c} = A_{(c,D)} \argp{\frac{u}{T^{\psi}}}^{\nu}
	, \quad \sigc{u \geq v_c} = \sigc{v_c}
\end{equation}
where $\nu=10$ and $\psi=1$ in \eqref{fullRSB_sigc_solution}.

The thermal roughness \eqref{Br_Bth} generalizes into:
\begin{equation}
 \label{Br_thermal_regime_mu}
	B_{\text{th}} (r)
	\equiv \frac{T}{c} \int \dbar^d q \frac{2(1-\cos(\vec{q} \vec{r}))}{\valabs{q}^{\mu}}
	\sim \frac{T}{c} q^{- 2 \frac{\mu-d}{2}} \vert_{q \sim 1/r}
\end{equation}
i.e. skipping the angular dependence in the cosine's argument:
\begin{equation}
	B_{\text{th}} (r) \sim \frac{T r^{2 \zeta_{\text{th}}}}{c}
	\quad \text{with} \quad \zeta_{\text{th}} = \frac{\mu - d}{2 } 
\end{equation}

As for the `disorder' roughness \eqref{Br_Bdis}, separating the RS ($u \geq v_c$) and full-RSB ($u \leq v_c$) segments of $\sigc{u}$ and rescaling the energies with respect to $\sigc{v_c}$ we can apply these two changes of variables:
\begin{align}
 \label{Br_chgmtvar1}
	\frac{c \valabs{q}^{\mu}}{\sigc{v_c}}
	= (r_0 \valabs{q})^{\mu}
	\equiv \valabs{\tilde{q}}^{\mu}
	& \quad \text{i.e.} \quad r_0 \equiv \argp{\frac{c}{\sigc{v_c}}}^{1/\mu} \\
 \label{Br_chgmtvar2}
	\frac{\sigc{v}}{\sigc{v_c}} = \argp{\frac{v/v_c}{\valabs{\tilde{q}}^{\mu/\nu}}}^{\nu}
	\equiv \tilde{v}^{\nu}
	& \quad \text{i.e.} \quad \tilde{v} \vert_{v=v_c} = \valabs{\tilde{q}}^{\mu/\nu}
\end{align}
We eventually obtain:
\begin{equation}
 \label{rescaledBr}
 \begin{split}
	B_{\text{dis}} (r)
	= \frac{T v_c^{-1}}{c} r_0^{2 \zeta_{\text{th}}}
	\int \dbar^d \tilde{q} \cdot \frac{2(1-\cos(\tilde{q} r/r_0))}{\valabs{\tilde{q}}^{\mu}} \\
	\cdot \argp{\frac{1-v_c}{\valabs{\tilde{q}}^{\mu} +1} + \valabs{\tilde{q}}^{-\mu/\nu} \cdot F(\valabs{\tilde{q}}^{-\mu/\nu})}
 \end{split}
\end{equation}
with the `crossover function':
\begin{equation}
 \label{Br_crossoverF}
	F(x)
 	\equiv \int^x_0 dv \frac{v^{\nu-2}}{1 + v^{\nu}}
\end{equation}
As for the cutoffs in the Fourier-modes integral, they are simply redefined with respect to the Larkin length $r_0$ and are thus harmless at least for the 1D-interface case.

In order to extract the behaviour at large lengthscales $r$, we consider the limit $q \to 0$, i.e. $\tilde{q} \to 0$:
\begin{equation}
	\valabs{\tilde{q}}^{-\mu/\nu} \to \infty
	\Longrightarrow
	\valabs{\tilde{q}}^{-\mu/\nu}  \cdot  \underbrace{F(\valabs{\tilde{q}}^{-\mu/\nu} )}_{\approx \text{cte}} \approx \valabs{\tilde{q}}^{-\mu/\nu} 
\end{equation}
Thus, by counting in this limit the scaling of $\tilde{q}$ in \eqref{rescaledBr} we can relate $\nu$ to the asymptotic exponent $\zeta_{\text{asympt}}$ predicted by GVM:
\begin{equation}
	\tilde{q}^{\: d-\mu - \mu/\nu} \equiv \tilde{q}^{\:  -2 \zeta_{\text{asympt}}}
	\Longrightarrow
	\frac{\mu}{\nu} = 2 (\zeta_{\text{asympt}} - \zeta_{\text{th}})
\end{equation}
We can indeed check that for a 1D interface ($d=m=1$) with short-range elasticity ($\mu=2$) and RB disorder ($\zeta_F = \frac{4-d}{4+m}$), the GVM prediction $\zeta_{\text{asympt}}=\zeta_F=3/5$ and $\zeta_{\text{th}}=1/2$ are coherent with $\nu=10$ of \eqref{Ansatz_fullRSB_sigmacrochet}.

This link between $\nu$ and $\zeta_{\text{asympt}}$ can actually be used to predict the temperature dependence of the asymptotic roughness, disregarding the numerical and $A_{(c,D)}$ contributions:
\begin{equation}
	B(r)
	\stackrel{r \to \infty}{\sim}
	\frac{T}{v_c} r_0^{2 \zeta_{\text{th}}} (r/r_0)^{2 \zeta_{\text{asympt}}}
	= \frac{T}{v_c} r_0^{-\mu/\nu} r^{2 \zeta_{\text{asympt}}}
\end{equation}
Combining \eqref{Br_chgmtvar1} and \eqref{Br_sigc_generic} we can make explicit the power-law $T$-dependence of $r_0$ and eventually of the roughness:
\begin{equation}
	B(r)
	\stackrel{r \to \infty}{\sim}
	T^{1- \psi} \cdot v_c^0 \cdot r^{2 \zeta_{\text{asympt}}}
\end{equation}
We can thus conclude immediately that the asymptotic roughness has completely forgotten the interface width $\xi$ since $v_c (\xi)$ has been washed out, and that $\psi=1$ as imposed by the GVM scheme in \eqref{fullGVM_determination_ansatz} leads to a $T$-independent prediction. The elastic constant $c$ has to be $T$-independent to ensure the correct small-lengthscales behaviour \eqref{Br_thermal_regime_mu}, so the only way to modify this last prediction would be to choose a $T$-\emph{dependent} effective strength of disorder $D(T)$.
This will be the case in the second GVM procedure presented in the next section.


\section{Rounded toy model: GVM, roughness and crossover lengthscales} \label{toymodel}

 The 1D interface, that we have tackled up to now as a
static object, can actually be mapped onto a growing, dynamical object,
namely the directed polymer (DP), which consists in an elastic string
starting from a basepoint $y=0$ and getting ahead in `time' $t$
in a random potential $\widetilde{V}(t,y)$ \cite{huse_henley_fisher_1985_PhysRevLett55_2924}.
The weight of a trajectory $y(t')$ of duration $t$ is given by
\begin{equation}
  \exp\Big\{-\beta \int_0^t dt'\argc{ \frac{c}{2} (\partial_{t'} y(t'))^2+ \widetilde{V}(t',y(t'))   } \Big\}
\end{equation}
so that through the correspondence $(t,y)\leftrightarrow(z,x)$ one
recovers the weight of an interface of `length' $z=t$.
The roughness at lengthscale $z$ is for instance recovered from the
statistical properties of the DP at finite time $t$.
See also Ref.~\onlinecite{dotsenko_arXiv:1007.0852v1} for another approach.

Our interest
goes to finding a simple \emph{approximate} description of the
endpoint $y(t)$ fluctuations, at fixed time $t$.
This allows to determine the average of observables depending on the
sole endpoint position $y(t)$ of the DP (such as the roughness
$\overline{\langle[y(t)-y(0)]^2\rangle}$), but of course not of
observables depending on the whole trajectory.
The idea is to focus on the effective disorder ruling the
extremity of a DP of fixed length.

\subsection{Definition of the rounded toy model}

At fixed disorder $\widetilde{V}$ and final time $t$, we approximate the weight
$Z_{\widetilde{V}}(t,y)=\int \gdD y' e^{-\beta \gdH \argc{y',\widetilde{V}}}$ of all
trajectories $y'(t)$ arriving in $y$ at time $t$ through an effective
free-energy $F_\eta(t,y) \approx -\beta^{-1} \log Z_{\widetilde{V}}(t,y)$ where
$\eta=\eta(y)$ is an effective disorder depending only on the arrival position $y$ of the DP:
\begin{equation}
 \label{eq:Fyt_toymodel}
	F_{\eta} (t,y)
	= \frac{c}{2 t} y^2  + \int_0^y dy' \cdot \eta \argp{y'} + \text{cte}_\eta (t)
\end{equation}
The first term $\frac{c}{2 t} y^2$ is simply the (exact) free-energy in the absence of disorder.
The $t$-dependent constant arises from normalization along $y$, while the $\int_0^y\eta$ represents
the effective potential experienced by the endpoint.
Ideally, one would infer the distribution of $\eta$ from that of $\widetilde{V}$, but this task
is \emph{a priori} extremely difficult at finite time.

However, in the infinite-time limit, it is known \cite{huse_henley_fisher_1985_PhysRevLett55_2924} that
the exact correlator $\langle [F(t,y_2)-F(t,y_1)]^2\rangle$ is proportional to $|y_2-y_1|$,
as provided e.g. by $\eta$ being a delta-correlated white noise.
For finite time, numerical evidence \cite{mezard_1990_JPhys51_1831} support that this
correlator still behaves as $|y_2-y_1|$ at short distance, and it has
been proposed \cite{parisi_1990_JPhys51_1595} \cite{bouchaud_1990_JPhysStat61_877} that taking a delta-correlated white noise for
$\eta$ in \eqref{eq:Fyt_toymodel} indeed provides an approximate but
good description of the DP endpoint statistics.

In order to tackle the question of the influence of finite disorder correlation length
on the properties of an interface, it is thus very natural to examine in details
the corresponding model in which the delta function correlated noise $\eta$
is now replaced by a white noise with Gaussian correlations of width $\tilde{\xi}$
and effective strength $\widetilde{D}$:
\begin{equation}
	\moydes{\eta (y)}=0, \quad \moydes{\eta (y) \eta (y')}
	= \widetilde{D} \cdot R_{\tilde{\xi}} (y-y') \vert_{D=1}
\end{equation}
(see \eqref{disorder_correlator} for the definition of the correlator $R_{\tilde\xi}$).
The limit $\tilde\xi\to 0$ yields a delta-correlated disorder and corresponds to the original so-called `toy model' \cite{ledoussal_2003_PhysicaA317_140}. A finite $\tilde{\xi}$ actually rounds the correlator of the free energy $F(t,y)$ around $y=0$.

These properties thus define a modified `rounded' toy model, in which the average of an observable $\gdO$ writes
\begin{equation}
	\moy{\gdO \argp{t}}_{\eta}
	=\frac {\int dy \cdot \gdO (t,y) \cdot e^{-\beta F_{\eta} (t,y)}}
		{\int dy \cdot e^{-\beta F_{\eta} (t,y)}}
\end{equation}

Although much more simple than the full interface model, since the fluctuations are now reduced to those of the endpoint instead of the total interface, our rounded toy model is still not tractable in an exact way, and we resort to the GVM approximation in the replica approach. We follow here the route used for a previous modified toy model in Ref.~\onlinecite{mezard_parisi_1992_JPhysI02_2231} (in which was assumed a power-law asymptotic decay of the disorder correlator). Having reduced the dimension of the model from the directed polymer to the toy model, we hope that the GVM method catches exact scaling exponents at small and large times, together with crossover lengthscales.

Introducing replicas and averaging over disorder one gets the analogue of \eqref{moyth_rep}:
\begin{equation}
	\moydes{\moy{\gdO (t)}}
	= \lim_{n \to 0} \int dy_1 \argp{\cdots} dy_n \cdot \gdO \argp{t,y_1} \cdot e^{- \beta \widetilde{F} \argp{t, \vec{y}}}
\end{equation}
with the effective free energy
\begin{equation}
	\widetilde{F} \argp{t, \vec{y}}
	= \frac{c}{2 t} \sum_{a=1}^n y_a^2 - \frac{\beta \widetilde{D}}{2} \sum_{a,b=1}^n \text{min}_{\tilde{\xi}} \argp{y_a,y_b}
\end{equation}
where at $\tilde{\xi}=0$ (i.e. for $\eta$ a delta-correlated white noise),
$\text{min}_{\tilde{\xi}}$ would be the minimum function.
At finite $\tilde{\xi}$ it actually writes:
\begin{equation}
	\text{min}_{\tilde{\xi}} \argp{y_a,y_b}
	= \frac{1}{2} \argp{y_a + y_b - \valabs{y_a - y_b}_{\tilde{\xi}}}
\end{equation}
$\valabs{y}_{\tilde{\xi}}$ is the absolute value function, rounded close to the origin. In Fourier representation, it reads:
\begin{equation}
	\valabs{y}_{\tilde{\xi}}
	= \int_{\gdR} \dbar \lambda \cdot e^{- \lambda^2 \tilde{\xi}^2} \cdot \frac{2 (1- \cos (\lambda y))}{\lambda^2}
\end{equation}
Note that we have not given a definite value to the strength $\widetilde{D}$
of the effective disorder $\eta(y)$. To fit the infinite-time exact result \cite{huse_henley_fisher_1985_PhysRevLett55_2924} for $\langle [F(t,y_2)-F(t,y_1)]^2\rangle$, one has to take
\begin{equation}
 \label{def_Dtilde}
		\widetilde{D} = \frac{c D}{T}
\end{equation}
This ensures in particular that the disorder strength $D$ of the full
model (see equation~\eqref{distrV})) has the same dimensions as the
constant $D$ of~\eqref{def_Dtilde} for the toy model.
In addition, the expression~\eqref{def_Dtilde} tells us that the disorder
contribution to the effective free-energy $F_\eta(t,y)$
\eqref{eq:Fyt_toymodel} \emph{depends on temperature},
which is in a way expected since it is aimed at representing
the contribution of all trajectories arriving at the endpoint,
each trajectory being sensitive to thermal fluctuations.

\subsection{GVM and determination of $\sigma (u)$ and $\sigc{u}$}

To determine the roughness at all times $t$ we follow
the lines of \sref{fullGVM_introGVM_Fourier}.
The main difference is that we work in direct space at fixed $t$
so that there are no Fourier transformations to consider, and the GVM solution depends on time $t$
(i.e. of the lengthscale $r$ in the 1D-interface formulation).
The equivalent of the trial Hamiltonian is provided by the quadratic trial free-energy:
\begin{equation}
	F_0 \argp{t,\vec{y}}
	= \frac{1}{2} \sum_{a,b=1}^n y_a \cdot G^{-1}_{ab} (t) \cdot y_b
\end{equation}
parametrized by a $n \times n$ hierarchical matrix whose elements are:
\begin{equation}
	G^{-1}_{ab} (t)
	= c/t \cdot \delta_{ab} - \sigma_{ab}
	\end{equation}
The diagonal term $c/t$ corresponds to the elastic part of the
free-energy, while the off-diagonal elements $\sigma_{ab}$ do not depend on $t$ since $\widetilde{F}$ is local in time.
The variational method yields the following disorder-independent connected parts
\begin{equation}
 \label{toymodel_Gc}
	G^{-1}_c (t) = c/t \quad \Longleftrightarrow \quad G_c (t) = t/c
\end{equation}
thanks to the statistical tilt symmetry as in \eqref{GVM-Gc-1} and \eqref{GVM-Gc-2}, together with self-consistent saddle point equations for the off-diagonal elements
\begin{equation}
	\sigma_{a \neq b}
	= \frac{ \widetilde{D}}{\sqrt{\pi} T} \argc{\tilde{\xi}^2 + T \argp{\widetilde{G} (t) - G_{a \neq b} (t)}}^{-1/2}
\end{equation}
The above is equivalent to the saddle point \eqref{saddle-pt-equation_u_powerlaw} of the full interface but with two important differences: i) there is no integration over Fourier modes $q$ since we work in a direct space representation; ii) the exponent is $-1/2$ instead of $-3/2$, and thus leads to a different asymptotic behaviour.

As in \sref{fullGVM_saddle_pt_equation} we use the more generic full-RSB formulation with the mapping parameter $u \in \argc{0,1}$:
\begin{equation}
 \label{toymodel_saddle-pt-equation_u_powerlaw}
	\sigma (u)
	= \frac{\widetilde{D}}{\sqrt{\pi} T} \argc{\tilde{\xi}^2 + T \argp{\widetilde{G} (t) - G (t,u)}}^{-1/2}
\end{equation}
Following the same procedure as in \sref{fullGVM_determination_ansatz}, we obtain:
\begin{equation}
 \label{toymodel_derivee_sigma_powerlaw}
	\sigma '(u)
	= \sigma ' (u) \cdot \frac{\argp{T \, \sigma (u)}^{3}}{\argp{G^{-1}_c (t) + \sigc{u}}^2} \cdot \frac{\pi}{2 \widetilde{D}^2}
\end{equation}
If we are not on a plateau $\sigma ' (u) =0$, we thus obtain:
\begin{equation}
 \label{toymodel_relation_sigma_sic}
	T \sigma (u)
	= \argp{\frac{2  \widetilde{D}^2}{\pi}}^{1/3} \cdot \argp{G^{-1}_c (t) + \sigc{u}}^{2/3}
\end{equation}
Taking again the derivative $\partial_u$ on this last expression, in order to reintroduce a $u$ dependence, and identifying $\sigma (u)$ with respect to $\argp{G^{-1}_c + \sigc{u}}$, we obtain
\begin{align}
 \label{toymodel_ansatz_sigma}
	\sigma (u) & = \frac{2^3}{3^2 \pi} \frac{\widetilde{D}^2}{T} \cdot \argp{u/T}^2 \\
 \label{toymodel_ansatz_sigma_crochet}
	\sigc{u} & = \frac{2^4}{3^3 \pi} \widetilde{D}^2 \cdot (u/T)^3 - G^{-1}_c (t)
\end{align}
where we recognize the same powerlaw structure $(u/T)^\nu$, imposed by the very GVM procedure, as in \eqref{fullRSB_sigc_solution} and \eqref{fullRSB_sigma_solution}, respectively for $\sigc{u}$ and $\sigma (u)$.
However the effective strength of disorder $\widetilde{D}$ given by \eqref{def_Dtilde} allows to modify this temperature dependence, and thus to circumvent the GVM artefact which leads to a $T$-independent asymptotic roughness for the 1D-interface GVM (see \sref{scaling_analysis}).
Moreover, contrarily to \eqref{fullRSB_sigma_solution} there is no additive constant to $\sigma (u)$, the GVM in direct space provides indeed both the solution for $\sigma (u)$ and $\sigc{u}$. In the Fourier representation we obtained first $\sigc{u}$ and then had to integrate it to recover $\sigma (u)$, which was leading to an additive constant. As a consequence the solution $\sigma (u)$ has to stick to these power-law functions, with possible steps and plateaux inbetween monotonous segments.
	
As for the GVM of the 1D interface, we look for a full-RSB \emph{continuous} solution, without any additional step.
In fact, by definition $\sigc{0}=0$ and consequently this imposes the existence of a plateau for $u$ below a first cutoff $u_* (t)$ which introduces the lengthscale dependence into the solution and later on into the roughness via the connected part \eqref{toymodel_Gc}:
\begin{align}
 \label{toymodel_def_ustar}
	& u_* (t) = \frac{3}{2} \argp{\frac{\pi c}{2 \widetilde{D}^2}}^{1/3} T \cdot t^{-1/3} \\
	& \sigc{u \leq u_*} \equiv \sigc{0} =0 \\
	& \sigma \argp{u \leq u_*} = \sigma (0) = \argp{\frac{2 \widetilde{D}^2 c^2}{\pi}}^{1/3} T^{-1} \cdot t^{-2/3}
\end{align}
The consistency of \eqref{toymodel_saddle-pt-equation_u_powerlaw} for $u \geq u_*$ imposes the existence of a second cutoff $u_c (\tilde{\xi})$ above which $\sigma (u)$ has a plateau, similarly to \eqref{equa_vc_xi-1}.
Indeed, using the inversion formula \eqref{A_replic_Gt-Gu1} for $u_* \leq u \leq u_c$ and taking care of the two plateaux in $\int^1_u dv$, the saddle point equation \eqref{toymodel_saddle-pt-equation_u_powerlaw}  requires again that the $\tilde{\xi}$-dependence cancels the $u_c$ terms as in \eqref{equa_vc_xi-2}.
We thus obtain the following polynomial equation for the second full-RSB cutoff $u_c (\tilde{\xi})$ similarly to \eqref{equa_vc_xi-2}:
\begin{equation}
 \label{toymodel_equa_vc_xi-2}
 \begin{split}
	& u_c^4 = \widetilde{A} \argp{3/4 - u_c} \\
	& \widetilde{A}
	\equiv \frac{3^3}{2^4} \pi \cdot \frac{T^4}{(\tilde{\xi} \widetilde{D})^2}
	= \frac{3^3}{2^4} \pi \cdot \frac{T^6}{(\tilde{\xi} c D)^2}
 \end{split}
\end{equation}
which is equivalent to substituting \eqref{toymodel_ansatz_sigma} and \eqref{toymodel_ansatz_sigma_crochet} into \eqref{toymodel_saddle-pt-equation_u_powerlaw} at $u=u_c$.
As for the GVM of the 1D interface, it can be shown that the factor $3/4$ is closely related to the GVM prediction for the asymptotic roughness exponent, which in the case of the toy model will be the exact random manifold exponent for a 1D interface: $u_c < 3/4 = (2 \zeta_{\text{RM}})^{-1}$.
	
An explicit analytical expression for $u_c \in \argc{0,3/4}$ can be obtained for the two opposite limits for $\widetilde{A}$, which at fixed $\tilde{\xi}, \widetilde{D} >0$ correspond respectively to $T \to 0$ and $T \to \infty$:
\begin{align}
 \label{toymodel_uc_small_T}	
	u_c
	& \stackrel{\widetilde{A} \to 0}{\approx}
		\frac{3 \pi^{1/4}}{2^{3/2}} \cdot \frac{T}{ (\tilde{\xi} \widetilde{D})^{1/2}}
	=\frac{3 \pi^{1/4}}{2^{3/2}} \cdot \frac{T^{3/2}}{ (\tilde{\xi} c D)^{1/2}}
	\to 0 \\
 \label{toymodel_uc_large_T}	
	u_c & \stackrel{\widetilde{A} \to \infty}{\approx} 3/4 + 0^-
\end{align}
The polynomial equation for $u_c (\tilde{\xi})$ \eqref{toymodel_equa_vc_xi-2} leads to the analogous definition of a `critical' temperature with respect to \eqref{Atilde_1_def_Tc} between the low and high temperature regime, using \eqref{def_Dtilde}:
\begin{equation}
 \label{toymodel_Atilde_1_def_Tc}
	\widetilde{A} =1
	\Longleftrightarrow
	\frac{T_c}{(\tilde{\xi} c D)^{1/3}}
	= \argp{\frac{2^4}{3^3 \pi}}^{1/6}
	\approx 0.76
\end{equation}
and this last constant is of order 1.
This criterion shows again that the limits of the different parameters, in particular $T \to 0$ and $\xi \to 0$ cannot be exchanged.
Moreover the scaling between the parameters $\arga{T,\tilde{\xi},D}$ is exactly the same as in \eqref{Atilde_1_def_Tc} for the GVM on the 1D interface.
	
Using the inversion formulas of hierarchical matrices in $\lim_{n \to 0}$, the physically relevant parameters are $\sigc{u}$ and $\sigma (0)$, so this full-RSB solution can be summarized by:
\begin{equation}
 \label{toymodel_ansatz_fullRSB}
 \begin{split}
	& \sigma_{\text{RSB}} (0) = \argp{\frac{2 \widetilde{D}^2 c^2}{\pi}}^{1/3} T^{-1} \cdot t^{-2/3}  \\
	& \argc{\sigma}_{\text{RSB}} (u \leq u_*) = \argc{\sigma}_{\text{RSB}} (u \geq u_c) = 0 \\
	& \argc{\sigma}_{\text{RSB}} (u_* \leq u \leq u_c) = \frac{2^4}{3^3 \pi} \widetilde{D} \cdot (u/T)^3 - c/t
 \end{split}
\end{equation}
To be consistent the Ansatz must also satisfy $0 \leq u_* (t) < u_c (\tilde{\xi}) \leq 3/4$, and the collapse of those two cutoffs actually defines the lowest time $t_c$ for the existence of a full-RSB segment in the solution:
\begin{equation}
 \label{toymodel_def_Larkin_tc}
	u_* (t_c) \equiv u_c (\tilde{\xi})
\end{equation}
This condition is equivalent to $\sigc{u_*(t_c)} = \sigc{u_c (\tilde{\xi})}$ and can be reformulated as:
\begin{equation}
 \label{toymodel_Larkin_tc_T_vc}
	t_c = \frac{3^3 \pi}{2^4} \frac{c}{\widetilde{D}^2} \cdot \argp{\frac{T}{u_c}}^3
		= \frac{3^4 \pi}{2^6} \frac{c T^3}{\widetilde{D}^2} u_c^{-4} - \frac{c \tilde{\xi}^2}{T}
\end{equation}
This crossover time can be made explicit in the two limits \eqref{toymodel_uc_small_T} and \eqref{toymodel_uc_large_T} plugged in \eqref{toymodel_def_ustar}. This gives respectively:
\begin{align}
 \label{toymodel_dev_tc_lowT}
	t_c & \stackrel{\widetilde{A} \to 0}{\approx} 2^{1/2} \pi^{1/4} \cdot \tilde{\xi}^{3/2} c \widetilde{D}^{-1/2} \\
 \label{toymodel_dev_tc_highT}
	t_c & \stackrel{\widetilde{A} \to \infty}{\approx} 4 \pi \frac{T^3 c}{\widetilde{D}^2}
\end{align}
For smaller lengthscales $t \leq t_c$ the two plateaux merge into a RS Ansatz. Following \eqref{A_RS-inversion-formulas} and \eqref{toymodel_Gc}, we have then $\tilde{G} (t) - G(t,u) = G_c (t)  = t/c$ and the saddle-point equation \eqref{toymodel_saddle-pt-equation_u_powerlaw} becomes:
\begin{equation}
 \label{toymodel_ansatz_RS}
 \begin{split}
	& \sigma_{\text{RS}} (u) = \sigma_{\text{RS}}  (0)
	   = \frac{\beta \widetilde{D}}{\sqrt{\pi}} \argp{\tilde{\xi}^2 + \frac{T t}{c}}^{-1/2} \\
	& \argc{\sigma}_{\text{RS}} (u) = 0
 \end{split}
\end{equation}

\begin{figure}[htbp]
 \subfigure[\label{fig_toymodel-sigmacrochet-totale_sigma0}]{\includegraphics[width=\figwidth]{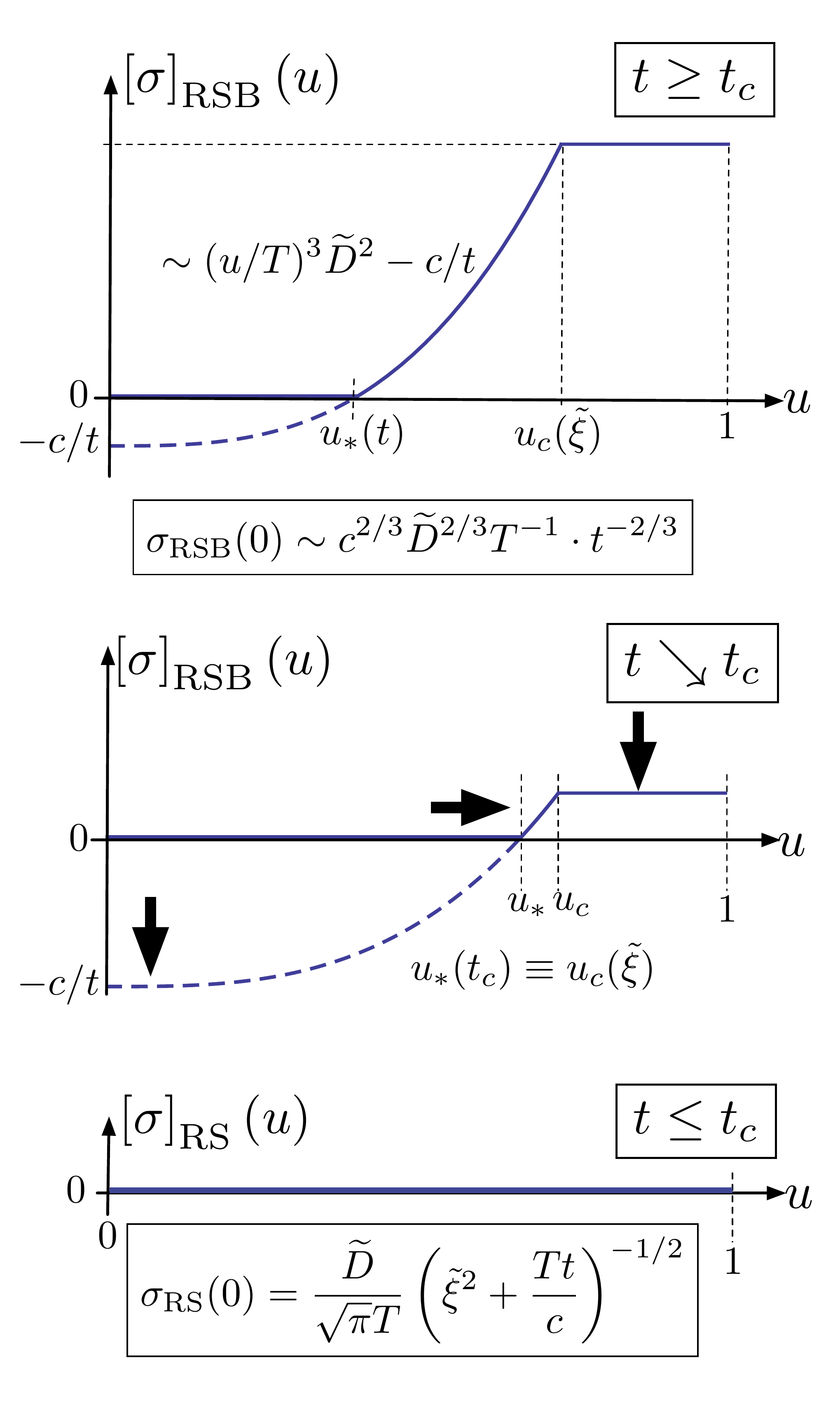}}
 \subfigure[\label{fig:toymodel_ucAtilde} ]{\includegraphics[width=\figwidth]{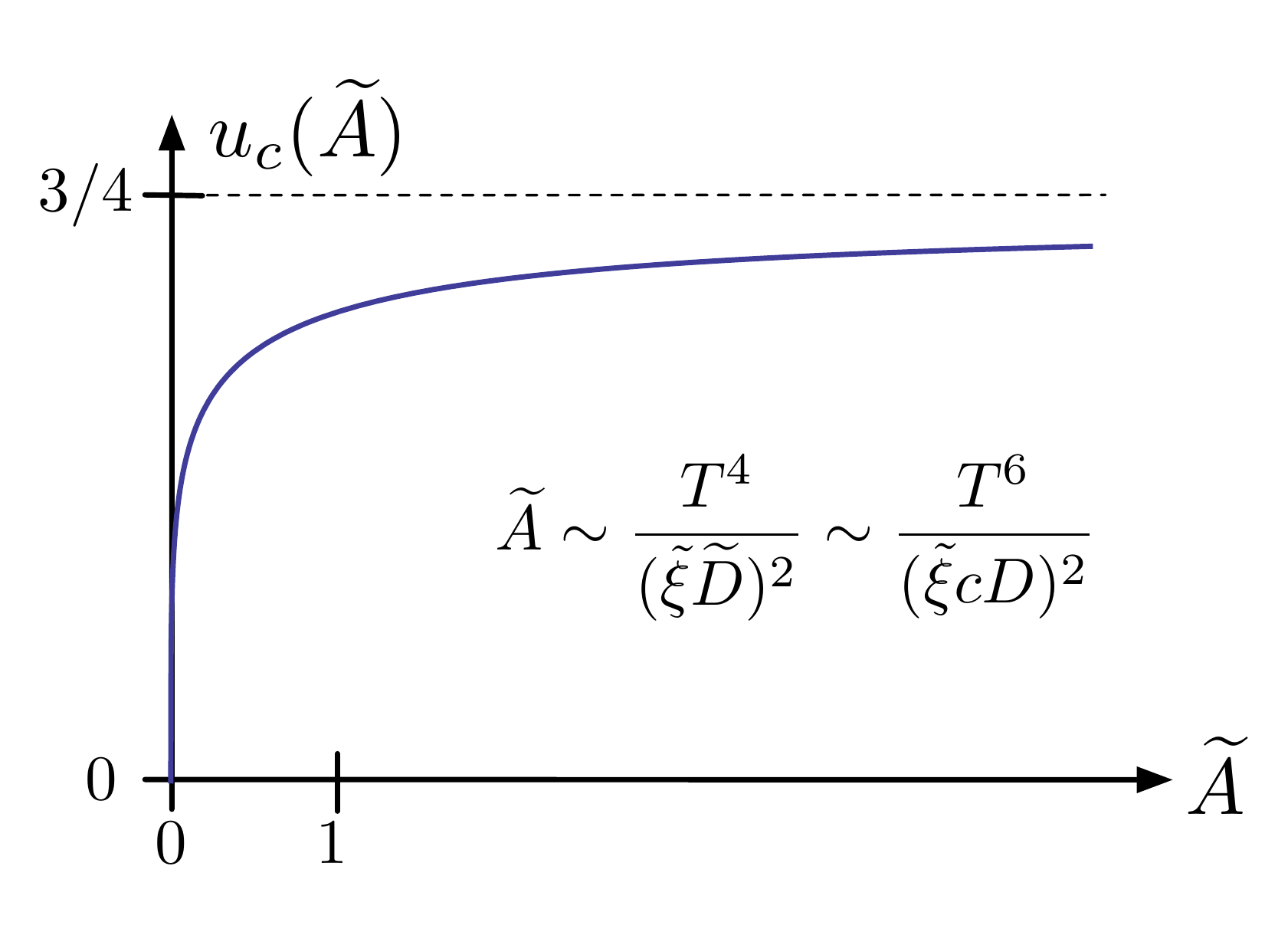}}
 \caption{\label{fig:toymodel_GVMsolution}
	GVM solution for the toy model.
	\subref{fig_toymodel-sigmacrochet-totale_sigma0} Self-energy $\sigc{u}$ given by \eqref{toymodel_ansatz_fullRSB}. At large times (large lengthscales) $t \geq t_c$ it is full-RSB with two plateaux which merge at $t=t_c$, as indicated by the black arrows, and at small times (small lengthscales) $t \leq t_c$ it is RS.
	\subref{fig:toymodel_ucAtilde} Full-RSB cutoff $u_c$ as a function of $\widetilde{A} \sim \frac{T^4}{(\tilde{\xi} \widetilde{D})^2} \sim \frac{T^6}{(\tilde{\xi} c D)^2}$, obtained by solving the polynomial equation \eqref{toymodel_equa_vc_xi-2}. It starts linearly at $\widetilde{A} \to 0$ and saturates to $3/4$ at $\widetilde{A} \to \infty$ similarly to \fref{fig:interface1D_vcAtilde}, but $\widetilde{D}= \widetilde{D}(T)$ will modify its low-temperature dependence \eqref{toymodel_uc_small_T}.
	}
\end{figure}

So the GVM in direct space of the DP toy model yields a solution of the saddle-point equation \eqref{toymodel_saddle-pt-equation_u_powerlaw} which depends explicitly on the time $t$, and consequently on the length of the DP; it is RS for $t \leq t_c$ and becomes continuously full-RSB with two plateaux for $t \geq t_c$.
This is summarized in \fref{fig:toymodel_GVMsolution}.
The breaking of the replica symmetry at $t=t_c (\tilde{\xi})$ has the same source as the full-RSB cutoff $v_c (\xi)$ in the GVM of the 1D interface: it marks the transition from a small-lengthscales regime where the DP starts by fluctuating thermally and crosses smoothly over to the large-lengthscales RM regime where the disorder-induced metastability is dominant \cite{mezard_parisi_1992_JPhysI02_2231}.
In the next section we compute the corresponding roughness, which also gives access to the actual crossover from the thermal to the RM regime depending on the explicit finite width $\tilde{\xi}$.

\subsection{Computation of the roughness} \label{toymodel_roughness}

The roughness of the 1D interface \eqref{def_roughness_Br} translates into the transverse fluctuations of the DP's endpoint $\moydes{\moy{(y(t)-y(0))^2}}$, and since $y(0) \equiv 0$ the corresponding GVM correlation function is given by:
\begin{equation}
 \label{def_roughness_toymodel}
	\moydes{\moy{y(t)^2}}
	= T \lim_{n \to 0} \widetilde{G} (t)
\end{equation}
Using the inversion formula \eqref{A_replic_Gt} we eventually obtain:
\begin{align}
	& t \geq t_c \, : \, \lim_{n \to 0} \widetilde{G} (t)
	= \frac{3}{2} \argp{\frac{2 \widetilde{D}^2}{\pi c^4} }^{1/3} t^{4/3} - \tilde{\xi}^2 \\
	& t \leq t_c \, : \, \lim_{n \to 0} \widetilde{G} (t)
	= \frac{T t}{c} + \frac{\widetilde{D}}{c^2 \sqrt{\pi}} \cdot t^2 \argp{\tilde{\xi}^2 + \frac{T t}{c}}^{-1/2}
	\end{align}
We can check that this quantity and its derivative are continuous, in particular at the junction of the RS and full-RSB solutions:
\begin{align}
 \label{toymodel_roughness_at_tc}
	& \lim_{n \to 0} \widetilde{G}_{\text{RS}} (t_c)
	= \lim_{n \to 0} \widetilde{G}_{\text{RSB}} (t_c)
	= \tilde{\xi}^2 \cdot \frac{3/2 + u_c}{3/4 - u_c} \\
 \label{toymodel_roughness_derivative_at_tc}
	& \partial_t \lim_{n \to 0} \widetilde{G}_{\text{RS}} (t) \vert_{t=t_c}
	= \partial_t \lim_{n \to 0} \widetilde{G}_{\text{RSB}}  (t) \vert_{t=t_c}
	= \frac{3 T}{u_c}
\end{align}

In order to compare the GVM predictions for the 1D interface and the rounded toy model, we use \eqref{def_Dtilde} to recover the physical temperature dependence of $\widetilde{D}$ and we identify:
\begin{equation}
	r_0 = t_c
	\, , \quad
	B(r) = \moydes{\moy{y(t)^2}}\vert_{t=r}
\end{equation}
Keeping track of the RS and full-RSB solution we obtain the roughness predicted by GVM on the toy model:
\begin{align}
 \label{toymodel_roughness_Br_RS}
	& r \geq r_0 \, : \, B_{\text{RSB}} (r)
	= \frac{3}{2} \argp{\frac{2 D^2}{\pi c^2 T^2} }^{1/3} r^{4/3} - \tilde{\xi}^2 \\
 \label{toymodel_roughness_Br_fullRSB}
	& r \leq r_0 \, : \, B_{\text{RS}} (r)
	= \frac{T r}{c} + \frac{D}{c T \sqrt{\pi}} \cdot r^2 \argp{\tilde{\xi}^2 + \frac{T r}{c}}^{-1/2}	
\end{align}
with the characteristic lengthscale associated to the full-RSB cutoff $u_c (\tilde{\xi})$ \eqref{toymodel_equa_vc_xi-2}:
\begin{equation}
 \label{toymodel_Larkin_r0_T_vc}
	r_0
	= \frac{3^3 \pi}{2^4} \frac{T^5}{c D^2} \cdot u_c^{-3}
\end{equation}
This length differs from \eqref{Larkin_r0_T_vc} essentially by the scaling of $u_c$.

\begin{figure}
 \subfigure[\label{fig:graph_roughness_lowT_toymodel} Low-temperature regime]{\includegraphics[width=\figwidth]{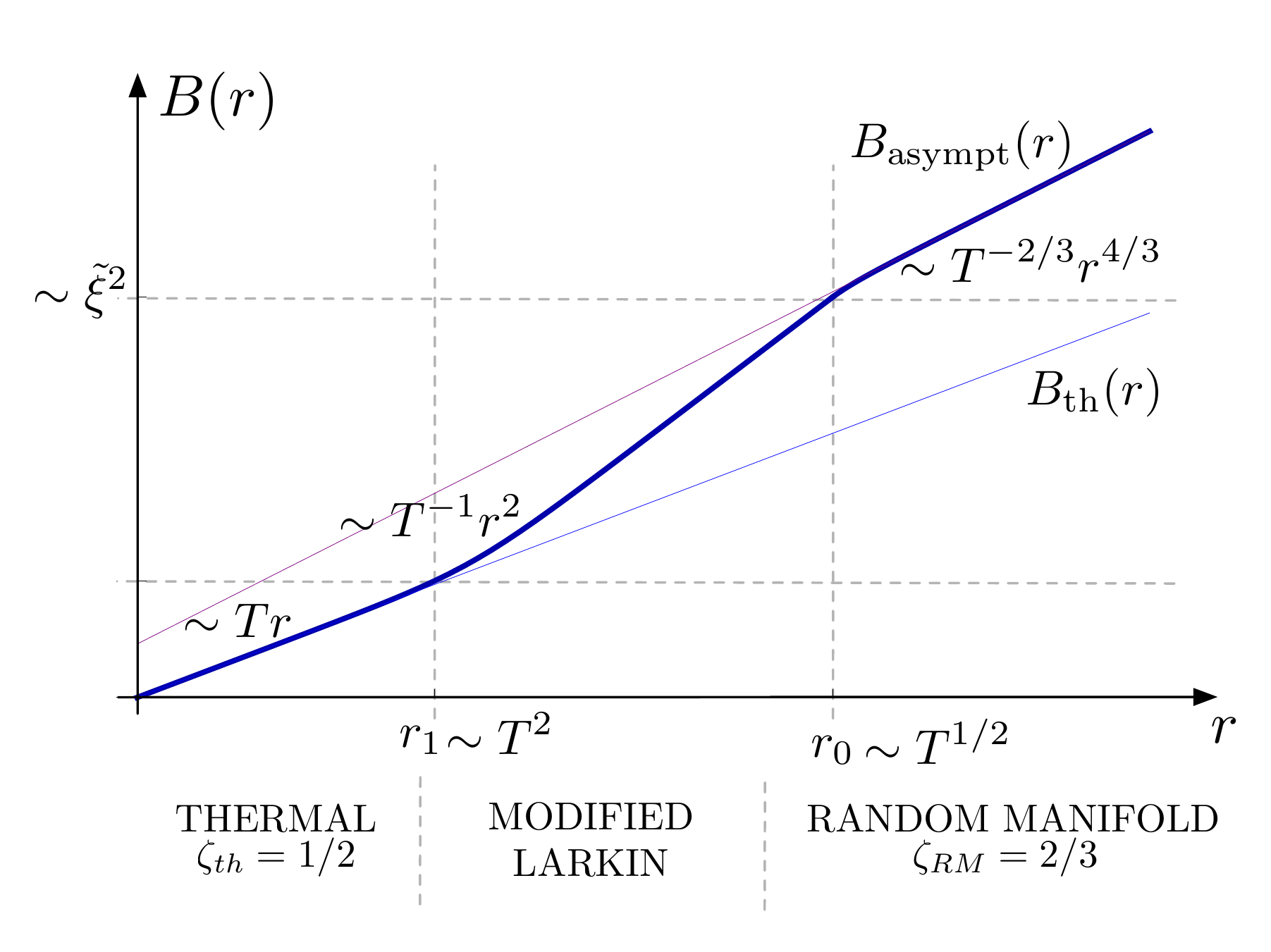}}
 \subfigure[\label{fig:graph_roughness_highT_toymodel} High-temperature regime]{\includegraphics[width=\figwidth]{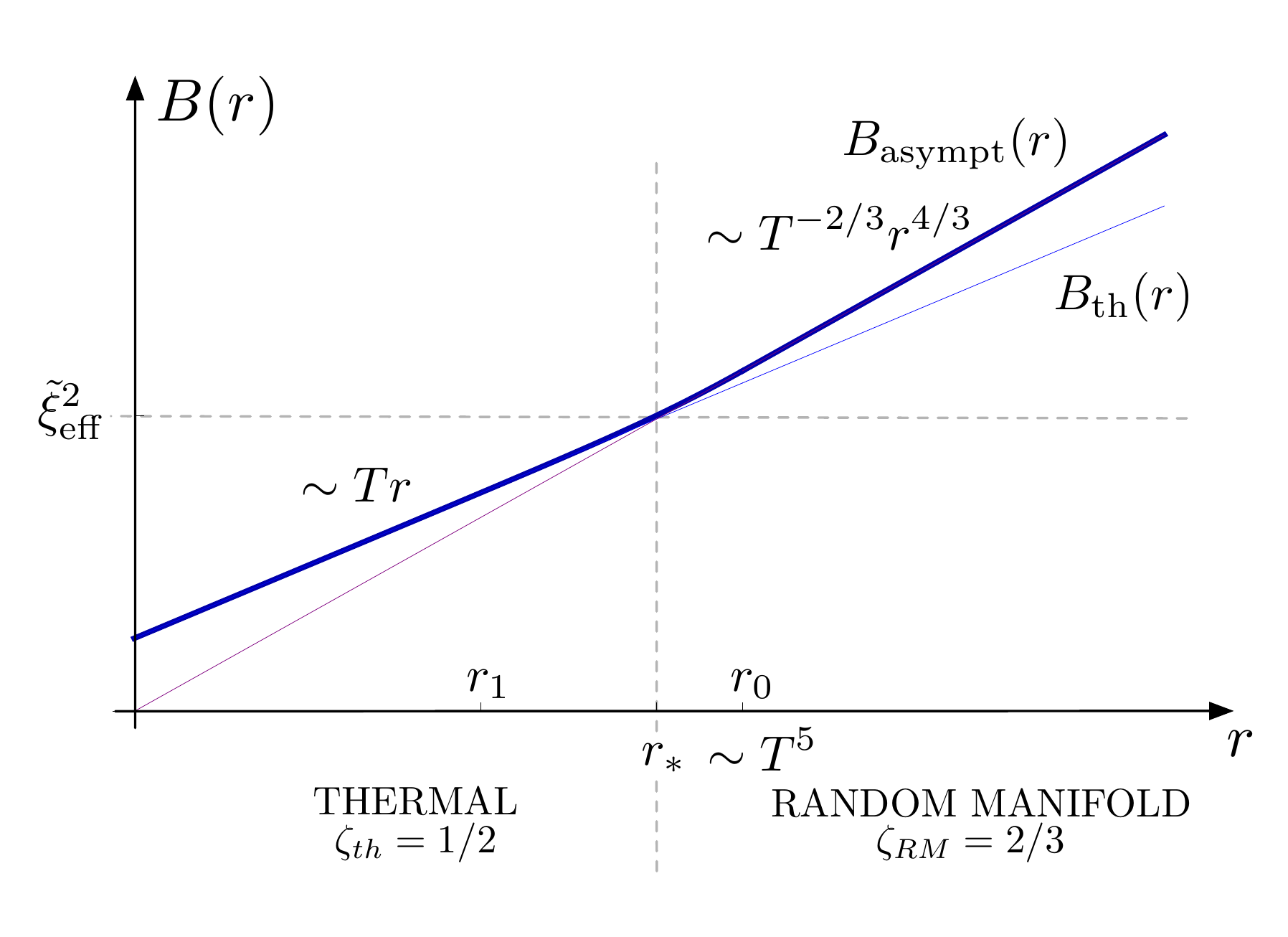}}
\caption{ \label{fig_graph_roughness_toymodel_compil-1}
 	GVM prediction for the toy-model static roughness $B(r)$, in $\log-\log$ representations; the slope of the curves corresponds to $2 \zeta(r)$ as defined by \eqref{def_roughness_exponent_loglog}. It can be compared to \fref{fig_graph_roughness_compil-1}.
 	($\tilde{\xi}=c=D=1$, \subref{fig:graph_roughness_lowT_toymodel}~$T=10^{-3}$, \subref{fig:graph_roughness_highT_toymodel}~$T=10$)}
\end{figure}

We show in \fref{fig_graph_roughness_toymodel_compil-1} two graphs illustrating the low versus high temperature regimes of $B(r)$.
A summary of the different roughness regimes along with their corresponding exponent $\zeta$ and their crossover lengthscales is given by \fref{fig:toymodel_roughness_summary}.

\subsection{Roughness regimes and crossover lengthscales} \label{toymodel_regimes_crossovers}

\begin{figure}
 \includegraphics[width=\figwidth]{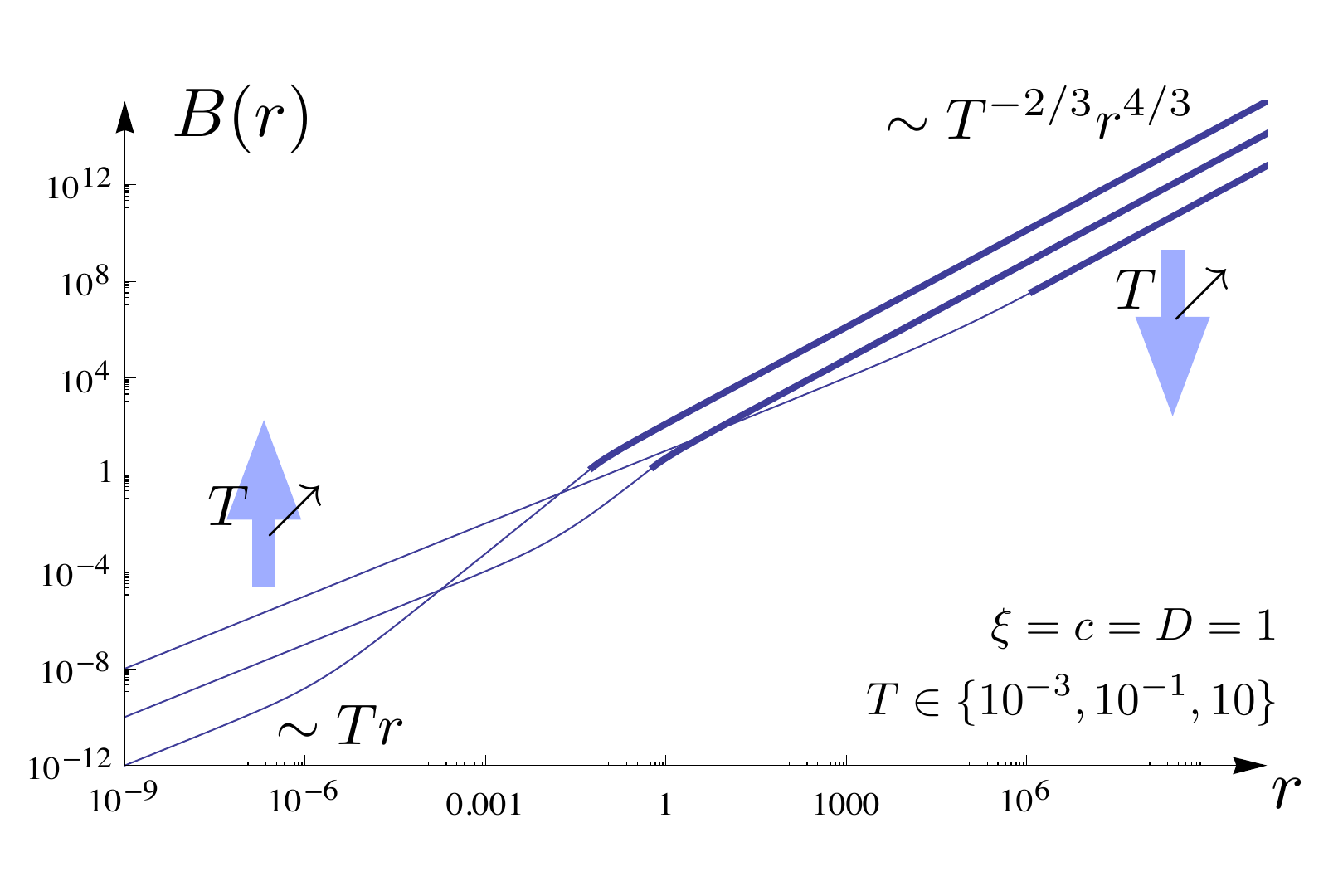}
 \caption{\label{fig:graph_roughness_fixedD_toymodel}
 	GVM roughness at fixed disorder for the toy model, on four orders of magnitude for the temperature: the thick curves correspond to $B_{\text{RSB}} (r)$, the thin ones to $B_{\text{RS}} (r)$,
and they connect at the Larkin length $r_0$. Unlike \fref{fig:graph_roughness_fixedD} the curves intersect because of the asymptotic scaling $\sim T^{-2/3}$ and become thus unphysical in the zero-temperature limit.}
\end{figure}

We recover at small lengthscales the same thermal regime of exponent $\zeta_{\text{th}}=1/2$ as in \sref{full_roughness_regime_thermal}, whereas at large lengthscales the power-law behaviour is directly given by \eqref{toymodel_roughness_Br_RS} which predicts the \emph{exact} RM exponent $\zeta_{RM}=2/3$ of a 1D interface instead of the Flory exponent $\zeta_F=3/5$ obtained in \sref{full_roughness_regime_RM}.
Note that $\zeta_{RM}=2/3$ coincides with an exact prediction for the original toy model ($\tilde{\xi}=0$) \cite{schulz_1988_JStatPhys51_1}.
This leads in particular to an asymptotic temperature-dependence $\sim T^{-2/3}$, which makes the $B(r)$ curves intersect with each other upon varying $T$, as illustrated by \fref{fig:graph_roughness_fixedD_toymodel}.

As in \sref{full_roughness_Larkin_xieff} the lengthscale $r_0=t_c$ corresponds to the Larkin length of the system (associated to the appearance of full RSB), at least if we define an effective width $\tilde{\xi}_{\text{eff}}$ at high temperatures.
Indeed, the expansions \eqref{toymodel_dev_tc_lowT} and \eqref{toymodel_dev_tc_highT} give the following analytical expressions for $r_0$:
\begin{align}
 \label{toymodel_larkin_r0_low_T}	
	r_0
	& \stackrel{T \to 0}{\approx}
	2^{1/2} \pi^{1/4} \cdot \tilde{\xi}^{3/2} c D^{-1/2} T^{1/2} \\
 \label{toymodel_larkin_r0_high_T}
	r_0
	& \stackrel{T \to \infty}{\approx}
	4 \pi \frac{T^5}{c D^2}
\end{align}	
which, when compared to \eqref{larkin_r0_low_T} and respectively to \eqref{larkin_r0_high_T} predicts exactly the same scaling of parameters at high temperatures, but introduces a temperature dependence $\sim T^{1/2}$ and slightly modifies the scaling-exponents values at low temperatures.
Combining \eqref{toymodel_roughness_at_tc} with \eqref{toymodel_equa_vc_xi-2} the roughness at $r_0$ is given by:
\begin{equation}
 \label{toymodel_roughness_at_larkin}
	B(r_0)
	= \tilde{\xi}^2 \cdot \frac{3/2 + u_c}{3/4 - u_c}
	= \frac{3^3}{2^4} \pi \frac{T^6}{(c D)^2} \cdot u_c^{-4}
\end{equation}
so we have respectively, in the two opposite temperature limits \eqref{toymodel_uc_small_T} and \eqref{toymodel_uc_large_T}:
\begin{align}
 \label{toymodel_roughness_r0_low_T}
	B(r_0)
	& \stackrel{T \to 0}{\approx}
	\frac{1}{2} \tilde{\xi}^2 \\
 \label{toymodel_roughness_r0_high_T}
	B(r_0)
	& \stackrel{T \to \infty}{\approx}
	\frac{2^4}{3} \pi \frac{T^6}{(c D)^2}
	\equiv \tilde{\xi}_{\text{eff}}^2
\end{align}
Remarkably, except for the numerical factors, both these expressions are strictly equivalent to \eqref{roughness_r0_low_T} and \eqref{roughness_r0_high_T} obtained in the previous GVM procedure.
This is compatible with a generic scaling analysis of the model (see \sref{scaling_analysis}).

Below the Larkin length the roughness is described by the RS solution:
\begin{equation}
 \begin{split}
	& B_{\text{RS}} (r)
	= B_{\text{th}} (r) + \frac{\xi^3 c D}{\sqrt{\pi} T^3} \cdot f(r/r_c) \\
	& r_c
	\equiv c \tilde{\xi}^2 / T
	\Leftrightarrow B_{\text{th}} (r_c)=\tilde{\xi}^2 \\
 \label{toymodel_crossover_function}
	& f(x) = \frac{x^2}{\sqrt{1+x}}
 \end{split}	
\end{equation}
where the dimensionless $f(x)$ describes the actual shape of the crossover along with the definition of $r_c$.
The thermal roughness obtained from the rounded toy model differs on two points from its counterpart for the 1D interface, as a consequence of the GVM in direct-space versus Fourier representation: i) the RS solution encodes both the thermal and the intermediate crossover regimes, instead of the thermal regime only, ii) the crossover function is given explicitly after the retrieval of the thermal roughness, whereas in \eqref{roughness_dis_full} it was given as a series.
Taylor expanding $f(x)$ around $x=0$, we have:
\begin{equation}
 \label{toymodel_crossover_Taylor}
	B(r)
	\stackrel{r \to 0}{\approx}	
	\frac{T r}{c} + r^2 \frac{D}{\sqrt{\pi} \tilde{\xi} c T} - r^3 \frac{D}{2 \sqrt{\pi} c^2 \tilde{\xi}^3} + \gdO \argp{Tr^4}
\end{equation}
so the end of the thermal regime can be defined similarly to \eqref{def_r1_vc} as the quadratic take-off $\sim r^2$:
\begin{equation}
	\frac{T r_1}{c}
	\equiv r_1^2 \frac{D}{\sqrt{\pi} \tilde{\xi} c T}
	\Leftrightarrow
	r_1 = \frac{T^2 \tilde{\xi}}{ \sqrt{\pi} D}
\end{equation}
The collapse of $r_1$ and $r_0$ yields the analogous criterion to \eqref{1Dinterface_collapse_r1r0} for a `critical' temperature:
\begin{equation}
 \label{toymodel_collapse_r1r0}
	r_1 = r_0
	\Longleftrightarrow
	u_c = \frac{3 \pi^{1/6}}{2^{4/3}} \frac{T}{(\tilde{\xi} c D)^{1/3}}
\end{equation}
which is compatible with the previous criterion \eqref{toymodel_Atilde_1_def_Tc} on the GVM solution itself.

As for the zero-temperature limit, it is ill-defined, since $r_0 \to 0$ and $B_{\text{dis}} (r_0)$ is finite \eqref{toymodel_roughness_r0_low_T}, whereas its derivative and $B_{\text{RSB}}(r)$ itself clearly diverge according to \eqref{toymodel_roughness_derivative_at_tc} and  \eqref{toymodel_roughness_Br_fullRSB}.
The toy model is actually an effective model at \emph{finite} temperature, and is thus non-physical for this particular limit.

Finally at high temperatures there is a single crossover lengthscale between the thermal and the RM regime, denoted $r_*$ and scaling as \eqref{def_rstar}:
\begin{equation}
 \label{toymodel_def_rstar}
	\frac{Tr_*}{c} \equiv \frac{3}{2} \argp{\frac{2 D^2}{\pi c^2 T^2}}^{1/3} r_*^{4/3}
	\Longleftrightarrow
	r_* = \frac{4 \pi}{27} \frac{T^5}{c D^2}
\end{equation}
which is equivalent up to a factor $1/27$ to the Larkin length $r_0$ at high temperatures and can be predicted by scaling arguments.

\section{Scaling analysis} \label{scaling_analysis}

In order to explain the universal expressions of $T_c$, $r_0$ and $\xi_{\text{eff}}$, 
we follow in spirit the scaling analysis presented in
Refs.~\onlinecite{calabrese_2010_EPL90_20002,bustingorry_arXiv:1006.0603}, but keeping $\xi$
finite.
We fully harness the scaling relations of the Hamiltonian and the roughness so as to derive identities between different scaling exponents, and also to show how the Flory exponent $\zeta_F$ arises naturally even if the so-called Flory argument (see \aref{A-Flory}) does not apply.
In particular, we account for the role of the GVM approximation in those
scalings, for both the 1D interface and the rounded toy model.

For the sake of generality, we consider a manifold of internal
dimension $d$ with $m$ transverse components, subjected to a
random-bond potential $\widetilde V$.  The weight of one
configuration reads (see \sref{model_model}):
\begin{equation}
   \int\mathcal Du
  \exp\Big\{-\beta\int d^dz\:\big[ \frac c2 (\nabla_zu)^2+\widetilde V(u(z),z)\big]\Big\}
  \label{eq:defWy1x1}
\end{equation}
The distribution of $\widetilde V$ is assumed to be Gaussian with correlations
\begin{equation}
  \overline{\widetilde V(u,z) \widetilde  V(u',z')}\ =\ D\: \frac 1{\xi^m} R\Big(\frac{u-u'}\xi\Big)\:\delta^{(d)}(z-z')
\end{equation}
$\xi$ being the microscopic disorder correlation length and $R$ a dimensionless function.
Let's perform the change of variable
\begin{equation}
  u= a \bar u
  \qquad
  \qquad
  z= b \bar z
  \label{eq:change-of-variable_uubar-zzbar}
\end{equation}
We note that the elastic part of the Hamiltonian rescales as:
\begin{equation}
  \int d^dz\: \frac c2 (\nabla_zu)^2
  \ =\
  b^{d-2} a^2 \ \int d^d\bar z\: \frac c2 (\nabla_{\bar z}\bar u)^2
  \label{eq:rescaling_Hel}
\end{equation}
while (in distribution) the disorder part rewrites:
\begin{equation}
   \int d^dz\: \widetilde V(u,z)
  \ =\
  b^{d/2} a^{-m/2} \
  \int d^d\bar z\: \bar V(\bar u,\bar z)
  \label{eq:rescaling_Hdis}
\end{equation}
where $\bar V$ has microscopic correlation length $\xi/a$.
Searching for a \emph{unique} exponent $\zeta$ such that $a=b^\zeta$
and such that the elastic and the disorder parts of the Hamiltonian both scale
in the same way, we find
\begin{equation}
 \zeta=\zeta_F=\frac{4-d}{4+m}
 \label{eq:zeta_F-scaling}
\end{equation}
Starting from the definition of the roughness
\begin{align}
  B(z)=
  \int\mathcal D\widetilde V\frac {P[\widetilde V]}{Z_{\widetilde V}} \int \mathcal Du\ \big(u(z)-u(0)\big)^2
  e^  {-\beta \mathcal H[u,\widetilde V]}
\label{eq:explicit_B-of-z}
\end{align}
and denoting
\begin{equation}
  \chi_F=2(\zeta_F - \zeta_{\text{th}})
  \label{eq:def_generic_chiF}
\end{equation}
(one has $\zeta_{\text{th}}=\frac{2-d}{2}$ in the short-range elastic case, and $\chi_F>0$),
one eventually obtains that the roughness obeys the following scaling relation
\begin{equation}
  B(z;c,D,T,\xi)
   =
  b^{2\zeta_F}
  B\big(b^{-1}z;c,D,b^{-\chi_F}T, b^{-\zeta_F} \xi\big )
 \label{eq:rescaling_B-b}
\end{equation}
Note that this scaling relation also holds in the GVM approach for
the interface: indeed, the replicated
Hamiltonian~\eqref{Ham_DES_effectif} and thus its quadratic GVM
equivalent~\eqref{def_quadr_repli_Ham} both rescale in the same way as
the original Hamiltonian
upon~\eqref{eq:change-of-variable_uubar-zzbar}, with $a=b^{\zeta_F}$
as implied by the scaling of the elastic contribution $G_c^{-1}(q) = cq^2$.

In a high-temperature regime the thermal fluctuations wash out the presence of $\xi$  and thus the roughness is expect to be $\xi$-independent. It follows that one can probe a scale-invariant behaviour (e.g. at small or large $z$) of the form
\begin{equation}
  B(z)\sim \text{cte\ }T^{2\text{\thorn}}z^{2\zeta}
  \label{eq:def-thorn}
\end{equation}
where we have isolated the $T$-dependence of the prefactor of
$z^{2\zeta}$, defining the \emph{thorn} exponent $\text{\thorn}$.
%
Choosing $b=T^{1/\chi_F}$ we obtain from~\eqref{eq:rescaling_B-b}
that in each regime where the scaling law~\eqref{eq:def-thorn} applies, the following relation holds
\begin{equation}
  \text{\thorn}=\frac{\zeta_F-\zeta}{\chi_F}
  \label{eq:zeta_and_thorn}
\end{equation}
This last relation is always
verified in the thermal regime (at small lengthscales $z$), where
$\zeta=\zeta_{\text{th}}=\frac{2-d}2$ which
from~\eqref{eq:def_generic_chiF} yields
$\text{\thorn}_{\text{th}}=\frac 12$ as expected.
Denoting by $\zeta_{\text{RM}}$ the roughness exponent at large $z$,
one has in particular
\begin{equation}
  \text{\thorn}_{\text{RM}}=\frac{\zeta_F-\zeta_{\text{RM}}}{\chi_F}
  \label{eq:zeta_and_thorn_RM}
\end{equation}
Note however that any relevance of $\xi$ would \textit{a priori} add a $T$-dependent prefactor to the power-law behavior \eqref{eq:def-thorn}.
In the large $z$ regime, the value of $\zeta$ depends on the model
details for disorder: we note for instance
that~\eqref{eq:zeta_and_thorn_RM} is verified in the 1D-interface GVM approach, where
as noted previously the roughness is then temperature-independent at large
scale ($\text{\thorn}_{\text{RM}}^{\text{1D}}=0$) and the roughness exponent
arising from the computation is
$\zeta_{\text{RM}}^{\text{1D}}=\zeta_F$,
whereas the exact value $\zeta_{\text{RM}}^{\text{exact}}=2/3$ predict for \eqref{eq:zeta_and_thorn_RM} $\text{\thorn}=-\frac{1}{3}$ which is compatible with the numerical simulations of Ref.~\onlinecite{bustingorry_arXiv:1006.0603}.

Our previous considerations are based on exact relations
-~forgetting about cutoffs in Fourier modes~- arising from scaling.
Let's now examine the (usually approximate) Flory argument and
determine how it fails to gives the correct $\zeta_{\text{RM}}$ for
the 1D interface.
It consists again in searching for an exponent $\zeta$ such that
$a=b^\zeta$ and such that the elastic~\eqref{eq:rescaling_Hel} and the
disorder~\eqref{eq:rescaling_Hdis} parts of the Hamiltonian both scale
in the same way, \emph{and} in assuming scale invariance i.e. that the roughness is
a unique power law $B(z)\sim a^{2\zeta_F}\sim z^{2\zeta_F}$ at all lengthscales, for $b=z$ absorbing
all $z$-dependence through~\eqref{eq:change-of-variable_uubar-zzbar}
--~see also \aref{A-Flory}.
We see however that indeed taking $b=z$ in~\eqref{eq:rescaling_B-b},
one would find $\zeta_{\text{RM}}=\zeta_F$ for large $z$ if it were
true that $B\big(1;c,D,z^{-\chi_F}T, z^{-\zeta_F} \xi\big )$ was
independent of $z$ for large $z$. This last assumption is however wrong
 since in general $\text{\thorn}\neq 0$.

The scaling arguments we have explicited above can also be extended as follows:
we now search in~\eqref{eq:rescaling_B-b} for $a$ and $b$ functions of $c$, $D$ and $T$
so as to absorb in $\beta \mathcal H$ all the dependence in $c$, $D$ and $T$.
We first notice that the random potential scales in distribution as
\begin{equation}
  \widetilde V(u,z) = a^{-\frac m2}b^{-\frac d2}\bar V(\bar u, \bar z)
\label{eq:rescaling_potential-V}
\end{equation}
where $\bar V$ is a random potential of Gaussian distribution with
$D$-independent variance and of microscopic correlation length $\bar\xi = \xi/a$,
or in other words
$
\overline{\bar V(\bar u,\bar z)V(\bar u',\bar z')}=
{\bar \xi}^{-m} R\big((\bar u-\bar u')/\bar\xi\big)\:\delta^{(d)}(\bar z-\bar z')
$.
Then one checks that
$
\beta \mathcal H[u,V]=\mathcal H[\bar u,\bar V]|_{c=D=1}
$
is fulfilled provided (see also Ref.~\onlinecite{bustingorry_arXiv:1006.0603}) that:
\begin{equation}
 a= \big(c^{-d} D^{d-2} T^{4-d}\big)^{\frac 1{(4+m)\chi_{\text F}}}
 \quad
 b= \big(c^{-m} D^{2} T^{4+m}\big)^{\frac 1{(4+m)\chi_{\text F}}}
 \label{eq:solution-a-b}
\end{equation}
%
This yields the scaling form
\begin{equation}
  B(z;c,D,T,\xi)
   =
   a^2
  B_1\big(b^{-1}z;a^{-1}\xi\big )
 \label{eq:rescaling_B-a-b}
\end{equation}
where $B_1\big(\bar z;\bar\xi\big )$ is the roughness at distance $\bar z$
with $c=D=T=1$ and disorder correlation length $\bar\xi$.
Let's examine the information contained in the scaling
form~\eqref{eq:rescaling_B-a-b}, which holds for both the interface
and the toy model with or without GVM, as directly checked.  We first
note that if $a^{-1}\xi$ is small enough, the existence of a small
disorder correlation length can be ignored; this defines a
characteristic temperature
\begin{equation}
T_c= c^{\frac d{4-d}} D^{\frac{2-d}{4-d}}\xi^{\frac{\chi_F}{\zeta_F}}
\label{eq:def_Tc-scaling}
\end{equation}
above which the effects of $\xi$ should be irrelevant. For the 1D
interface ($d=m=1$), one recovers indeed the same characteristic
temperature $T_c\sim(\xi c D)^{\frac 13}$ of~\eqref{Atilde_1_def_Tc}
and~\eqref{toymodel_Atilde_1_def_Tc} respectively for the 1D-interface and rounded toy-model GVM approaches.

Besides, in the now well-defined high-temperature regime $T>T_c$ the roughness should
scale as $ B(z;c,D,T) = a^2 B_1\big(b^{-1}z) $ independently of $\xi$,
and display two asymptotic regimes with no intermediate regime (see
Ref.~\onlinecite{bustingorry_arXiv:1006.0603} for a numerical study).
The Larkin length $r_0$ at which the `thermal' and `random manifold'
regimes connect can be directly inferred
from~(\ref{eq:solution-a-b}-\ref{eq:rescaling_B-a-b}) and is remarkably \emph{independent} on
the actual values of $\zeta_{\text{th}}$ and $\zeta_{\text{RM}}$.
Indeed $r_0$ is solution of $a (b^{-1}r_0)^{2 \zeta_{\text{th}}} = a
(b^{-1}r_0)^{2 \zeta_{\text{RM}}}$ which gives $r_0=b$. One recovers the same $\frac{T^5}{c D^2}$ behaviour
of~\eqref{expr_rstar_1Dinterface} for the GVM of the 1D interface and
of~\eqref{toymodel_def_rstar} for GVM of the rounded toy model.
Similarly the value $\xi_{\text{eff}}^2\equiv B(r_0)$ of the roughness at
this length is independent of both exponents:
$\xi_{\text{eff}}=a$. This yields
$\xi_{\text{eff}}=\frac{T^3}{cD}$ as indeed obtained both in the interface~\eqref{def_width_effective} and toy
model~\eqref{toymodel_roughness_r0_high_T} GVM approaches, even though
they do not share the same value for $\zeta_{\text{RM}}$.
In the low temperature phase $T<T_c$ however,
having more than one characteristic lengthscale, the previous argument
does not apply and indeed we have observed that the low-temperature Larkin length $r_0$ differ between the two GVM predictions for the interface and the toy model.

One has to make an important observation about applying to the toy
model the scaling arguments exposed at the beginning of this section.
The Flory exponent ensuring that both parts of the Hamiltonian scale
identically is $\zeta_F^{\text{toy}}=\frac 23$ instead of~\eqref{eq:zeta_F-scaling}
($\zeta_F^{\text{toy}}=\frac 23$ as assumed in the formal argument of \aref{A-Flory}).
Applying blindly~\eqref{eq:zeta_and_thorn_RM}, one would infer that
$\text{\thorn}^{\text{toy}}=0$, whereas the thorn exponent for the toy
model has the correct value $\text{\thorn}^{\text{toy}}=-\frac 13$.
To understand this, one has to remark that~\eqref{eq:rescaling_B-b}
is true if we replace $D$ by $\widetilde D=\frac{cD}{T}$,
which induces an additional $T$-dependence
ensuring $\text{\thorn}^{\text{toy}}=-\frac 13$.
In a nutshell, we just pointed that it is the $T$-dependence of
  the effective disorder $\widetilde D$ seen by the DP endpoint in the
  toy model which allows for the Flory argument and the GVM method to
  yield the correct $\zeta_{\textnormal{RM}}$ and
  $\text{\thorn}_{\textnormal{RM}}$ exponents.
Note that nevertheless~\eqref{eq:solution-a-b}
and~\eqref{eq:rescaling_B-a-b} hold without restriction for the
toy model.


\section{Discussion} \label{discussion}

\begin{figure}[htbp]
 \subfigure[\label{fig:1Dinterface_roughness_summary} 1D interface]{\includegraphics[width=\figwidth]{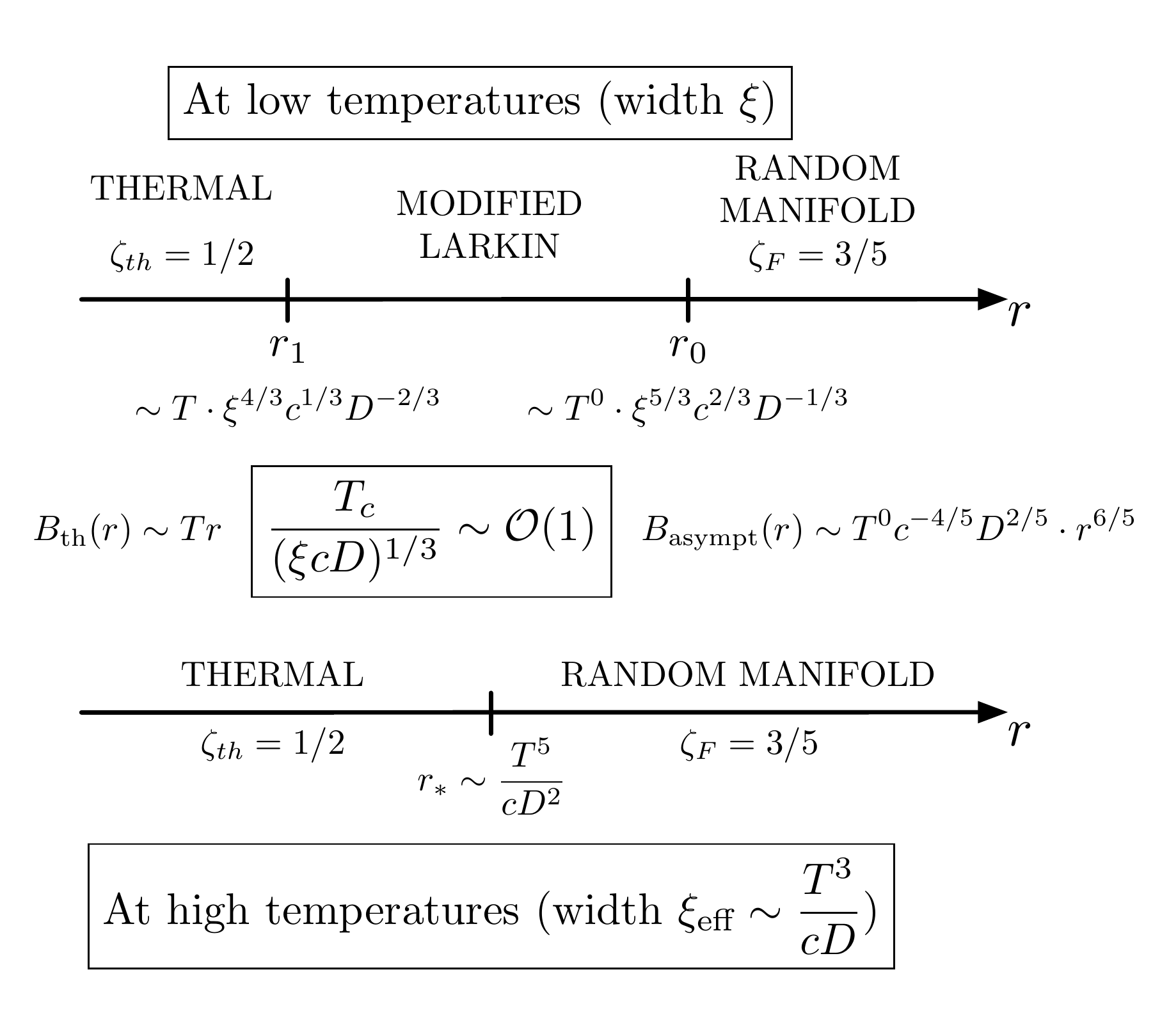}}
 \subfigure[\label{fig:toymodel_roughness_summary} Toy model]{\includegraphics[width=\figwidth]{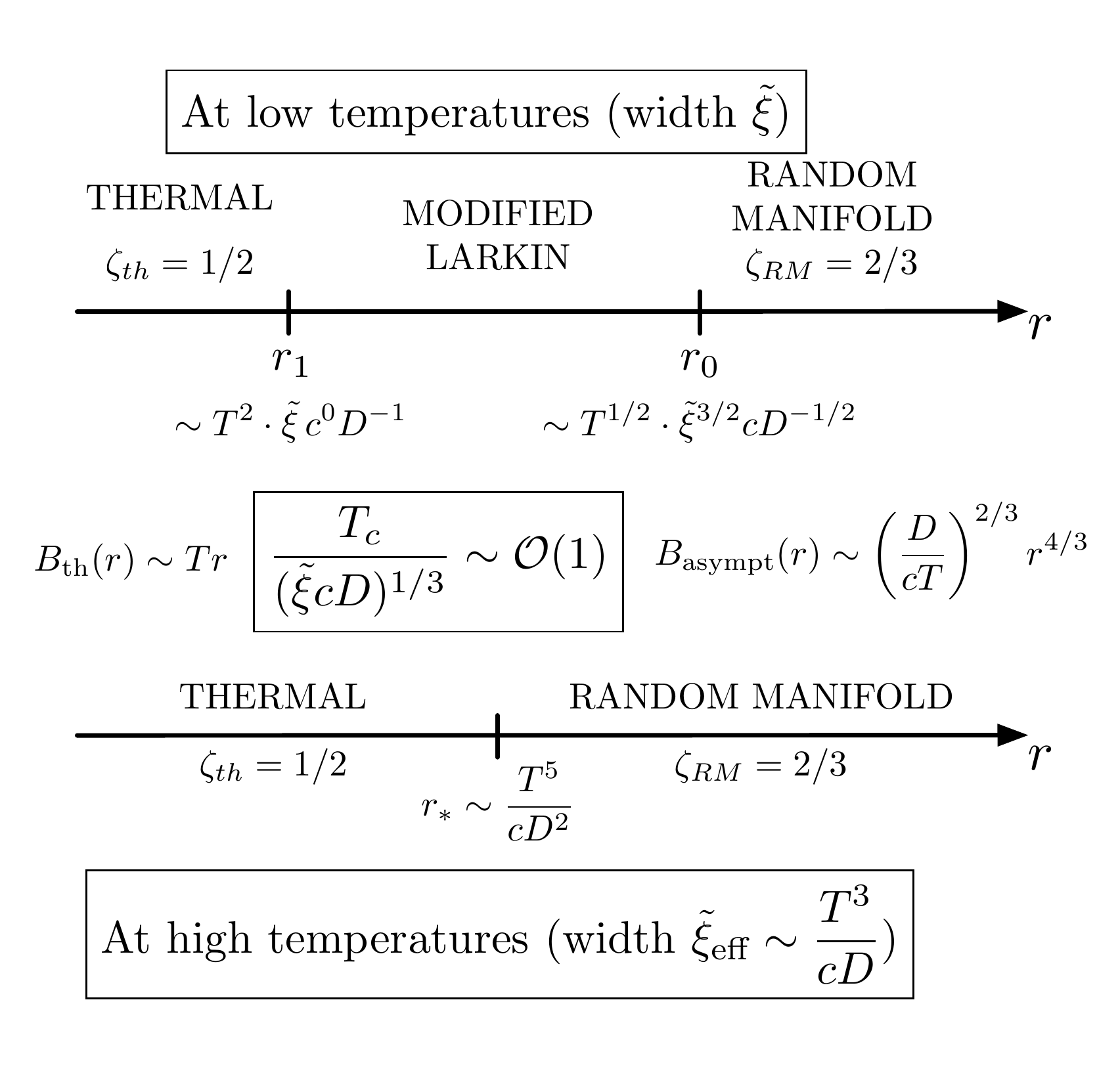}}
 \caption{Summary of the GVM predictions for the 1D-interface and the toy-model roughness about their roughness regimes and their crossover lengthscales, at low versus high temperatures.}
\label{fig:roughness_summary}
\end{figure}

\subsection{Low and high temperature regimes}
\label{sec:Discussion_LowHighT}

In the previous sections we have determined under varied approximation
schemes the expression of the roughness $B(r)$, not only in the
asymptotic regimes but also on the whole range of scales $r$.
We have kept the disorder correlation length $\xi$ finite, and it is
now time to fit the pieces of the puzzle together and analyze its
role.

An important quantity that has come to light is the characteristic
`critical' temperature (defined here disregarding numerical factors)
\begin{equation}
  T_c\equiv (\xi c D)^{1/3}
\end{equation}
for the one-dimensional interface ($d=m=1$;
see~\eqref{eq:def_Tc-scaling} for generic $d$ and $m$).  It is the
temperature at which the intermediate lengthscales $r_0$ and $r_1$
collapse --~we recall that $r_1$ marks the end of the thermal regime
while $r_0$ marks the beginning of the RM regime,
see \fref{fig:roughness_summary}.
Strikingly, the same $T_c$ holds for the
interface~\eqref{1Dinterface_collapse_r1r0} and the
toy model~\eqref{toymodel_collapse_r1r0}.
Moreover, in their corresponding variational computation $T_c$
separates the two regimes for the RSB cutoff in $[\sigma](u)$,
that is, corresponds to having the parameter $\widetilde A$ of
order unity, see Eqs.~\eqref{Atilde_1_def_Tc}
and~\eqref{toymodel_Atilde_1_def_Tc}.
The generic scaling arguments of \sref{scaling_analysis}
indicate it is certainly no coincidence
that these criteria, albeit disparate,
in fact all lead to the same $T_c$.

Let's now depict each of the temperature regime, having in mind to
compare our two models at hand (the results are summarized in
\fref{fig:roughness_summary}).

\emph{(i)} The \emph{high temperature regime} $T\gg T_c$ is
characterized by the existence of only two roughness regimes, the
thermal and the RM ones, which connect at a lengthscale
$r_*\sim\frac{T^5}{cD^2}$ common to both GVMs
(see~\eqref{expr_rstar_1Dinterface} and~\eqref{toymodel_def_rstar}).
We have seen in~\sref{scaling_analysis} how this unique
$\xi$-independent length~$r_*$ was imposed by the Hamiltonian generic
scaling properties, independently of the actual values of the
asymptotic exponent $\zeta_{\text{RM}}$.
For $r<r_*$, $B(r)\sim Tr$ is thermal, disorder plays no role and
thermal fluctuations enhance roughness.
On the other hand, for $r>r_*$ the roughness scales as $B(r)\sim
T^{2\text{\thorn}}r^{2\zeta_{\text{RM}}}$, meaning that it now
\emph{decreases} as temperature increases ($\text{\thorn}=-\frac 13$ by
scaling, see also Refs.
~\onlinecite{nattermann_1990_PhysRevB42_8577,bustingorry_arXiv:1006.0603,calabrese_2010_EPL90_20002}) --
physically, it means that thermal fluctuations inhibit the interface
excursions to spread too widely in the random potential.
We also remark that for the toy model, the roughness and thorn exponent
obtained by GVM are the exact ones ($\zeta_{\text{RM}}=\frac 23$,
$\text{\thorn}_{\text{RM}}=-\frac 13$ ) while for the interface one
gets $\zeta_{\text{asympt}}=\frac 35$ and
$\text{\thorn}_{\text{RM}}=0$.
In the high temperature regime the rounded toy model, coupled to the GVM, 
thus provides a reasonable description of the interface and its crossovers. 
It is a more efficient mapping than to use directly the GVM on the variational
Hamiltonian. 

\medskip

\emph{(ii)} In the \emph{low temperature regime} $T\ll T_c$, $\xi$ 
becomes relevant and two lengthscales $r_0$ and $r_1$ now define an
intermediate regime separating the thermal and RM ones.
The GVM has allowed us to compute the actual crossover function
of the roughness \eqref{rescaledBr} and \eqref{Br_crossoverF}, versus \eqref{toymodel_crossover_function}.
It is instructive to compare the value of $r_0$ to the Larkin length
$L_c$ obtained in the Larkin model or through FRG. Using for instance
the notation of Ref.~\onlinecite{chauve_2000_ThesePC_PhysRevB62_6241} (equation (4.17) with the
random force strength $\Delta(0)=D \xi^{-3}$ and the disorder correlation
length $r_f=\xi$) one has,
\begin{equation}
  L_c\sim \left(\frac{c^2\xi^2}{D\xi^{-3}}\right)^{1/3}
    \sim \left(\frac{c^2\xi^5}{D}\right)^{1/3}
    \label{eq:Lc_lowT}
\end{equation}
which matches exactly the low-temperature limit result
\eqref{larkin_r0_low_T} $r_0\sim \xi^{5/3} c^{2/3} D^{-1/3}$ for the
1D interface.  On the other hand, the toy model exhibits a different scaling
$r_0\sim \xi^{3/2} c D^{-1/2}T^{1/2}$,
see~\eqref{toymodel_larkin_r0_low_T}.
In the low-temperature regime, the direct GVM
results for the full 1D interface provide a better picture than the use of the rounded toy model. 
We can thus use whatever mapping is more appropriate depending on the temperature regime we are interested 
in. 

Let us emphasize that depending on their order, the limits $\xi\to 0$ and $T \to 0$ lead
to different physical regimes, since $T_c$ crucially depends on $\xi$.
The question of determining whether experimentally one lies in the low
or high temperature regime is thus of particular interest.
Note that on the numerical side, the high-$T$
regime can be probed through the directed polymer\cite{bustingorry_arXiv:1006.0603}
for which the roughness rescales as in \eqref{eq:rescaling_B-a-b}.

It is also instructive to inspect the zero temperature limit in
detail: on the one hand the toy model has again the correct
$\zeta_{\text{RM}}=\frac 23$, but since
$\text{\thorn}_{\text{RM}}^{\text{toy}}<0$ the toy model also predicts
a non-physical divergence of $B(r)$ as $T\to 0$ in the random manifold
regime, although the roughness should remain finite and not depend on
$T$, as predicted e.g. from the FRG zero-temperature fixed
point\cite{chauve_2000_ThesePC_PhysRevB62_6241}.
One may conjecture that the behaviour $B(r)\sim T^{-2/3}r^{4/3}$
remains valid for $r>r_\star$ only for a finite range of temperature,
below which the scaling would become $B(r)\sim T^{0}r^{4/3}$ with an
effective thorn exponent $\text{\thorn}_{\text{RM}}=0$.
To describe this behaviour in the toy model approach, one would need to introduce
a large scale cutoff  (possibly \emph{not} being the system size)
to prevent the roughness to diverge in the limit $T\to 0$.

\subsection{Consequences for the dynamics: the quasistatic creep regime}
\label{sec:Discussion_Creep}

Although satisfactory from a theoretical point of view, the
determination of the crossover lengths $r_0$ and $r_1$ may not yield
directly observable predictions, since in many instances the
experimental resolution for $B(r)$ is insufficient to probe such small
lengthscales.
A context however where an indirect measurement could be achieved is
that of interfaces driven by a small force $F$ (compared to the `depinning' force $F_c$). Instead of following
a mere linear response, the velocity $v(F)$ of the interface is given
by the `creep law',\cite{ioffe_vinokur_1987_JPhysC20_6149,feigelman_1989_PhysRevLett63_2303,nattermann_1990_PhysRevLett64_2454,chauve_2000_ThesePC_PhysRevB62_6241}
archetypal of glassy systems\cite{blatter_1994_RevModPhys66_1125,giamarchi_2006_arXiv:0503437}
\begin{equation}
  v(F)\sim \exp{\left[-\beta U_c\left(\frac{F_c}{F}\right)^\mu\right]}
  \label{eq:creeplaw}
\end{equation}
where $\mu=\frac{d-2+2\zeta}{2-\zeta}$ is the creep exponent,
$F_c=c\xi/L_c^2$ is a characteristic depinning force and
$U_c=c\xi^2/L_c$ an energy scale (see~Ref.~\onlinecite{giamarchi_2006_arXiv:0503437} for a review).
In the previous expressions, $\xi$ represents the effective width of
the interface, and its value depends on whether one lies in the low-
or high-temperature regime --~which is in general not known in
practice.
A way to discriminate between the two regimes is to use $U_c$ itself
(readily attainable when fitting the creep law \eqref{eq:creeplaw} on
numerical or experimental data)
since its $c,D,T$-dependence is
directly related to that of the width $\xi$ and of the Larkin length
$L_c$.

In the high temperature regime, taking $U_c=c\xi_{\text{eff}}^2/r_*$
directly leads to a linear behaviour $U_c=T$, a result at first sight
compatible e.g. with the numerical simulations of
Ref.~\onlinecite{kolton_2005_PhysRevLett94_047002} where the measured value of
$U_c$ is indeed proportional to $T$. However, the value of $F_c$ used
in Ref.~\onlinecite{kolton_2005_PhysRevLett94_047002} is the zero temperature
critical depinning force, not scaling like the characteristic force
$F_c=c\xi_{\text{eff}}/r_*^2=c^2D^3T^{-7}$.  The interpretation of the
temperature dependence of $U_c$ observed in numerical models thus
remains to be clarified.

In the low temperature regime, one has
$U_c=c\xi^2/L_c$ where $\xi$ is the microscopic width of the interface,
and $L_c$ is given by~\eqref{eq:Lc_lowT}. This yields
$U_c=(\xi c D)^{\frac 13}=T_c$
which is now temperature-independent.
In
experiments\cite{lemerle_1998_PhysRevLett80_849,metaxas_2007_PhysRevLett99_217208,repain_2004_EurPhysLett68_460}
on magnetic DWs, quantities such as the roughness or the response to
small force have been measured, and a question is to determine whether
the low temperature regime describes these results.
Before examining this, we first point out a possible caveat: it
is not clear whether the interface is at equilibrium or not; though
the measured roughness $\zeta_{\text{RM}}\approx 0.66$ is compatible
with the theoretical value $\zeta_{\text{RM}}=\frac 23$, it could also
be that the interface is stuck in a very slowly relaxing state,
reached due to the applied external field.
In that case, the measured depinning roughness exponent is also
compatible with the result $\zeta^{\text{dep}}\approx 0.63$ for an
interface with harmonic $(\nabla_z u)^2$ and non-harmonic $(\nabla_z
u)^4$ contributions\cite{kolton_2009_PhysRevB79_184207} (this value of $\zeta^{\text{dep}}$ is more
relevant for experiments than the ill-defined result
$\zeta^{\text{dep}}\approx 1.2$ of the purely harmonic elasticity).

Assuming nevertheless that the observed interface is at equilibrium we
can examine whether the analysis we have put forward applies.
A first evidence that experiments on magnetic domain wall (DW) fall
into the low-$T$ regime is that the observed value of $U/T$ actually
depends on the combination of $c$ and $\xi$ (see e.g. Table~I in Ref.~\onlinecite{metaxas_2007_PhysRevLett99_217208}
where $U/T$ is denoted $T_{\text{dep}}/T$) as opposite to high-$T$ behaviour
$U_c=T$.
In another experiment on magnetic DWs\cite{repain_2004_EurPhysLett68_460}, one
can estimate $T_c$ from the elastic constant $c\simeq 2.4\cdot
10^{-12}~\text{J~m}^{-1}$, the domain wall width $\xi\simeq
8~\text{nm}$ and the Larkin length $L_c\simeq 40~\text{nm}$ (estimated
from small force measurements). Using our expressions to compute
$T_c$, one arrives at $T_c\simeq 325\text{~K}$ which is indeed
slightly above the room temperature $T\simeq 300\text{~K}$ of the
experiment, suggesting that the interface is indeed in the low temperature regime. Further evidence could be provided by a direct evaluation of $\text{\thorn}_{\text{RM}}$ which should be zero in that case.


\section{Conclusion} \label{conclusion}

In this paper we have examined the static roughness of a one
dimensional interface subjected to a random-bond disorder.
Using a Gaussian Variational Method we have determined the various
regimes due to the finite width $\xi$ of the interface (or
equivalently to the finite correlation length of the disorder).
We obtained that at large temperature, there is one universal
$\xi$-independent crossover length from small to large lengthscales regimes,
whereas at low temperature two $\xi$-dependent crossover lengths
describe a more complex the intermediate regime. Results summarized in
\fref{fig:1Dinterface_roughness_summary} and show that the low
temperature regime is correctly tackled only if one keeps $\xi$
finite.

To discuss the scaling of the different roughness regimes, we have compared the results on the
original model with the ones on a modified `rounded' toy model, which
happens to capture the correct random-manifold exponent but turns out
to be ill-defined in the $T\to 0$ limit.  The high temperature results
match those of the interface, while at low $T$ the scalings of the
intermediate lengths differ, due to a distinct RM exponent --~see
\fref{fig:toymodel_roughness_summary} for a summary.

We have described the existence of two temperature regimes above and
below a characteristic temperature $T_c=(\xi cD)^{\frac 13}$.
On the numerical side simulations\cite{bustingorry_arXiv:1006.0603} show
for the directed polymer clear evidence supporting the existence of
a high-temperature regime with $B(r)\sim T^{2\text{\thorn}}r^{2\zeta_{\text{RM}}}$ with $\text{\thorn}=-\frac 13$.
However the observation on the experimental side of a negative thorn
exponent $\text{\thorn}$ or of a $T$-dependent Larkin length remains to be done.
One first question would be to know whether the temperature $T_c$ is
physically relevant or not.
We have estimated that in creep experiments on magnetic DWs, the value
of $T_c$ is above, but of the order of $T$, and it would be
interesting clarify the question e.g by a direct measurement of
$\text{\thorn}$ in both regimes.

The analysis we have presented also raises several interesting
questions and suggests extensions deserving further inquest: one can
wonder to what extent the variational method induces artefacts in 
the crossover lengthscales scalings we put forward, and try to investigate
this question by use of the FRG, tackling in particular the low- and
high-temperature regimes --~in other words, one would need to
reconcile the zero-temperature fixed point with the observed
$T$-dependence.
On the other hand, the modified toy model proved to be an attractive
simpler version of the problem, rich enough to encompass many correct
scalings of the interface, except for the $T\to 0$ limit. One could
for instance refine the toy model by introducing a large-scale cutoff that
would constrain the roughness to remain finite in this limit.
An exact rather than GVM solution may also be attainable.
Finally, given the success of this model in describing the static
properties, it would be interesting to ascertain how much the dynamics
of the interface could be fairly approximated by the Langevin dynamics
corresponding to the rounded toy model.

\begin{acknowledgments}
  We would like to thank Sebastian Bustingorry, Alejandro Kolton,
  Pierre Le Doussal, Markus M\"{u}ller, Alberto Rosso and Valerii
  M. Vinokur for interesting discussions.  TG would like to thank
  KITP, where the final part of this work was done, and NSF Grant
  No. NSF PHY05-51164 for support.  This work was supported in part by
  the Swiss NSF under MaNEP and Division II.
\end{acknowledgments}

\appendix

\section{`Flory-like' scaling arguments} \label{A-Flory}

Simple back-of-the-napkin power-counting leads to the so-called Flory (or Imry-Ma)
estimation for the roughness exponent: this is what we relate in that
appendix, for the interface and for its toy model.  Assuming that at
distance $z=L$ the interface presents typical excursions of extension
$u=u(L)$ in the transverse direction, we write that the elastic and disorder
contributions to the Hamiltonian scale according to
\cite{mezard_parisi_1991_replica_JournPhysI1_809}
\begin{align}
  \gdHel \argc{u} &=
  \frac{c}{2} \int_{\gdR} dz \cdot \argp{\nabla_z u(z)}^2
  \sim L^{d-2} u^2
\\
  \gdHdis \big[u,\widetilde V\big] &= \int_{\gdR} dz \cdot \widetilde{V} \argp{z,u(z)}
  \sim L^{d/2} u^{-m/2}
\end{align}
where we have used that
$
\moydes{\widetilde{V} (z,x) \, \widetilde{V}(z',x')} = \delta^{(d)} (z-z') \, R_{\xi}(x-x')\sim
L^{-d}u^{-m}
$
to determine the scaling of~$\widetilde{V}$.
Assuming that in the RM regime both
contributions scale in the same way $\gdHel\sim\gdHdis$,
we obtain
\begin{equation}
  u(L)\sim L^{\zeta_F}\ ,
\quad
  \zeta_F=\frac{4-d}{4+m}
\end{equation}
We analyze in \sref{scaling_analysis} why this Flory-like argument
does not yield the correct RM exponent
(e.g. $\zeta_F=3/5$ instead of $\zeta_{\text{RM}}=2/3$ for the 1D
interface). It is instructive to observe that on the contrary the same
argument actually \emph{works} for the toy model of~\sref{toymodel}.
%
%
Assuming now that the endpoint of the directed polymer $y(t)$ has excursions
of order $Y$ at time $t$, one gets the following scaling for the
contributions to the effective free-energy~\eqref{eq:Fyt_toymodel}
\begin{align}
  F_{\text{el}}(y,t) &= \frac{c}{2t}y^2
  \sim t^{-1}Y^2
\\
  F_{\text{dis}}(y) &= \int_0^y dy' \cdot \eta \argp{y'}
  \sim Y^{\frac{1}{2}}
\end{align}
where we have used that
$
\moydes{\eta (y) \eta (y')}
 = \widetilde{D} \cdot R_{\tilde{\xi}} (y-y')
 \sim Y^{-1}
$
to determine the scaling of $\eta$.
Imposing
$ F_{\text{el}}\sim F_{\text{dis}}$
one obtains
\begin{equation}
  Y\sim t^{\zeta_F^{\text{toy}}} \ ,
\quad
 \zeta_F^{\text{toy}}=\frac{2}{3}
\end{equation}


\section{Hierarchical matrices} \label{A-RSB_inversion}

In this appendix we recall some properties of the hierarchical matrices. Both to fix the notations and for convenience for the reader we also give the inversion formulas of replica-symmetric (RS) and full-replica-symmetry-breaking (full-RSB) Ansatz directly in the limit $n \to 0$,
using extensively the formulas and derivations of M\'ezard and Parisi in Ref.~\onlinecite{mezard_parisi_1991_replica_JournPhysI1_809,book_beyond-MezardParisi}.

The replica trick constrains the structure of the $n \times n$ matrix $G^{-1}_{ab}$ which must be symmetric and equivalent under permutation over the replica indices. The matrix has to be inverted in the limit $n \to 0$. A hierarchical matrix can be reconstructed by all the permutations of its first line, which can be chosen as the reference sequence in which the coefficients are classified monotonously. Such a generic $n \times n$ matrix $\widehat{G}$ and its corresponding inverse matrix $\widehat{G}^{-1}$ (also hierarchical) are thus defined by:
\begin{equation} \begin{split}
	\widehat{G} & \equiv \left(
		\begin{array}{ccc}
		\widetilde{G} & & G_{a \neq b} \\
		& \ddots & \\
		G_{a \neq b} & & \widetilde{G}
		\end{array}
		\right) \\
	& \quad \Longleftrightarrow \quad
	\widehat{G}^{-1} \equiv \left(
		\begin{array}{ccc}
		\widetilde{G}^{-1} & & G_{a \neq b}^{-1} \\
		& \ddots & \\
		G_{a \neq b}^{-1} & & \widetilde{G}^{-1}
		\end{array}
		\right)
\end{split}
\end{equation}
or in the more compact way:
\begin{equation}
	\widehat{G} \equiv \argp{\widetilde{G}, G_{a \neq b}}
	\Longleftrightarrow
	\widehat{G}^{-1} \equiv \argp{\widetilde{G}^{-1}, G^{-1}_{a \neq b}}
\end{equation}
where $G_{aa} = \widetilde{G}$ and $G^{-1}_{aa} = \widetilde{G}^{-1}$ $\forall a$.

\begin{figure}[htbp]
 \begin{center}
 \subfigure[$k=2^1-1=1$]{\includegraphics[width=0.45\columnwidth]{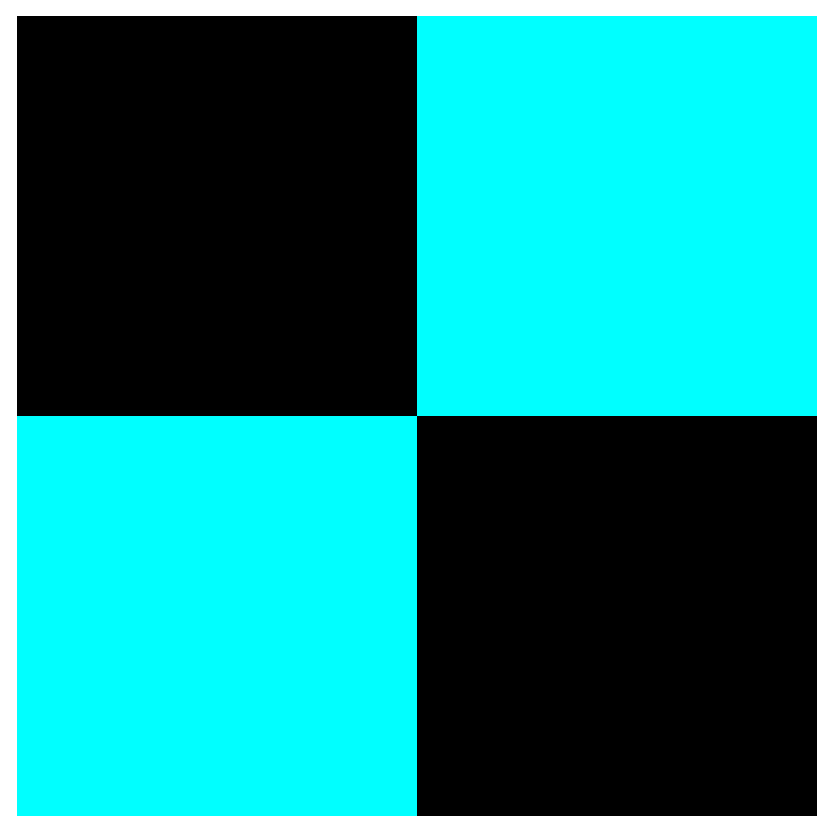}}
 \subfigure[$k=2^2-1=3$]{\includegraphics[width=0.45\columnwidth]{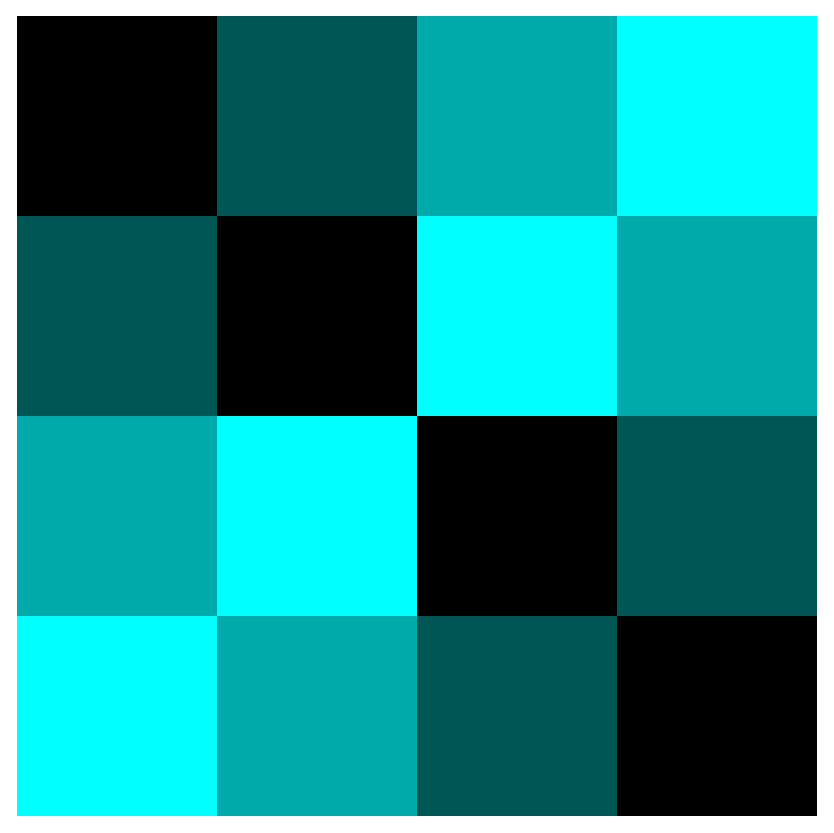}}
 \subfigure[$k=2^6-1=63$]{\includegraphics[width=0.45\columnwidth]{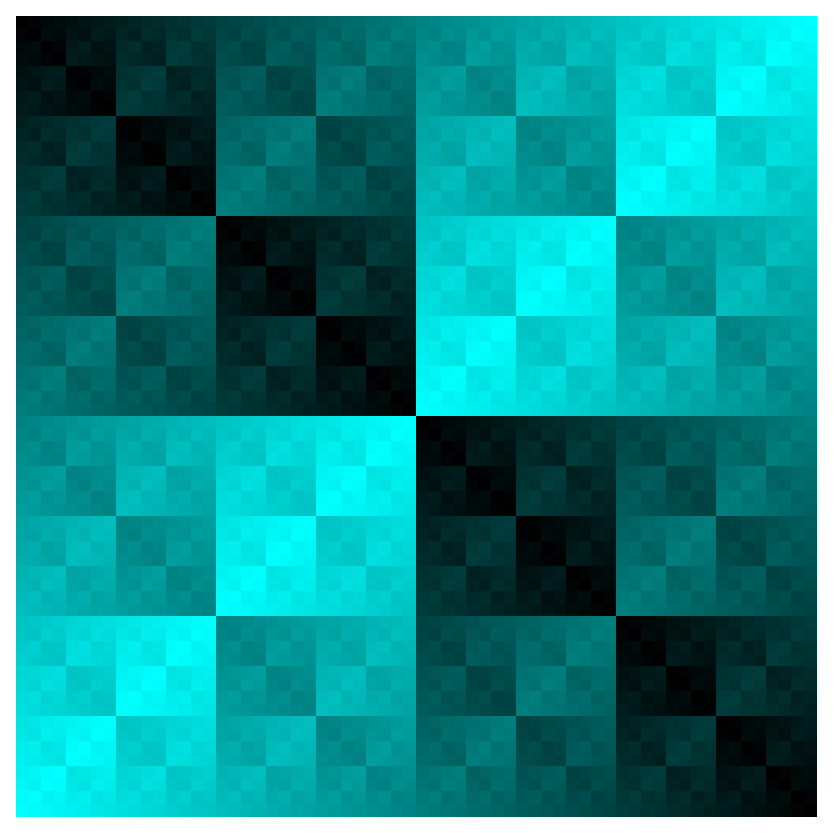}}
 \caption{\label{fig-eg-hierarchical-matrix}
	Examples of hierarchical matrices of $n \times n$ blocks, with increasing integer $k$ of replica-symmetry breaking. Each shade of color corresponds to a different value for the coefficients of the matrices.}
 \end{center}
\end{figure}

This allows the definition of the \emph{connected} part of these matrices, i.e. the sum of the coefficients on any line or column $G_c = \sum_a G_{ab} = \sum_b G_{ab}$, a conserved quantity which satisfies for two inverse matrices the relation
\begin{equation}
 \label{A_inversion_connected_part}
	G_c \cdot G^{-1}_c =1
\end{equation}

The simplest case of a hierarchical matrix is the \emph{replica-symmetric} (RS) Ansatz, in which all the off-diagonal coefficients of the matrix are equal:
\begin{equation}
	\widehat{G}^{-1} \equiv \argp{\widetilde{G}^{-1}, G^{-1}}
	\Longleftrightarrow
	\widehat{G} \equiv \argp{\widetilde{G}, G}
\end{equation}
and in the limit $n \to 0$:
\begin{equation}
 \label{A_RS-inversion-formulas}
 \begin{split}
	G^{-1}_c \equiv \widetilde{G}^{-1} - G^{-1}
	\, , \quad
	G_c \equiv\widetilde{G} - G \\
	G = - \frac{G^{-1}}{\argp{G_c^{-1}}^2}
	\, , \quad
	\widetilde{G} = \frac{1}{G^{-1}_c} \argp{1 - \frac{G^{-1}}{G^{-1}_c}}
 \end{split}
\end{equation}

If the off-diagonal terms count at least two different values $\arga{g_0,\dots,g_k}$, we have a \emph{replica-symmetry breaking} (RSB) Ansatz, the integer $k$ counting the number of such breakings (see \fref{fig-eg-hierarchical-matrix} for generic examples of this structure). The integer $n$ being arbitrarily large, in the limit $k \to \infty$ the monotonous sequence of coefficients on the first line of the matrix is more adequately described by a monotonous function $G(u)$, depending on a mapping parameter $u \in \argc{0,1}$. Thus a \emph{full RSB} hierarchical matrix is defined by
\begin{equation}
	\widehat{G} = \argp{\widetilde{G}, G(u)} \text{ with } u \in \argc{0,1}
\end{equation}
The full-RSB Ansatz is actually the most generic description of a hierarchical matrix (in regards to the limit $n \to 0$), since the replica-symmetric and $k$-RSB Ansatz can be recovered using step-functions for $G(u)$. The peculiar symmetries of hierarchical matrices allow to determine generic inversion formulas directly in the limit $n \to 0$\cite{mezard_parisi_1991_replica_JournPhysI1_809}, once the first line of the matrix is given. Thereafter they have been adapted to the following definition of full-RSB hierarchical matrices:
\begin{equation}
	\widehat{G}^{-1} (q) \equiv \argp{G_c^{-1}-\tilde{\sigma}, -\sigma (u)}
	\Longleftrightarrow \widehat{G} (q) \equiv \argp{\widetilde{G} (q), G(q,u)}
\end{equation}
$\sigma(u)$ being defined as a monotonous function on the interval $\argc{0,1}$, the discrete sums of matrix operations are replaced by integrals. Note that the definition of the connected part $G^{-1}_c$ implies that
\begin{equation}
 \label{A_sigma_tilde}
	\tilde{\sigma} = - \int_0^1 du \cdot \sigma (u)
\end{equation}

\begin{figure}[htbp]
 \includegraphics[width=\columnwidth]{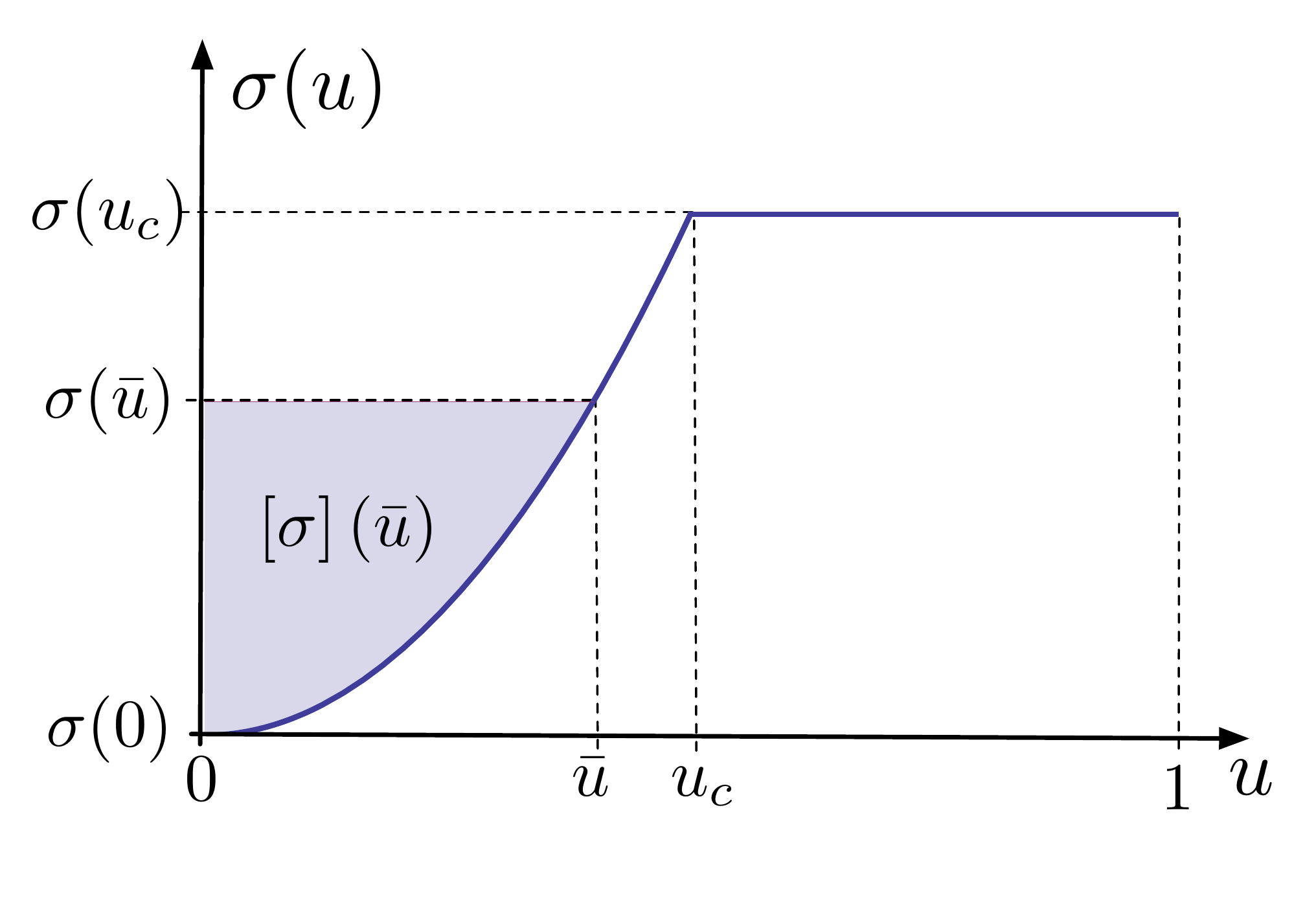}
 \caption{\label{fig_sigma_crochet_generic}
	The mass term $\sigc{\bar{u}}$ corresponds to the shaded area between $\sigma (\bar{u})$ and the $\sigma (u)$ curve, and thus saturates at the cutoff $u=u_c$.}
\end{figure}

We define the following self-energy, illustrated in \fref{fig_sigma_crochet_generic}:
\begin{equation}
 \label{A_replic_sigc}
	\sigc{u} \equiv  u \cdot \sigma (u) - \int^u_0 dv \cdot \sigma (v)
\end{equation}
This definition implies in particular that $\argc{\sigma}' (u) = u \cdot \sigma ' (u)$.

This self-energy acts in fact as a mass term in the propagators of the following inversion formulas:
\begin{align}
 \label{A_replic_Gt}
	& \widetilde{G}
	= \frac{1}{G^{-1}_c} \argp{ 1 + \int_0^1 \frac{dv}{v^2} \cdot \frac{\sigc{v}}{G^{-1}_c + \sigc{v}} + \frac{\sigma (0)}{G^{-1}_c} } \\
 \label{A_replic_Gt-Gu1}
	& \widetilde{G} - G(u)
	= \frac{1}{u} \cdot \frac{1}{G^{-1}_c + \sigc{u}} - \int^1_u \frac{dv}{v^2} \cdot \frac{1}{G^{-1}_c +\sigc{v}} \\
 \label{A_replic_Gt-Gu2}
	& \widetilde{G} - G(u)
	= \frac{1}{G^{-1}_c + \sigc{1}} + \int^1_u dv \cdot \frac{\sigma ' (v)}{\argp{G^{-1}_c + \sigc{v}}^2} 	
\end{align}
The relation \eqref{A_replic_Gt-Gu2} can be obtained from \eqref{A_replic_Gt-Gu1} by a simple integration by parts.

From the full-RSB point of view the RS Ansatz is the particular case when $\sigma (u)$ is a constant:
\begin{equation}
	\sigma (u) = \sigma_0 \, \forall u \in \argc{0,1}
	\Rightarrow\sigc{u}=0 \, \forall u \in \argc{0,1}
\end{equation}
in which case the previous inversion formulas collapse indeed on the RS case \eqref{A_RS-inversion-formulas}.

In the context of a GVM in a Fourier-space representation (as the one described in \sref{fullGVM}), those inversion formulas have still to be integrated over the Fourier modes $q$ in order to deal with the saddle-point equation or the computation of the roughness. Inverting conveniently the order of integration over the Fourier modes and over the RSB parameter $u$, and pushing the ultra-violet cutoff to $\infty$, the following identities for propagators are useful ($A >0$ is typically a self-energy $\sigc{u}$) for the case of a 1D interface:
\begin{align}
 \label{A_propagator1}
	\int_{\gdR} \dbar q \cdot \frac{1}{c q^2 + A}
	&= \frac{A^{-1/2}}{2 \sqrt{c}} \\
 \label{A_propagator2}
	\int_{\gdR} \dbar q \cdot \frac{1}{\argp{c q^2 +A}^2}
	&= \frac{A^{-3/2}}{4 \sqrt{c}}
\end{align}
and
\begin{equation}
 \label{A_propagator_cos}
 \begin{split}
	\int_{\gdR} \dbar q \cdot & \frac{2 \argp{1- \cos \argp{qr}}}{c q^2 \argp{c q^2 +A}} \\
	&= \frac{1}{A \sqrt{A c}} \argp{e^{-\sqrt{A/c}\cdot r} -1 + \sqrt{A/c} \cdot r} \\
	&= \frac{1}{A \sqrt{A c}} \sum_{k=2}^{\infty} \frac{\argp{- \sqrt{A/c} \cdot r}^k}{k !}
 \end{split}
\end{equation}
at first for $\int_{\gdR} \dbar q \cdot \argp{\widetilde{G} (q) - G(q,u)}$ in the manipulation of the saddle-point equation and secondly for $\int_{\gdR} \dbar q \cdot \argp{1- \cos (qr)} \cdot \widetilde{G} (q)$ in the computation of the roughness.

For the sake of completeness here is finally the formula for the trace of hierarchical matrices in the limit $n \to 0$, adapted from (AII.11) of Ref.~\onlinecite{mezard_parisi_1991_replica_JournPhysI1_809} to our conventions (with $\argc{G} (u)$ being defined similarly to $\sigc{u}$ \eqref{A_replic_sigc}):
\begin{equation}
 \begin{split}
	&\lim_{n \to 0} \frac{1}{n} \text{Tr} \log \widehat{G}^{-1} \\
	&= \log \argp{G^{-1}_c} - \frac{\sigma (0)}{G^{-1}_c} - \int^{1}_{0} \frac{du}{u^2} \log \argp{\frac{G^{-1}_c + \sigc{u}}{G^{-1}_c}}
 \end{split}
\end{equation}
\begin{equation}
 \begin{split}
	&\lim_{n \to 0} \frac{1}{n} \text{Tr} \log \widehat{G} \\
	&= \log \argp{G_c} + \frac{G (0)}{G_c} - \int^{1}_{0} \frac{du}{u^2} \log \argp{\frac{G_c - \argc{G} (u)}{G_c}}
 \end{split}
\end{equation}
The quantity $\frac{1}{n} \text{Tr} \log \widehat{G}$ is actually to be used in the GVM for the computation of the free energies $\gdF_0$ and $\gdF_{\text{var}}$ associated to a trial Hamiltonian $\gdH_0$, see Appendix C in Ref.~\onlinecite{giamarchi_ledoussal_1995_PhysRevB52_1242} for an example.



\end{document}